\documentclass{IEEE-Con-Sys-mag}
\pdfoutput=1

\title{Formal Verification and Control with Conformal Prediction\stitle{Practical Safety Guarantees for Autonomous Systems}}
\author{\hspace{-0.1cm} Lars Lindemann, Yiqi Zhao, Xinyi Yu, George J. Pappas, and Jyotirmoy V. Deshmukh\\}
\affil{\hspace{-0.1cm} L. Lindemann (\href{mailto:llindemann@ethz.ch}{llindemann@ethz.ch}) is with the Automatic Control Laboratory, ETH Zürich, Zürich, Switzerland. \\
Y. Zhao (\href{mailto:yiqizhao@usc.edu}{yiqizhao@usc.edu}), X. Yu (\href{mailto:xinyi.yu12@usc.edu}{xinyi.yu12@usc.edu}), and J. V. Deshmukh (\href{mailto:jdeshmuk@usc.edu}{jdeshmuk@usc.edu}) are with the Thomas Lord Department of Computer Science, University of Southern California, Los Angeles, CA, USA.\\
G. J. Pappas (\href{mailto:pappasg@seas.upenn.edu}{pappasg@seas.upenn.edu}) is with the  Department of Electrical and Systems Engineering, University of Pennsylvania, Philadelphia, PA, USA.}

\usepackage{amsthm}
\usepackage{amssymb}
\usepackage{amsmath}
\usepackage{mathrsfs}
\usepackage{hyperref}
\usepackage{makecell}
\usepackage{mathtools}
\usepackage{slashbox}
\allowdisplaybreaks

\usepackage[most]{tcolorbox}
\usepackage{graphicx}
\hypersetup{
    unicode=false,          
    pdftoolbar=true,        
    pdfmenubar=true,        
    pdffitwindow=false,     
    pdfstartview={FitH},    
    pdfnewwindow=true,      
    colorlinks=true,       
    linkcolor=blue,          
    citecolor=blue,        
    filecolor=magenta,      
    urlcolor=red,           
    breaklinks=true
}
\usepackage{cleveref}
\usepackage{xcolor}
\usepackage[sort, numbers, compress]{natbib}
\usepackage{here}
\usepackage{caption}
\usepackage{subcaption}
\usepackage{graphicx}

\usepackage{tikz}
\usetikzlibrary{shapes.geometric, arrows}
\tikzstyle{arrow} = [thick,->,>=stealth]
\tikzstyle{arrow_nohead} = [thick,-,>=stealth]
\tikzstyle{startstop} = [rectangle, rounded corners, minimum width=1cm, minimum height=2.5cm,text centered, text width=3cm, draw=black, fill=red!30]

\usepackage{tikz}
\usetikzlibrary{automata, positioning, arrows}
\tikzset{
->, 
>=latex,
node distance=3.5cm, 
every state/.style={thick, fill=blue!15}, 
initial text=$ $, 
}

\newcommand{\spacing}{\vspace{0.1 cm}}

\newtheorem{theorem}{Theorem}
\newtheorem{lemma}{Lemma}
\newtheorem{assumption}{Assumption}

\newtheorem{example}{Example}
\newtheorem{corollary}{Corollary}
\newtheorem{remark}{Remark}
\newtheorem{probl}{Problem}

\newtheorem{problem}{Sidebar}

\begin{document}

\maketitle

\begin{figure*}
    \centering
        \includegraphics[scale=0.25]{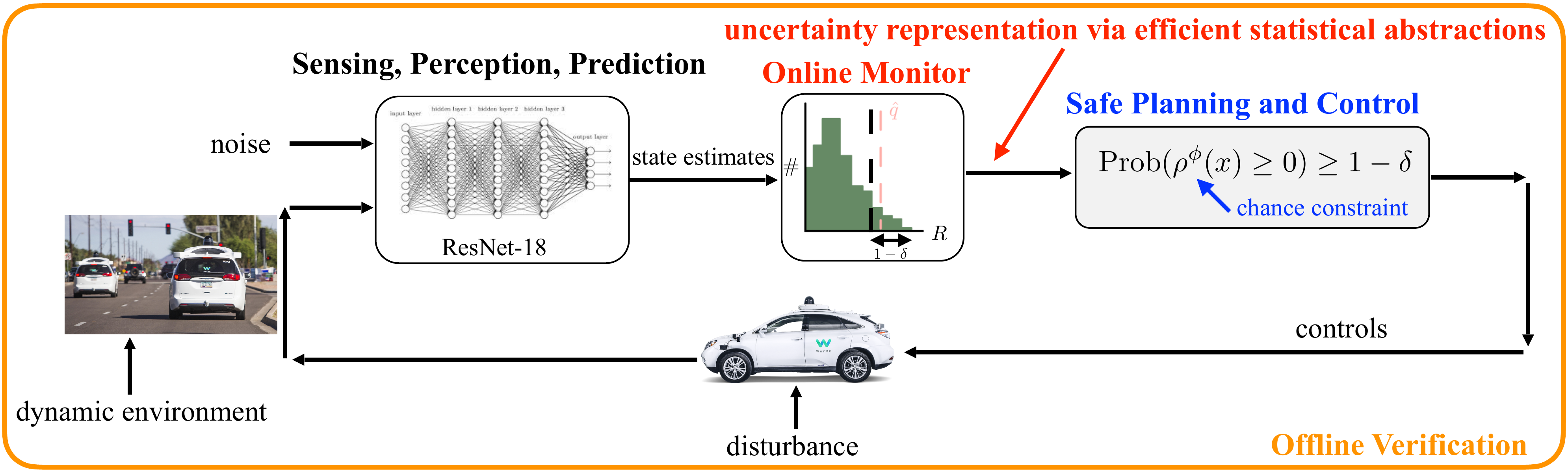}
        \caption{Formal verification and control of LEASs. In this survey, we show how to design offline verification algorithms and how to integrate online monitors obtained via efficient statistical abstractions into the control design.  }
        \label{fig:idea}
\end{figure*}

\section{Introduction}

\hspace{0.5cm}
\chapterinitial{T}he design of autonomous systems, which become increasingly learning-enabled, has attracted much attention within the research community. Research in this area promises to enable many future technologies such as autonomous driving, intelligent transportation, and robotics. Over the past years, much progress was made in the design of learning-enabled components (LECs), e.g., with neural networks  for perception tasks such as object detection \cite{zhao2019object,liu2020smoke}, localization and state estimation \cite{pavlakos2018learning,bowman2017probabilistic}, trajectory prediction \cite{hochreiter1997long,vaswani2017attention,alahi2016social}, for decision-making tasks such as motion and behavior planning \cite{liu2020robot,mnih2013playing}, and for low-level control \cite{arulkumaran2017deep,coulson2019data,robey2020learning}. However, the integration of LECs into safety-critical autonomous systems is limited by their fragility and can result in unsafe  behavior, e.g., inaccurate and non-robust object detectors in self-driving cars. The fragility of LECs is a result of highly nonconvex learning problems, distribution shifts from training to deployment domain, and lack of model robustness \cite{szegedy2013intriguing,goodfellow2014explaining}. Unfortunately, these safety challenges are further amplified by the  complexity of modern autonomous systems that operate in uncertain and dynamic environments where traditional approaches for localization and mapping may fail to provide guarantees, e.g., simultaneous localization and mapping (SLAM) techniques \cite{durrant2006simultaneous,bowman2017probabilistic} or Kalman/particle filters  \cite{wan2000unscented,djuric2003particle,simon2006optimal}.

For these reasons, we require verifiable frameworks for the  integration of LECs into the perception-action loops of autonomous systems, ultimately resulting in safe learning-enabled autonomous systems (LEASs).  Towards this goal, this survey focuses on designing  \textbf{formal verification and control algorithms for learning-enabled autonomous systems with practical safety guarantees using conformal prediction}. By practical, we mean that these algorithms:  (1) are applicable and scale to complex LEASs, (2) come with formal  guarantees, and (3) are easy to understand and extend, even for a novice in the field.

Conformal prediction is a lightweight statistical tool for uncertainty quantification, originally introduced by Shafer and Vovk  in \cite{shafer2008tutorial,vovk2005algorithmic}, which has recently attracted  attention within the machine learning community, see e.g., \cite{angelopoulos2021gentle,fontana2023conformal} for up-to-date tutorials.  In this survey, we use conformal prediction for designing safe controllers, for integrating online monitors into the perception-action loops of LEAS, and for performing offline verification, see Figure \ref{fig:idea}.

The main objective behind conformal prediction and uncertainty quantification can be summarized as follows. 
 \begin{tcolorbox}[
 colframe=red!70!white,
 colback=red!9!white,
 arc=8pt,
 breakable,
 left=1pt,right=1pt,top=1pt,bottom=1pt,
 boxrule=0.3pt,
 ]
\textbf{Uncertainty Quantification with Conformal Prediction.} Given $K+1$ independent and identically distributed (or exchangeable) random variables $R^{(0)}, R^{(1)},\hdots,R^{(K)}$, conformal prediction aims to quantify the uncertainty of $R^{(0)}$ based on $R^{(1)},\hdots,R^{(K)}$. Formally, given a failure probability $\delta\in (0,1)$, we want to construct a probabilistically valid prediction region $C:\mathbb{R}^K\to\mathbb{R}$ so that
\begin{align}\label{eq:CP_goal}
    \text{Prob}(R^{(0)}\le C(R^{(1)},\hdots,R^{(K)}))\ge 1-\delta.
\end{align}
 \end{tcolorbox}

The variable $R^{(i)}$ for $i\in\{0,\hdots,K\}$ is usually referred to as the nonconformity score. In regression, it may be defined as the prediction error $R^{(i)}:=\|y^{(i)}-\mu(u^{(i)})\|$ where a predictor $\mu$ attempts to predict an output  $y^{(i)}$ based on an input $u^{(i)}$.  A small (large) nonconformity score thus indicates a good (poor) predictive model. A larger part of our discussion will be on constructing informative nonconformity scores for verification and control. In practice, with equation \eqref{eq:CP_goal} we aim to find a probabilistic upper bound of a test datapoint $R^{(0)}$ from a calibration dataset $R^{(1)},\hdots,R^{(K)}$. Conformal prediction provides a simple and efficient procedure to compute the upper bound $C(R^{(1)},\hdots,R^{(K)})$ that is valid with a confidence of $1-\delta$. For simplicity, we will often simply use $C$ in the remainder and  omit the function arguments $R^{(1)},\hdots,R^{(K)}$.

\textbf{Goals.} In this survey, we leverage the conformal prediction guarantees in equation \eqref{eq:CP_goal} to achieve the following four objectives:
\begin{enumerate}
    \item designing verification algorithms for LECs to reason about input-output properties of LECs, 
    \item designing control algorithms for LEASs that come with probabilistic safety guarantees,
    \item designing offline verification algorithms for LEASs  to verify safety requirements of LEASs, and 
    \item designing online verification algorithms for LEAS that predict system failures with high confidence.
\end{enumerate}

Before starting with the main parts of this survey paper, we refer the reader to Sidebar \textbf{Spectrum of Formal Verification and Control Techniques} where we formally distinguish offline and online approaches to verification and control and discuss computational bottlenecks.

\begin{sidebar}{Spectrum of Formal Verification and Control Techniques}
\section[Spectrum of Formal Verification and Control Techniques]{}\phantomsection
   \label{sidebar-proof-CP}
\setcounter{sequation}{0}
\renewcommand{\thesequation}{S\arabic{sequation}}
\setcounter{stable}{0}
\renewcommand{\thestable}{S\arabic{stable}}
\setcounter{sfigure}{0}
\renewcommand{\thesfigure}{S\arabic{sfigure}}
The control community distinguishes between open-loop and closed-loop controllers. An open-loop controller computes a sequence of control inputs offline before the system is  deployed. On the other hand, a closed-loop (or feedback) controller computes control inputs from online information when the system is operating. Open-loop controllers are usually not used for systems that  require the controller to be robust against uncertainty and noise. Instead, a combination of offline (often referred to as planning) and online control is used \cite{belta2007symbolic,matni2024quantitative,laumond1998robot}. At each control level, the designer has to carefully trade off the \textbf{computational complexity} of the control problem to be solved with the \textbf{model fidelity} (abstract vs. concrete physical models) as well as the \textbf{task complexity} (high-level specifications vs. motion primitives). In safety-critical control, another trade-off arises between the type of \textbf{formal safety guarantee} one is interested in (e.g., deterministic, probabilistic, robust) and model fidelity as well as  task complexity. Another  distinction can be made between model-based control design, e.g., model predictive control \cite{garcia1989model,chen1998quasi,morari1999model} or robust control \cite{zhou1998essentials}, and data-driven control techniques, e.g., data-driven model predictive control \cite{coulson2019data,berberich2020data} or scenario optimization \cite{calafiore2006scenario,campi2009scenario}. In many ways, however, these lines are (and have always been) blurred, see \cite{dorfler2023datab,dorfler2023dataa} for more references and a detailed discussion.

Less familiar to the control community but well known within the formal methods community,  one can make similar analogies in verification. Broadly speaking, verification aims at checking the correctness of a system against a system specification expressed in a mathematical logic, e.g.,  in predicate or temporal logics (see Sidebars \textbf{LEC Specifications in Predicate Logic} and \textbf{LEAS Specifications in Signal Temporal Logic}).  Offline verification (which we simply refer to as verification in the remainder) aims at verifying the system as a whole by exhaustively checking all possible system behaviors. For this purpose, automated verification tools were developed, e.g.,  model checking \cite{baier,clarke1997model} or theorem proving \cite{shoukry2017smc,sheeran2000checking}. Motivated by the computational challenges that these methods face, especially when applied to learning-enabled autonomous systems,  statistical verification techniques were proposed as an efficient alternative that allows checking a $1-\delta$ fraction of the system behaviors \cite{younes2006statistical,legay2019}. In online verification, in contrast, one has already observed a partial execution of the system, e.g., a partial trajectory of a car navigating traffic, and our task is to check if all possible future behaviors satisfy the specification \cite{bauer2011runtime,leucker2009brief,deshmukh2017robust}. Online verification is often also referred to as runtime verification. In online verification, the verification answer can be inconclusive, i.e., neither satisfied nor violated.  Predictive online verification algorithms instead  predict future system behaviors and can thus provide verification answers more reliably and quickly \cite{babaee2018mathcal,yoon2019predictive}. Ultimately, as for control design, a combination of offline and online verification should be used in practice.

In the illustration below, we present an incomplete (and potentially biased) set of verification and control techniques. We list these techniques according to their scalability and verifiability properties. We specifically delineate between techniques that can (or have been) applied to learning-enabled autonomous systems. Conformal prediction falls into this category.
\begin{minipage}{.47\textwidth}
    \vspace{0.25cm}
        \includegraphics[scale=0.3]{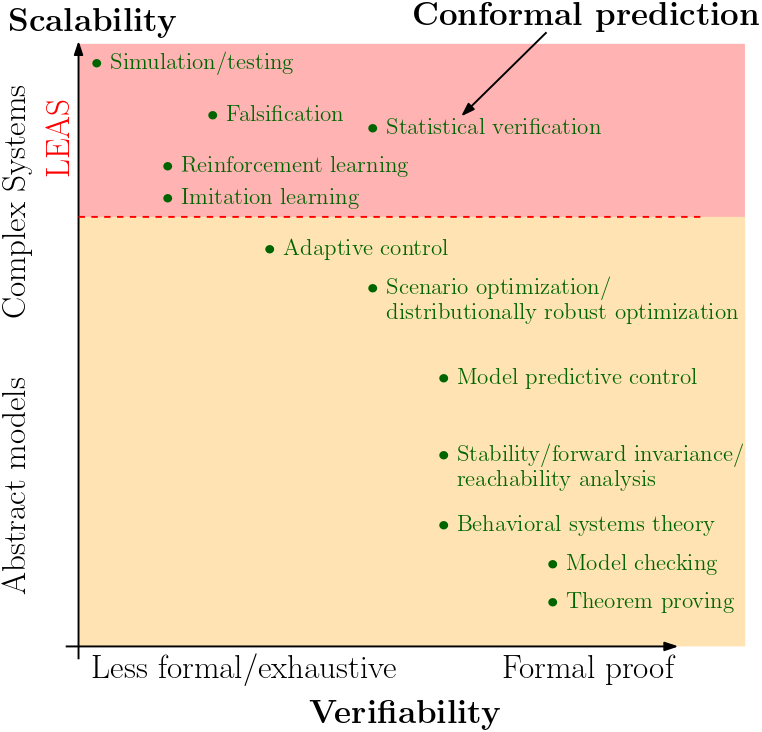}
\end{minipage}
\end{sidebar}

\begin{summary}
\summaryinitial{W}e present recent advances in formal verification and control for autonomous systems with practical safety guarantees enabled by conformal prediction (CP), a statistical tool for uncertainty quantification. This survey is particularly motivated by learning-enabled autonomous systems (LEASs), where the complexity of learning-enabled components (LECs) poses a major bottleneck for applying traditional model-based verification and control techniques. To address this challenge, we advocate for CP as a lightweight alternative and demonstrate its use in formal verification, systems and control, and robotics. CP is appealing due to its  simplicity (easy to understand, implement, and adapt), generality (requires no assumptions on learned models and  underlying data distributions), and efficiency (real-time capable and accurate). 

This survey provides an  accessible introduction to CP for non-experts  interested in applying CP to autonomy problems. We particularly show how CP can be used for formal verification of LECs and the design of safe control as well as offline and online verification algorithms for LEASs. We present these techniques within a unifying framework that addresses the complexity of LEASs. Our exposition spans simple specifications, such as robot navigation tasks, to complex mission requirements expressed in temporal logic. Throughout the survey, we contrast CP with other statistical techniques, including scenario optimization and PAC-Bayes theory, highlighting  advantages and limitations for verification and control. Finally, we outline open problems and promising directions for future research.
\end{summary}

\subsection{Related Survey Articles}
 Survey articles on the topic of conformal prediction were presented in \cite{angelopoulos2021gentle,shafer2008tutorial,fontana2023conformal}. However, the material in these articles is presented through the lens of a statistician and without reference to the use of conformal prediction in formal verification and control as we do. 

There exist various survey articles at the intersection of machine learning, formal verification, and control that are related to the topic of this survey. The authors in  \cite{ziemann2023tutorial} survey statistical methods for system identification of linear systems with finite sample guarantees. Utilizing such system identification methods, the same authors present a survey article on statistical learning theory for control design in \cite{tsiamis2023statistical}. In contrast to \cite{ziemann2023tutorial,tsiamis2023statistical}, however, our focus is more broadly on formal verification and control design of complex LEASs. This includes reasoning over LECs, e.g., for perception and prediction, dynamic and uncertain environments, and it requires the use of formal specification languages and design techniques. In this regard, we point the reader to the brief survey in \cite{seshiatoward} where the need of formal methods in verifiable autonomy is stressed and grand challenges are discussed.  Similarly to \cite{ziemann2023tutorial,tsiamis2023statistical}, we are interested in obtaining finite sample guarantees, but here by using conformal prediction which can provide both marginal and calibration conditional coverage guarantees (details  follow in the next section). Also related to our article is the survey in \cite{ganai2024hamilton} that reviews learning-based Hamilton-Jacobi reachability techniques for safe control design. In contrast, we are here concerned about dealing with the complexity of LECs within an LEAS and the associated verification and control challenges. The authors in \cite{yin2024formal} provided a detailed survey of formal control synthesis techniques for safety-critical autonomous systems; however, without a focus on using statistical methods for LEAS as we do. The survey article in \cite{cairoli2023learning} focuses on learning-enabled predictive online verification techniques using conformal prediction, similar to Section \textbf{Online Verification of LEAS with Conformal Prediction} in this survey. Our focus is again broader in scope considering offline and online formal verification as well as control design techniques for LEASs. Additionally, the online predictive verification techniques presented here are different in flavor than the ones surveyed in \cite{cairoli2023learning} (details in Section \textbf{Online Verification of LEAS with Conformal Prediction}). Lastly, we mention \cite{kapinski2016simulation,corso2021survey} which are survey articles on simulation-based approaches for autonomous system verification. We note that  \cite{kapinski2016simulation} provides a short survey advocating for a data-centric approach as we do, but that we provide an in-depth treatment of the topic with a strong focus on techniques and algorithms along with providing formal guarantees via the use of statistical tools. In contrast to \cite{corso2021survey}, which is similar in scope to what we present, the reader will find a different and complementary set of techniques and algorithms in our survey article, e.g., conformal prediction guarantees  and compositional verification and control (recall Figure \ref{fig:idea}).

Different from the coverage guarantees in equation~\eqref{eq:CP_goal} that we pursue here, risk metrics have appeared as an alternative to capture distinct properties of a random variable, e.g., the  tail behavior of a random variable that encodes the loss of a system. The use of risk metrics in robotics was advocated for in \cite{majumdar2020should}, and their use in verification and control was surveyed in \cite{wang2022risk} and \cite{akella2024risk}.

\section{Conformal Prediction In a Nutshell}\label{sec:nutshell}
Let  $R^{(0)},\hdots,R^{(K)}$ be $K+1$ independent and identically distributed random variables that follow a distribution $\mathcal{R}$, i.e., that are such that $R^{(0)},\hdots,R^{(K)}\sim \mathcal{R}$. We note that the theory of conformal prediction also holds, and in fact was originally invented, for exchangeable random variables $R^{(0)},\hdots,R^{(K)}$.\footnote{The random variables $R^{(0)},\hdots,R^{(K)}$ are exchangeable if the joint distribution of $R^{(0)},\hdots,R^{(K)}$ is equivalent to the joint  distribution of $R^{(\sigma(0))},\hdots,R^{(\sigma(K))}$ for any permutation $\sigma$ on $\{0,\hdots K\}$. It is easy to see that independent and identically distributed random variables are exchangeable, but not vice versa.} For simplicity, and as not  needed in this survey article, we avoid explicitly stating the underlying probability space of $R^{(0)},\hdots,R^{(K)}$. 

\subsection{Vanilla Conformal Prediction}
Based on a simple quantile argument, we can compute a probabilistic prediction region as in equation \eqref{eq:CP_goal}, i.e., we can compute an upper bound $C$ for $R^{(0)}$ from calibration data $R^{(1)},\hdots,R^{(K)}$ that holds with high confidence. We proceed as follows: for a given failure probability $\delta\in(0,1)$, we obtain the upper bound  $C:\mathbb{R}^K\to\mathbb{R}$ as
\begin{align}\label{eq:C_computation}
    C(R^{(1)}, \hdots, R^{(K)}):=\text{Quantile}_{1-\delta}( R^{(1)}, \hdots, R^{(K)}, \infty ),
\end{align}
which is the $(1-\delta)$th quantile (formally defined in equation \eqref{eq:emp_quantile}) over the empirical distribution of the values $R^{(1)},\hdots,R^{(K)}$ and $\infty$, where $\infty$ is added as a correction to obtain finite sample guarantees. While the definition of the $(1-\delta)$th quantile $\text{Quantile}_{1-\delta}( R^{(1)}, \hdots, R^{(K)}, \infty )$ in equation \eqref{eq:emp_quantile} appears difficult to parse at first, it turns out that we can efficiently compute $\text{Quantile}_{1-\delta}( R^{(1)}, \hdots, R^{(K)}, \infty )$. Therefore, define $R^{(K+1)}:=\infty$ and assume, without loss of generality, that $R^{(1)},\hdots,R^{(K)}$ are sorted in non-decreasing order, i.e., by re-ordering. We  now  obtain $C=R^{(p)}$ with $p:=\lceil (K+1)(1-\delta)\rceil$ where $\lceil \cdot\rceil$ is the ceiling function, i.e., $C$ is the $p$th smallest nonconformity score. \textcolor{black}{Effectively, the quantile computation in equation \eqref{eq:C_computation} amounts to computing the $p$-th order statistic of $R^{(1)}, \hdots, R^{(K)}$ and is thus related to order statistics \cite{david2004order}. This computation can be done efficiently using standard sorting algorithms which usually have $\mathcal{O}(K \log K)$ time complexity. For our purposes, the strength of this result lies in the freedom of choosing a suitable nonconformity score -- we will encounter various examples in this survey.} In Sidebar \textbf{Empirical Quantiles as Linear Programs}, we further illustrate that the empirical quantile can be approximated  by a linear program. This will later be useful in case that the nonconformity scores are parameterized by a free parameter that we want to optimize over. Let us  summarize our results so far before providing more explanations and insights. 
\begin{sidebar}{Empirical Quantiles as Linear Programs}
\section[Empirical Quantiles as Linear Programs]{}\phantomsection
   \label{sidebar-proof-CP}
\setcounter{sequation}{0}
\renewcommand{\thesequation}{S\arabic{sequation}}
\setcounter{stable}{0}
\renewcommand{\thestable}{S\arabic{stable}}
\setcounter{sfigure}{0}
\renewcommand{\thesfigure}{S\arabic{sfigure}}
\textcolor{black}{Following \cite{Koenker1978}, we can  over-approximate the empirical quantile  of the nonconformity scores $R^{(1)}, \hdots, R^{(K)}$ at level $1-\delta$ by the solution of the following linear program
\begin{subequations}\label{eq:quantileAsLP}
\begin{align} 
    q^*:= \text{argmin}_{q} &\quad \sum_{i=1}^{K} \left( (1-\delta) e_i^+ + \delta e_i^- \right) \label{eq:quantileAsLPa}  \\
    \text{s.t.} & \quad e_i^+-e_i^- = R^{(i)} - q, \; i=1,\hdots,K \label{eq:quantileAsLPb}  \\
    & \quad e_i^+,e_i^- \geq 0, \; i=1,\hdots,K. \label{eq:quantileAsLPc}
\end{align}
\end{subequations}
The intuition for equation \eqref{eq:quantileAsLP} is as follows. The variables $e^+_i$ and $e^-_i$ together with constraints \eqref{eq:quantileAsLPb} and \eqref{eq:quantileAsLPc} represent deviations of $R^{(i)}$ from the variable $q$ from above and from below. The objective function \eqref{eq:quantileAsLPa} penalizes points above $q$ with weight $1-\delta$ and points below $q$ with weight $\delta$. Minimizing this weighted sum of deviations forces the solution to “push” just enough nonconformity scores  $R^{(i)}$  above and below $q$ to enforce that $\text{Quantile}_{1-\delta}( R^{(1)},\hdots,R^{(K)})\le q^*$.}

\textcolor{black}{From Lemma \ref{lem:1} and equation \eqref{eq:C_computation}, we recall next that $\infty$ has to appear in the empirical quantile $\text{Quantile}_{1-\delta} (R^{(1)}, \hdots, R^{(K)}, \infty)$. However,  we cannot encode $\infty$ in the linear program \eqref{eq:quantileAsLP} in practice. Nonetheless, note that it holds that 
\begin{align*}
    \text{Quantile}_{1-\delta}&( R^{(1)},\hdots,R^{(K)},\infty)\\
    &=\text{Quantile}_{(1+1/{K})(1-\delta)}(R^{(1)},\hdots,R^{(K)})
\end{align*}
if $(1+1/{K})(1-\delta)\in (0,1)$ so that we can simply replace $1-\delta$ by $(1+1/{K})(1-\delta)$ in equation \eqref{eq:quantileAsLP}.\footnote{Alternatively, note that 
$\text{Quantile}_{1-\delta} (R^{(1)}, \hdots, R^{(K)}, M)=\text{Quantile}_{1-\delta} (R^{(1)}, \hdots, R^{(K)}, \infty)$
for a sufficiently large constant $M\ge \text{Quantile}_{1-\delta} (R^{(1)}, \hdots, R^{(K)}, \infty)$ so that we could instead use $M$ in the linear program \eqref{eq:quantileAsLP}.} If we do exactly this,  we obtain $\text{Quantile}_{1-\delta}( R^{(1)},\hdots,R^{(K)},\infty)\le q^*$. Indeed, one can show that equality, i.e., $\text{Quantile}_{1-\delta}( R^{(1)},\hdots,R^{(K)},\infty)=q^*$, holds if $(1-\delta)K\not\in\mathbb{N}$, see \cite{zhao2024conformal} for a proof.}

This linear program naturally lends itself to an optimization approach when we would like to optimize the empirical quantile (and thus the upper bound $C$) over a parameter, e.g., when we would like to minimize $\text{Quantile}_{1-\delta}(R^{(1)}_a, \hdots, R^{(K)}_a,\infty)$ over a design parameter $a$ that the nonconformity scores $R^{(1)}_a, \hdots, R^{(K)}_a$ depend on. A practical example would be when we aim to minimize the size of prediction regions, see Section \textbf{Control Synthesis for LEAS with Conformal Prediction}. 
\end{sidebar}

\begin{sidebar}{Proof: Conformal Prediction}
\section[Proof: Conformal Prediction]{}\phantomsection
   \label{sidebar-proof-CP}
\setcounter{sequation}{0}
\renewcommand{\thesequation}{S\arabic{sequation}}
\setcounter{stable}{0}
\renewcommand{\thestable}{S\arabic{stable}}
\setcounter{sfigure}{0}
\renewcommand{\thesfigure}{S\arabic{sfigure}}
We define the $(1-\delta)$th quantile  of a random variable $Z$ as
\begin{align*}
    \text{Quantile}_{1-\delta}(Z):=\inf\{z\in\mathbb{R}|\text{Prob}(Z\le z)\ge 1-\delta\}.
\end{align*}
We then define the $(1-\delta)$th quantile of $R^{(0)},R^{(1)},\hdots, R^{(K)}$ via the random variable $Z_0\sim \sum_{i=0}^K \delta_{R^{(i)}}/(K+1)$ where $\delta_{R^{(i)}}$ is the Dirac distribution centered at ${R^{(i)}}$. With slight abuse of notation,  let hence the $(1-\delta)$th quantile of $R^{(0)},R^{(1)},\hdots, R^{(K)}$   be
\begin{align*}
    \text{Quantile}_{1-\delta}(R^{(0)},R^{(1)}, \dots, R^{(K)}):=\text{Quantile}_{1-\delta}(Z_0).
\end{align*}
Similarly, let the $(1-\delta)$th quantile of $R^{(1)},\hdots, R^{(K)},\infty$ be
\begin{align}\label{eq:emp_quantile}
    \text{Quantile}_{1-\delta}(R^{(1)}, \dots, R^{(K)},\infty):=\text{Quantile}_{1-\delta}(Z_\infty)
\end{align}
with $Z_\infty\sim(\sum_{i=1}^K \delta_{R^{(i)}}+\delta_{\infty})/(K+1)$. 

\textbf{Fact 1:} Following these definitions, note the following. If 
\begin{align*}
	R^{(0)}>\text{Quantile}_{1-\delta}(R^{(0)},R^{(1)}, \dots, R^{(K)})=:q,
\end{align*}  
then we can change $R^{(0)}$ to arbitrary values without changing $\text{Quantile}_{1-\delta}(R^{(0)},R^{(1)}, \dots, R^{(K)})$ as long as $R^{(0)}>q$, i.e., for $M>q$ we have that $\text{Quantile}_{1-\delta}(R^{(0)},R^{(1)}, \dots, R^{(K)})=\text{Quantile}_{1-\delta}(M,R^{(1)}, \dots, R^{(K)})$. This implies that $R^{(0)}> \text{Quantile}_{1-\delta}(R^{(0)},R^{(1)}, \dots, R^{(K)})$ is equivalent to $R^{(0)}> \text{Quantile}_{1-\delta}(R^{(1)}, \dots, R^{(K)},\infty)$. Consequently, it  holds that
\begin{align*}
	&R^{(0)}\le \text{Quantile}_{1-\delta}(R^{(1)}, \dots, R^{(K)},\infty)\\
	\Leftrightarrow \;\;\;\;\; &R^{(0)}\le\text{Quantile}_{1-\delta}(R^{(0)},R^{(1)}, \dots, R^{(K)}).
\end{align*} 

\textbf{Fact 2:} Let $z$ take one of the values of $R^{(0)},R^{(1)}, \dots, R^{(K)}$, and let the $p_z$-th order statistic correspond to this value.\footnote{If there are ties, let $p_z$ be the largest possible value, e.g., let $p_z:=3$ for $z:=5$ and  $R^{(1)}:=1,R^{(2)}:=R^{(3)}:=5,R^{(4)}:=7$.}  Then, it holds that $\text{Prob}(Z_0\le z)=p_z/(K+1)$. By definition, we know that  $ \text{Quantile}_{1-\delta}(Z_0)$ is equivalent to the minimum value $z$ from $R^{(0)},R^{(1)}, \dots, R^{(K)}$ such that $\text{Prob}(Z_0\le z)\ge 1-\delta$.  This implies that $p_z$ is such that $p_z/(K+1)\ge 1-\delta$, and since we desire the minimum  $p_z$ also  that  $p_z= \lceil (K+1)(1-\delta)\rceil$. Hence, we know that $R^{(0)}\le \text{Quantile}_{1-\delta}(R^{(0)},R^{(1)}, \dots, R^{(K)})$ is equivalent to $R^{(0)}$ being among the $\lceil (K+1)(1-\delta)\rceil$-th smallest values of $R^{(0)},R^{(1)}, \dots, R^{(K)}$.

\textbf{Putting it all together:} From Facts 1 and 2, it follows that $R^{(0)}\le \text{Quantile}_{1-\delta}(R^{(1)}, \dots, R^{(K)},\infty)$ is equivalent to $R^{(0)}$ being among the $\lceil (K+1)(1-\delta)\rceil$-th smallest values of $R^{(0)},R^{(1)}, \dots, R^{(K)}$. As $R^{(0)},R^{(1)}, \dots, R^{(K)}$ are independent and identically distributed, the latter (and thus the former) holds with a probability of at least  $\lceil (K+1)(1-\delta)\rceil/(K+1)\ge 1-\delta$.
\end{sidebar}

\begin{lemma}[Marginal Coverage \cite{shafer2008tutorial,vovk2005algorithmic}]\label{lem:1}
    Let  $R^{(0)},\hdots,R^{(K)}$ be $K+1$ independent and identically distributed random variables. Then, it holds that the probabilistic prediction region in equation \eqref{eq:CP_goal} is valid by the choice of $C$ in equation \eqref{eq:C_computation}. 
\end{lemma}

For the interested reader, we provide a proof of this result in Sidebar \textbf{Proof: Conformal Prediction}. This proof is inspired and mainly taken from  \cite{tibshirani2019conformal}. We remark that it can further be shown that the probabilistic prediction region is tight, i.e., that it holds that
\begin{align*}
    \text{Prob}(R^{(0)}\le C(R^{(1)},\hdots,R^{(K)}))\le 1-\delta+\frac{1}{K+1}
\end{align*}
under mild assumptions on $R^{(0)},R^{(1)},\hdots,R^{(K)}$, i.e., that the joint distribution of $R^{(0)},R^{(1)},\hdots,R^{(K)}$ is continuous \cite{lei2018distribution}.

\textbf{How much calibration data is needed?} We may now ask the question how much data is needed to be able to compute a nontrivial upper bound $C$ with equations \eqref{eq:CP_goal} and \eqref{eq:C_computation}. If $\lceil (K+1)(1-\delta)\rceil>K$, we note that $C$ attains the value of a trivial and uninformative upper bound $C=\infty$.\footnote{Note that computing $\text{Quantile}_{1-\delta}(R^{(1)}, \hdots, R^{(K)},\infty)$ is equivalent to selecting the $p:=\lceil (K+1)(1-\delta)\rceil$-th smallest value among $R^{(1)}, \hdots, R^{(K+1)}$ where $R^{(K+1)}=\infty$. Correspondingly, if $p>K$ then the $p$-th smallest value among $R^{(1)}, \hdots, R^{(K+1)}$ is $R^{(K+1)}=\infty$.} To obtain a nontrivial upper bound, we hence require that $\lceil (K+1)(1-\delta)\rceil\le K$. From here, it is easy to see that   
\begin{align*}
	K\ge \frac{1-\delta} { \delta} 
\end{align*}
provides such a lower bound on the number of data required to obtain a nontrivial upper bound $C$.  

\textbf{Marginal coverage guarantees.}  We note that the guarantees in equation \eqref{eq:CP_goal} are ``marginal''\footnote{We are here using the standard terminology from the conformal prediction community, not to be confused with a marginal distribution of a set of random variables.} over the randomness in the test and calibration datapoints $R^{(0)}, R^{(1)}, \hdots, R^{(K)}$ as opposed to being conditional on the calibration datapoints $R^{(1)}, \hdots, R^{(K)}$. In other words, $\text{Prob}(\cdot)$ captures randomness in the draw over all random variables $R^{(0)}, R^{(1)}, \hdots, R^{(K)}$.   Formally, the probability measure $\text{Prob}(\cdot)$ in equation \eqref{eq:CP_goal} is a $K+1$ product measure of the probability measure associated with $\mathcal{R}$, see \cite{durrett2019probability}.\footnote{\textcolor{black}{We  choose a simplified notation here to not overcomplicate the matter and make the survey accessible to a larger readership. Specifically, we use the notation $\text{Prob}(\cdot)$ in a liberal way throughout the article where $\text{Prob}(\cdot)$ can capture randomness over either the test data $R^{(0)}\sim\mathcal{R}$ or both test and calibration data  $R^{(0)},  R^{(1)}, \hdots,  R^{(K)}\sim\mathcal{R}$. The only exception to this rule is when we discuss calibration conditional guarantees where we use  $\text{Prob}_K(\cdot)$ to capture randomness over calibration data $R^{(1)}, \hdots,  R^{(K)}\sim\mathcal{R}$.}}

How can we now empirically validate  marginal guarantees of the form  \eqref{eq:CP_goal}  to show that the theoretical guarantees of conformal prediction indeed align with experimental observations? We do so by  performing the following experiment $N$ times: we sample one test datapoint $R^{(0)}$ and $K$ calibration datapoints $R^{(1)},\hdots,R^{(K)}$ from $\mathcal{R}$. For each experiment, we check if $R^{(0)}\le C(R^{(1)},\hdots,R^{(K)})$ is satisfied. We then compute the empirical coverage, which is the ratio between the number of times $R^{(0)}\le C(R^{(1)},\hdots,R^{(K)})$ being satisfied and $N$. As $N\to \infty$, we expect (and will observe) that the empirical coverage converges to $1-\delta$. Formally, for $R^{(0)}_n,R^{(1)}_n,\hdots,R^{(K)}_n\sim\mathcal{R}$ with $n\in\{1,\hdots,N\}$, we compute the empirical coverage as
\begin{align}\label{eq:emp_coverage}
    EC&:=\sum_{n=1}^N \mathbb{1}\big(R^{(0)}_n\le C_n\big)/N
\end{align}
with $C_n:=C(R^{(1)}_n,\hdots,R^{(K)}_n)$ denoting the $n$-th experiment and where $\mathbb{1}(R^{(0)}_n\le C_n)$ is the indicator function that evaluates to $1$ if  $R^{(0)}_n\le C_n$ and $0$ otherwise.\footnote{\textcolor{black}{To compute the empirical coverage \eqref{eq:emp_coverage} in practice (similar to the conditional empirical coverage  \eqref{eq:cond_emp_coverage}), we have to observe the realizations of $R^{(0)}_n,R^{(1)}_n,\hdots,R^{(K)}_n\sim\mathcal{R}$ for each $n\in\{1,\hdots,N\}$. }}

  Let us think about a practical example to get some intuition of how we can interpret the marginal guarantees in equation \eqref{eq:CP_goal}. Assume that we have designed a control algorithm for a robot, and that $R^{(0)},R^{(1)},\hdots,R^{(K)}$ denote the performance of the robot over $K+1$ independent runs under the designed controller. In this sense, equation \eqref{eq:CP_goal} gives us guarantees for the control algorithm that hold on average over the $K+1$ runs rather than for an individual run of the robot.\footnote{Specifically, we can obtain an equivalent interpretation as $\text{Prob}(R^{(0)}\le C(R^{(1)},\hdots,R^{(K)}))=\mathbb{E}(\mathbb{1}(R^{(0)}\le C(R^{(1)},\hdots,R^{(K)}) ))$.} This is a common characteristic of statistical  techniques with finite sample guarantees, and similar (yet slightly different) interpretations can be found in other statistical (learning) techniques that provide calibration conditional coverage guarantees which we discuss next.

\textbf{Calibration conditional coverage guarantees.} In the aforementioned robot example, we would ideally like to obtain conditional guarantees of the form $\text{Prob}(R^{(0)}\le C(R^{(1)},\hdots,R^{(K)}) | R^{(1)},\hdots,R^{(K)})$ that tell us more about an individual run of the robot.\footnote{\textcolor{black}{The nonconformity scores $R^{(0)}, R^{(1)}, \hdots, R^{(K)}$ are generally treated as random variables that we would like to reason about. With the statement $\text{Prob}(R^{(0)}\le C(R^{(1)},\hdots,R^{(K)}) | R^{(1)},\hdots,R^{(K)})$ we condition over observed realizations of $R^{(1)},\hdots,R^{(K)}$.}} It is known that such conditional guarantees cannot be obtained in this setting without additional assumptions. However, it is known that this conditional probability is a random variable  that follows a beta distribution centered around $1-\delta$ if the distribution of $\mathcal{R}$ is continuous, with decreasing variance as $K$ increases, see \cite{vovk2012conditional} and \cite[Section 3.2]{angelopoulos2021gentle}  for details or \cite{duchi2025few} for newer results and a summary.  Formally, if the distribution of $\mathcal{R}$ is continuous, then 
\begin{align}\label{eq:cond_CP}
    \text{Prob}(R^{(0)}\le C(R^{(1)},\hdots,R^{(K)}) | R^{(1)},\hdots,R^{(K)})\sim \text{Beta}(\beta_1,\beta_2)
\end{align}
where $\text{Beta}(\beta_1,\beta_2)$ is a beta distribution with parameters $\beta_1:=K+1-\lfloor (K+1)\delta \rfloor$ and $\beta_2:=\lfloor (K+1)\delta \rfloor$. This result tells us that by increasing the size of the calibration data $K$, we can reduce the variance in our conditional coverage. 

How can we  empirically validate conditional guarantees of the form  \eqref{eq:cond_CP}? We do so by performing the following experiment $N$ times: we sample $J$ test datapoints $R^{(0)}_1,\hdots,R^{(0)}_{J}$ and $K$ calibration datapoints $R^{(1)},\hdots,R^{(K)}$ from $\mathcal{R}$. For each experiment,   we compute the ratio between the number of times $R^{(0)}_1,\hdots,R^{(0)}_{J}$  satisfying $R^{(0)}_j\le C(R^{(1)},\hdots,R^{(K)})$ and $J$. We then plot the histogram of these ratios of the $N$ experiments, which  will resemble $\text{Beta}(\beta_1,\beta_2)$ as $N,J\to \infty$. Formally, for $R^{(0)}_{nj},R^{(1)}_{n},\hdots,R^{(K)}_{n}\sim\mathcal{R}$ with $n\in\{1,\hdots,N\}$ and $j\in\{1,\hdots, J\}$, compute the conditional empirical coverage as
\begin{align}\label{eq:cond_emp_coverage}
CEC_n&:=\sum_{j=1}^{J} \mathbb{1}\big(R^{(0)}_{nj}\le C_n\big)/{J}
\end{align}
and  plot the histogram of $CEC_n$ over the $N$ experiments. For more detailed discussion on the effects of increasing/decreasing the parameters $K$, $N$, and $J$, we refer the reader to \cite[Appendix C]{angelopoulos2021gentle}. 

This was a lot already! In the Sidebar \textbf{Robot Navigation under Sensor Uncertainty} we illustrate everything that we learned so far in a simple academic example. For the controls engineer, this example provides a first connection between conformal prediction and dynamical systems.

\begin{figure*}
    \centering
    \begin{subfigure}[t]{0.31\textwidth} 
        \includegraphics[width=\textwidth]{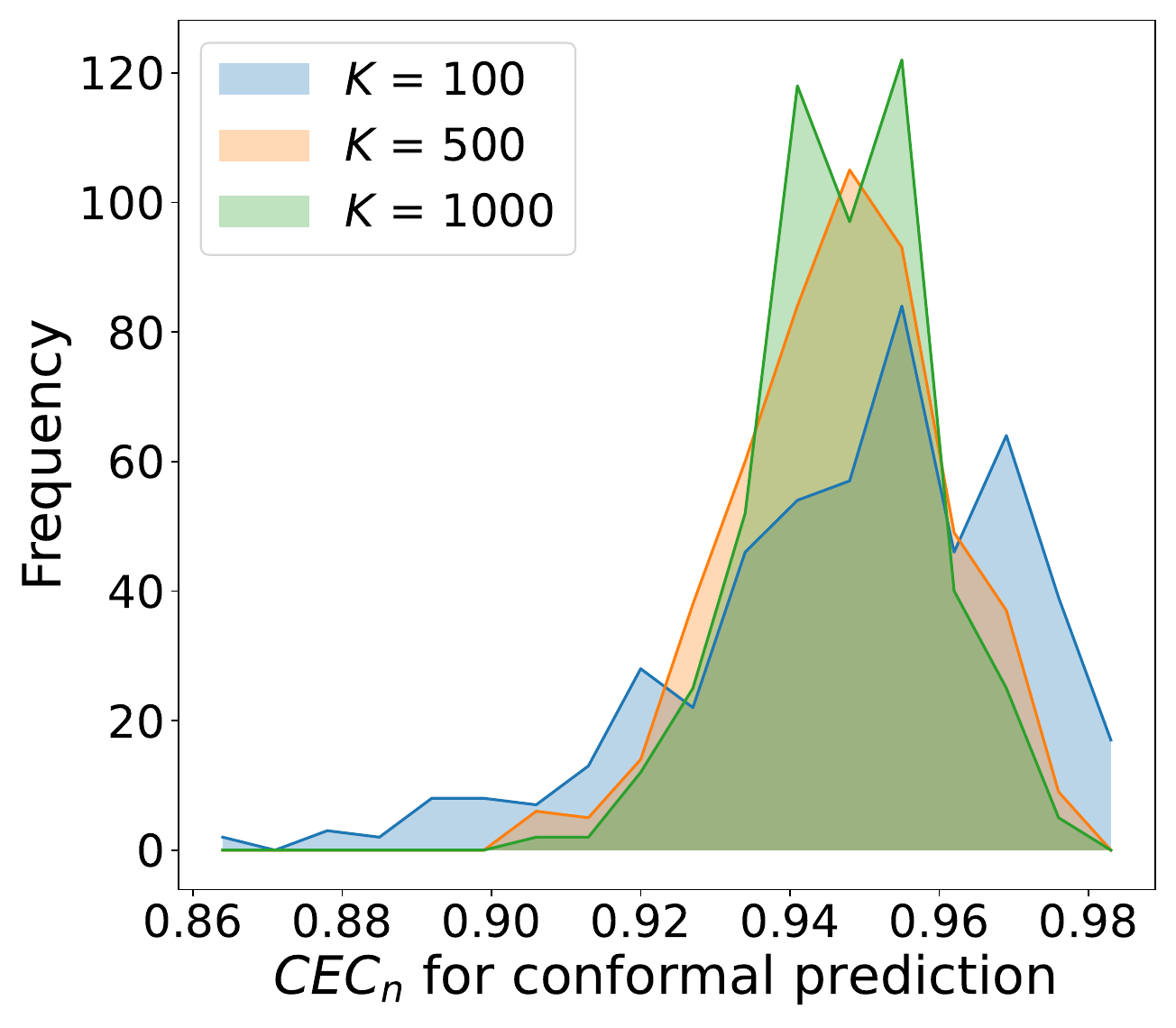}
        \caption{Histogram of the conditional empirical coverage $CEC_n$ over all $N$ experiments.}
        \label{fig:case_1_coverage_cp}
    \end{subfigure}
    \hspace{2mm}
    \begin{subfigure}[t]{0.31\textwidth}
        \includegraphics[width=\textwidth]{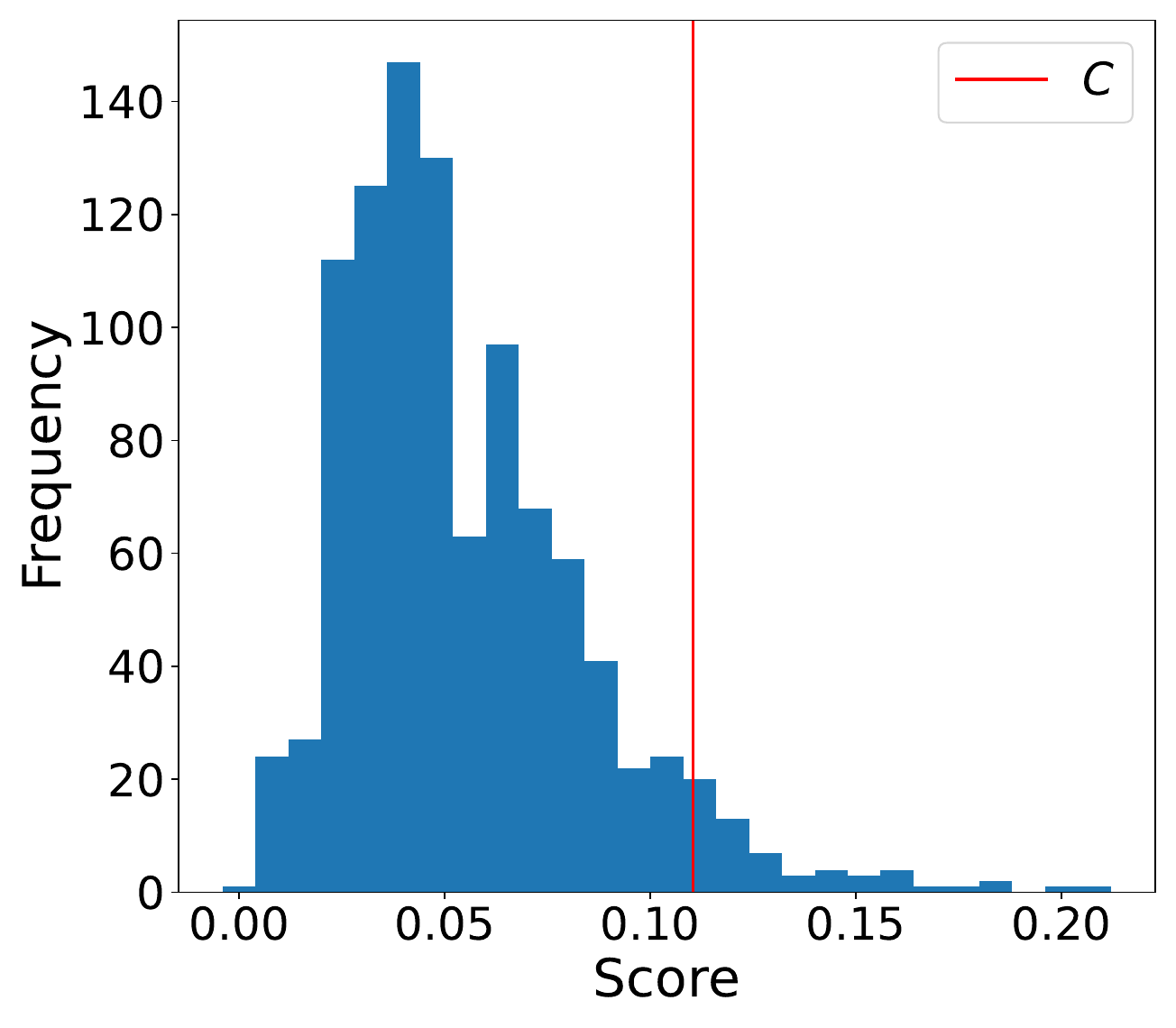}
        \caption{Histogram  of the nonconformity score $R^{(i)}$ over $K:=1000$ calibration datapoints (for one of the $N$ experiments).}
        \label{fig:case_1_nonconformities}
    \end{subfigure}
    \hspace{2mm}
    \begin{subfigure}[t]{0.31\textwidth}
        \includegraphics[width=\textwidth]{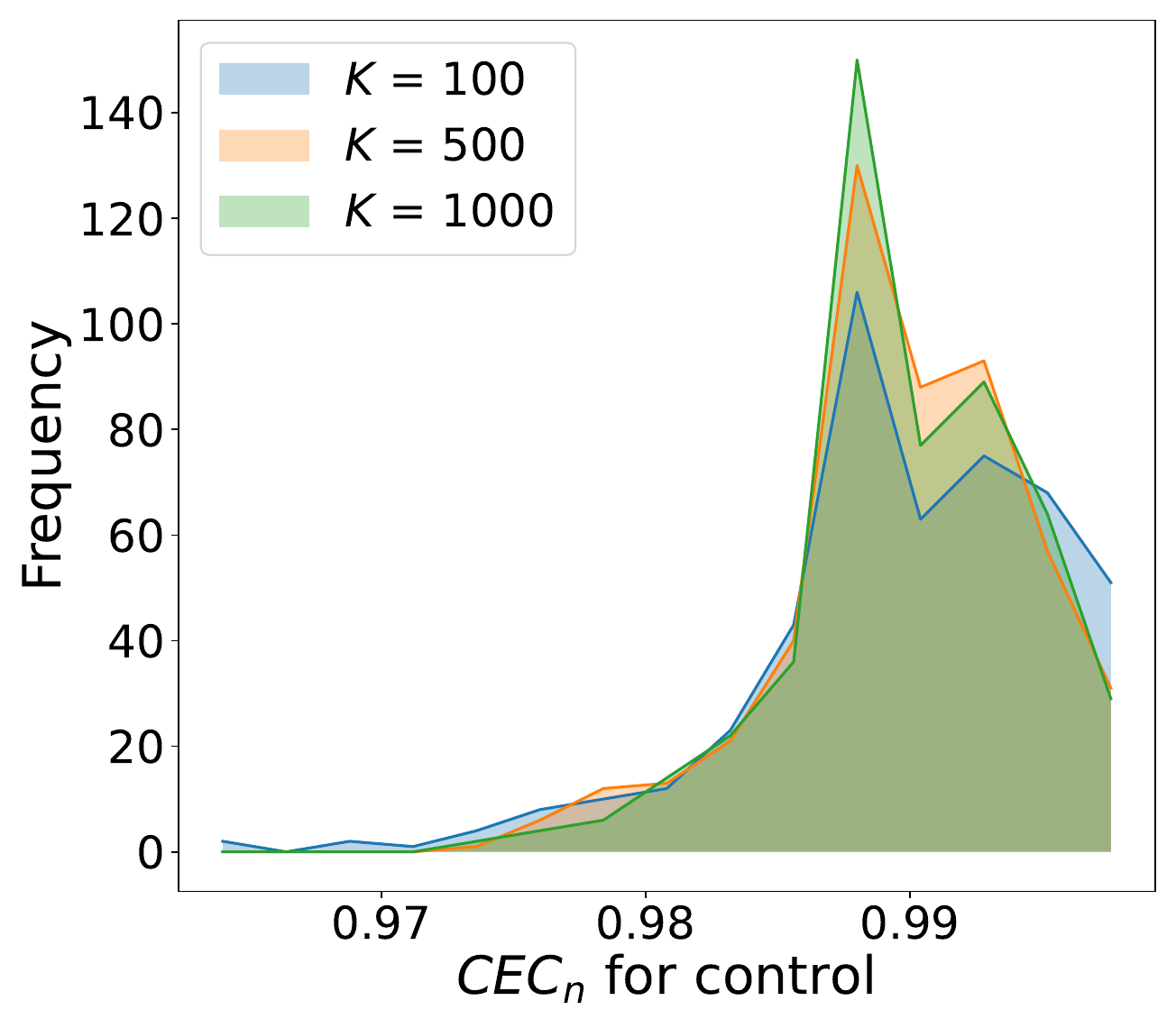}
        \caption{Histogram of the conditional empirical coverage of \eqref{eq:case_1_goal} over all $N$ experiments. }
        \label{fig:case_1_coverage_controller}
    \end{subfigure}
    \caption{Empirical validation for the example in Sidebar \textbf{Robot Navigation under Sensor Uncertainty}.}
    \label{fig:case_1_cp}
\end{figure*}

\begin{figure*}
    \centering
    \includegraphics[width=0.9\textwidth]{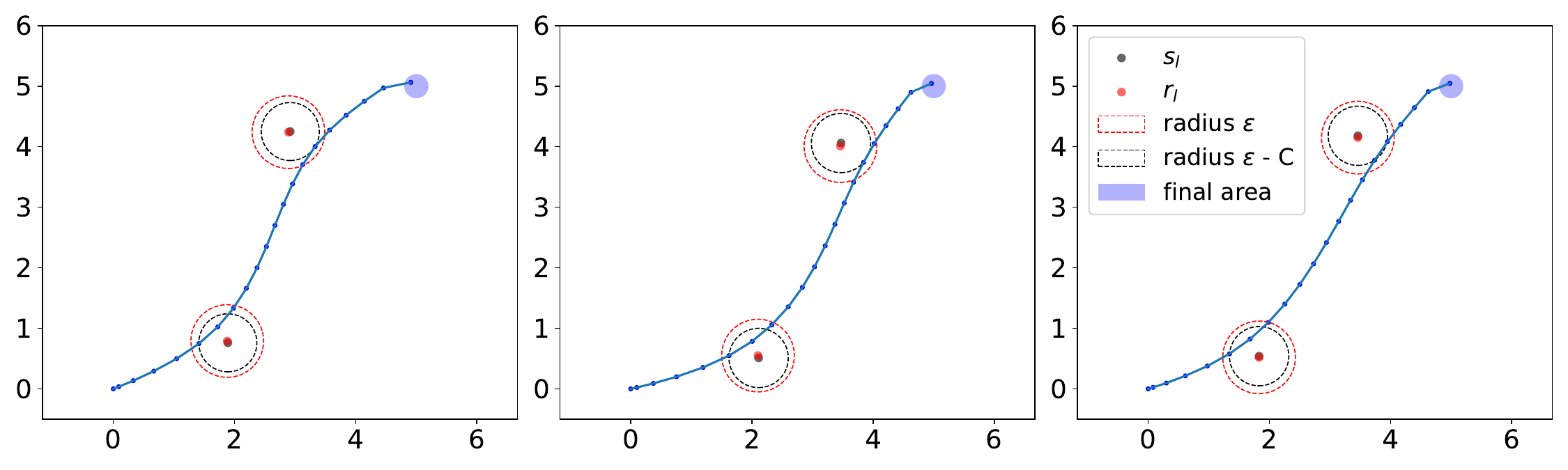}
    \caption{Three robot trajectories for the example in Sidebar \textbf{Robot Navigation under Sensor Uncertainty} example.}
    \label{fig:case_1_controller_results}
\end{figure*}

\begin{sidebar}{Robot Navigation under Sensor Uncertainty}
\section[Robot Navigation under Sensor Uncertainty]{}\phantomsection
   \label{sidebar-proof-CP}
\setcounter{sequation}{0}
\renewcommand{\thesequation}{S\arabic{sequation}}
\setcounter{stable}{0}
\renewcommand{\thestable}{S\arabic{stable}}
\setcounter{sfigure}{0}
\renewcommand{\thesfigure}{S\arabic{sfigure}}
Consider a robot described by  discrete-time double-integrator dynamics $x_{t + 1} = f(x_t, u_t)$ where $x_t:=(p_t,v_t)$ denotes the two-dimensional position $p_t$ and velocity $v_t$. There are two unknown locations $r_l := (r_{lx}, r_{ly})$ for $ l \in \{1, 2\}$ that follow uniform distributions $r_1 \sim \mathcal{U}([1.5, 2.5] \times [0.5, 1])$ and $r_2 \sim \mathcal{U}([2.5, 3.5] \times [4, 4.5])$. To estimate their location, we assume to have noisy sensors  $s_l = (s_{lx}, s_{ly}) \sim \mathcal{L}(r_l, 0.025)$ where $\mathcal{L}$ denotes the Laplace distribution. The distributions of $r_l$ and $s_l$ are unknown, but for calibration purpose we assume to have access to $K$ calibration samples from the sensors $(s_1^{(i)}, s_2^{(i)})$ along with ground truth locations $(r_1^{(i)}, r_2^{(i)})$ for $i\in\{1,\hdots, K\}$.

\textbf{Task.} At test time, the robot should reach the locations $r_1^{(0)}$ and $r_2^{(0)}$ at times $5$ and $15$, respectively, with a probability no less than  $1 - \delta := 0.95$, i.e., for precision $\epsilon := 0.6$ we want that
\begin{align}
    \label{eq:case_1_goal}
        \text{Prob}(\max(\|p_5 - r_1^{(0)}\|, \|p_{15} - r_2^{(0)}\|) \le \epsilon) \ge 1 - \delta.
\end{align}
Furthermore, the robot should arrive at location $(5, 5)$ with precision $0.2$ at time $T:=20$, i.e., $\|p_T - (5, 5)^\top\| \leq 0.2$.

\textbf{Sensor calibration.} We define the nonconformity score
\begin{align*}
        R^{(i)} \coloneqq \max_{l \in \{1,2\}}(\|s_l^{(i)} - r_l^{(i)}\|).
\end{align*}
Following Lemma \ref{lem:1}, we know that $\text{Prob}(\max_{l \in \{1,2\}}(\|s_l^{(0)} - r_l^{(0)}\|)\le C)\ge 1-\delta$, and consequently we also know that 
\begin{align*}
    \text{Prob}(\|s_l^{(0)} - r_l^{(0)}\|\le C, \forall l \in \{1,2\})\ge 1-\delta
\end{align*}
for sensor reading $s_l^{(0)}$ and unknown ground truth location $r_l^{(0)}$.

To perform an empirical validation of the previous statement, we conduct $N \coloneqq 500$ experiments for calibration set sizes of $K \in \{100, 500, 1000\}$. The obtained empirical coverage $EC$ in equation \eqref{eq:emp_coverage} was $0.963$, $0.947$, and $0.957$, respectively. We also computed the conditional empirical coverage $CEC_n$ in equation \eqref{eq:cond_emp_coverage} with $J := 500$.
We plot the histogram of $CEC_n$ in Figure \ref{fig:case_1_coverage_cp} along with the histogram of the nonconformity scores of the calibration data from one experiment in Figure \ref{fig:case_1_nonconformities}.

\textbf{Control synthesis.} We solve the optimization problem
\begin{align*}
    \min_{u_0,\hdots,u_{T-1}} \quad &   \sum_{t=0}^{T-1} 0.4u_t^2 +  0.6\|p_T - (5, 5)^\top\|\\
    \textrm{subject to} \quad & x_{t + 1} = f(x_t, u_t), \forall t \in \{1, \hdots, T - 1\},\\
    \quad & -1 \le u_t \le 1, \forall t \in \{0, \hdots, T - 1\},\\
    \quad & \max(\|p_5 - s_1^{(0)}\|,\|p_{15} - s_2^{(0)}\|)  \le  \epsilon - C,\\
    \quad & \|p_T - (5, 5)^\top\| \leq 0.2. 
\end{align*}
Note that a feasible solution ensures that \eqref{eq:case_1_goal} is satisfied. To see why, note that by triangular inequality we have that 
\begin{align*}
    \|p_5 - r_1^{(0)}\| &\le \|p_5 - s_1^{(0)}\| + \|s_1^{(0)} - r_1^{(0)}\| \\
    &\le \epsilon-C + \max(\|s_1^{(0)} - r_1^{(0)}\|, \|s_2^{(0)} - r_2^{(0)}\|). 
\end{align*}
The same upper bound can be derived for the expression $\|p_{15} - r_2^{(0)}\|$. Since we know that
$\text{Prob}(\max(\|s_1^{(0)} - r_1^{(0)}\|, \|s_2^{(0)} - r_2^{(0)}\|) \le C)\ge 1-\delta$, we can conclude that equation \eqref{eq:case_1_goal} is satisfied.  Importantly, we want to remind the reader that the obtained guarantees  are marginal, i.e.,  that in this case the probability measure $\text{Prob}(\cdot)$ is  defined over the randomness of test and calibration data $\{(r_1^{(i)},r_2^{(i)},s_1^{(i)},s_2^{(i)})\}_{i=0}^K$. 

For validation of the open-loop control policy, we compute the conditional empirical coverage of the constraint $\max(\|p_5 - r_1^{(0)}\|, \|p_{15} - r_2^{(0)}\|) \le \epsilon$ (in the same way $CEC_n$ is defined in equation \eqref{eq:cond_emp_coverage}), and we plot the histogram in Figure \ref{fig:case_1_coverage_controller}.
Finally, we show three robot trajectories under the controller in Figure~\ref{fig:case_1_controller_results}. 
\end{sidebar}

\textcolor{black}{\textbf{How to select nonconformity scores. } In the previous example, we have selected a particular nonconformity score to quantify sensor uncertainty. Indeed, the choice of the nonconformity score is essential in obtaining accurate uncertainty estimates. While this choice gives us engineers great flexibility, it also raises the question on how to select it optimally to solve a given problem. Unfortunately, there is no general rule on how to do so. What we present in this survey are choices of nonconformity scores that have worked well and have appeared in the literature. We will provide literature pointers whenever needed. }

\textcolor{black}{\textbf{Full and split conformal prediction. } There are two prominent variants of conformal prediction which differ in the way the nonconformity score is set up. In this survey, we will only use split conformal prediction. Nonetheless, we next briefly explain full conformal prediction and illustrate its computational challenges in practice. Therefore, assume that we are given a dataset of input-output pairs $(u^{(i)},y^{(i)})\sim \mathcal{D}$ for $i\in\{0,\hdots,L+K\}$ where $L,K>0$. We  assume scalar inputs and outputs for simplicity.}

\textcolor{black}{In split conformal prediction, we split the dataset into a training dataset of size $L$ and a calibration dataset of size $K$. Specifically, we (1) train/construct a predictor $\mu:\mathbb{R}\to\mathbb{R}$ using only the training data $(u^{(i)},y^{(i)})$ where $i\in\{K+1,\hdots,K+L\}$, and (2) define a nonconformity score $R^{(i)}$ that measures the prediction error between the predictor $\mu$ and the calibration data $(u^{(i)},y^{(i)})$ where $i\in\{1,\hdots,K\}$. For instance, a naive choice could be 
\begin{align*}
    R^{(i)}:=|y^{(i)}- \mu(u^{(i)})|.
\end{align*}
Due to the dataset splitting, it is easy to see that the nonconformity scores $R^{(i)}$ computed over $i\in\{0,\hdots,K\}$ are independent and identically distributed so that we obtain $\text{Prob}(R^{(0)}\le C(R^{(1)},\hdots,R^{(K)}))\ge 1-\delta$ by Lemma~\ref{lem:1}. This guarantee can be rewritten in the form
\begin{align*}
    \text{Prob}\big( y^{(0)} \in\mathcal{Y}\big(u^{(0)}\big)\big) \ge 1-\delta
\end{align*}
where the prediction set $\mathcal{Y}\big(u^{(0)}\big)\subseteq \mathbb{R}$ is defined as
\begin{align*}
    \mathcal{Y}\big(u^{(0)}\big):=\{y\in\mathbb{R}||y- \mu(u^{(0)})|\le  C(R^{(1)},\hdots,R^{(K)})\}.
\end{align*}
In other words, the set $\mathcal{Y}\big(u^{(0)}\big)$ contains the test output $y^{(0)}$ with a probability no less than $1-\delta$. Additionally, note that the set $\mathcal{Y}\big(u^{(0)}\big)$ is easy to compute in closed-form.}

\textcolor{black}{Full conformal prediction, on the other hand, is computationally more expensive, but does not require dataset splitting. For now, fix the pair $(u,y)\in\mathbb{R}^2$ and train/construct a predictor $\mu_{(u,y)}:\mathbb{R}\to\mathbb{R}$ using the augmented dataset consisting of the pair $(u,y)$ and the full dataset $(u^{(i)},y^{(i)})$ where $i\in\{1,\hdots,K+L\}$.\footnote{\textcolor{black}{We here assume that the training of $\mu_{(u,y)}$ is deterministic and a symmetric function of the training dataset, i.e., swapping datapoints within the training dataset does not change $\mu_{(u,y)}$.}} Similar to before, define the nonconformity scores \begin{align*}
    R^{(i)}_{(u,y)}:=|y^{(i)}- \mu_{(u,y)}(u^{(i)})|.
\end{align*} 
Let us now assume that $u=u^{(0)}$ and $y=y^{(0)}$. Then, note  that $R^{(i)}_{(u^{(0)},y^{(0)})}$ computed over $i\in\{0,\hdots,K+L\}$ are not independent and identically distributed. Luckily, due to the symmetry in $R^{(i)}_{(u^{(0)},y^{(0)})}$, one can check that $R^{(i)}_{(u^{(0)},y^{(0)})}$ are exchangeable so that we can still invoke Lemma \ref{lem:1}, i.e.,  
\begin{align*}
    \text{Prob}\Big(R^{(0)}_{(u^{(0)},y^{(0)})}\le C\Big(R^{(1)}_{(u^{(0)},y^{(0)})},\hdots,R^{(K)}_{(u^{(0)},y^{(0)})}\Big)\Big)\ge 1-\delta.
\end{align*}
This guarantee can again be rewritten in the form
\begin{align*}
    \text{Prob}\big( y^{(0)} \in\mathcal{Y}\big(u^{(0)}\big)\big) \ge 1-\delta
\end{align*}
where the prediction set $\mathcal{Y}\big(u^{(0)}\big)\subseteq \mathbb{R}$ is now defined as
\begin{align*}
    \mathcal{Y}\big(u^{(0)}\big):=\Big\{y\in\mathbb{R}||y- &\mu_{(u^{(0)},y)}(u^{(0)})|\le \\&C(R^{(1)}_{(u^{(0)},y)},\hdots,R^{(K)}_{(u^{(0)},y)})\Big\}.
\end{align*}
Importantly,  the set $\mathcal{Y}\big(u^{(0)}\big)$ is now not easy to compute anymore as it requires us to train/construct a new predictor $\mu_{(u^{(0)},y)}$ for each potential output $y\in\mathbb{R}$. This is why, in this survey, we focus on split conformal prediction. The interested reader can learn more about this in \cite{angelopoulos2024theoretical}.}

\subsection{Different Variants of Conformal Prediction }

Various extensions of the vanilla conformal prediction framework were presented over the last two decades. While it is not within scope of this article to review all of these extensions, we next present a selection of results that we believe are of great interest to the reader.

\textbf{Calibration conditional conformal prediction.} As we noted before, calibration conditional guarantees of the form $\text{Prob}(R^{(0)}\le C(R^{(1)},\hdots,R^{(K)}) | R^{(1)},\hdots,R^{(K)})$ cannot be obtained. However, the authors in \cite{vovk2012conditional} present a conformal prediction variant that provides calibration conditional coverage guarantees of the form 
\begin{align}\label{eq:CP_goal_cond}
    \text{Prob}_K\big(\text{Prob}(R^{(0)}\le \bar{C}(R^{(1)},\hdots,R^{(K)}))\ge 1-\delta\big)\ge 1-\beta
\end{align}
where $\beta\in (0,1)$ is a user-defined  failure probability over the calibration data. We intuitively interpret the statement in equation \eqref{eq:CP_goal_cond} as ``with a probability no less than $1-\beta$ over the draw of calibration datapoints $R^{(1)},\hdots,R^{(K)}$, it holds that $R^{(0)}\le \bar{C}(R^{(1)},\hdots,R^{(K)})$ with a probability no less than $1-\delta$ over the draw of a test datapoint $R^{(0)}$. Consequently, the outer probability measure $\text{Prob}_K(\cdot)$ is defined over the randomness in $R^{(1)},\hdots,R^{(K)}$, while the inner probability measure $\text{Prob}(\cdot)$ is defined over the randomness in $R^{(0)}$. If $\beta\in (0,1)$ is chosen to be very small, then one can approximately obtain calibration conditional guarantees. We next summarize how to compute the prediction region $\bar{C}:\mathbb{R}^K\to\mathbb{R}$ in \eqref{eq:CP_goal_cond} with calibration conditional conformal prediction.

\begin{lemma}[Calibration Conditional Coverage \cite{vovk2012conditional}]\label{lem:2}
    Let  $R^{(0)},\hdots,R^{(K)}$ be $K+1$ independent and identically distributed random variables. Then, it holds that the probabilistic prediction region in equation \eqref{eq:CP_goal_cond} is valid by the choice of
\begin{align*}
    \bar{C}:=\text{Quantile}_{1 - \delta + \sqrt{\frac{\ln(1/\beta)}{2K}}}( R^{(1)}, \hdots, R^{(K)}, \infty ).
\end{align*}
\end{lemma}

To summarize this result: if we are interested in calibration conditional guarantees of the form \eqref{eq:CP_goal_cond}, then we can simply use a slightly modified version of the vanilla conformal prediction algorithm. In fact, instead of computing the empirical quantile at a confidence level of $1-\delta$, we compute the empirical quantile at a corrected confidence level of $1-\delta + \sqrt{\frac{\ln(1/\beta)}{2K}}$. \textcolor{black}{We remark that Lemma \ref{lem:2} was derived in \cite{vovk2012conditional} using Hoeffding's inequality, a popular concentration inequality. As such, there is a close connection to concentration inequalities, see Section \textbf{Other Uncertainty Quantification Techniques} and Sidebar \textbf{Calibration Conditional Conformal Prediction and Hoeffding's Inequality} for more details. Nonetheless, this result shows that  calibration conditional  guarantees can be derived by computing an empirical quantile.   }

\textcolor{black}{There are two other variants of calibration conformal prediction that do not use Hoeffding's inequality and improve upon Lemma \ref{lem:2}. One variant is stated next.
\begin{lemma}[Calibration Conditional Coverage \cite{vovk2012conditional}]\label{lem:2______}
    Let  $R^{(0)},\hdots,R^{(K)}$ be $K+1$ independent and identically distributed random variables. Then, it holds that the probabilistic prediction region in equation \eqref{eq:CP_goal_cond} is valid by the choice of
\begin{align*}
    \bar{C}:=\text{Quantile}_{1 - \delta + \sqrt{\frac{2\delta\ln(1/\beta)}{K}}+\frac{2\ln(1/\beta)}{K}}( R^{(1)}, \hdots, R^{(K)}, \infty ).
\end{align*}
\end{lemma}
This  variant, compared to the first, is beneficial for small $\delta$. The last variant requires optimizing over an incomplete beta function to obtain the corrected confidence level that achieves \eqref{eq:CP_goal_cond}. We refer the interested reader to \cite{vovk2012conditional, angelopoulos2024theoretical}. }

Throughout this survey article, we will use conformal prediction with marginal coverage guarantees as per Lemma \ref{lem:1}. However, we emphasize that all proposed algorithms can be modified to provide calibration conditional coverage guarantees as per Lemma \ref{lem:2}. Although the marginal and calibration conditional guarantees in equations \eqref{eq:CP_goal}  and \eqref{eq:CP_goal_cond} are semantically different, \cite{vovk2012conditional} shows that one can convert \eqref{eq:CP_goal} into \eqref{eq:CP_goal_cond}, and vice versa. 

\textbf{Extensions of conformal prediction.}  Our exposition was naturally biased by the goals that we pursue with this survey, and we only scratched the surface of results from almost two decades of active research. For the interested reader, we summarize some of these results that we find important and potentially useful within formal verification and control in additional Sidebars. In Sidebar \textbf{Heteroskedasticity and Conformal Prediction}, we present different ways in which one can deal with heteroskedasticity in the underlying data distribution.\footnote{Heteroskedasticity happens when the variability of a  variable depends on another independent variable.} For instance, in the \textbf{Robot Navigation under Sensor Uncertainty} example one could imagine that the variability of the sensor measurements depends on the location of the regions of interest, i.e., the farther away a region is  from the robot, the larger the measurement error will be. In Sidebar \textbf{Conformal Prediction under Distribution Shift}, we further discuss how one can deal with the case when the datapoints $R^{(0)},R^{(1)},\hdots,R^{(K)}$ are not independent and identically distributed (and not exchangeable). There may be various reasons why our data may not be independent and identically distributed in practice, e.g., consider the sim2real gap of a system where design conditions are different from test/deployment conditions. 

\begin{sidebar}{Heteroskedasticity and Conformal Prediction}
\section[Heteroskedasticity and Conformal Prediction]{}\phantomsection
   \label{sidebar-proof-CP}
\setcounter{sequation}{0}
\renewcommand{\thesequation}{S\arabic{sequation}}
\setcounter{stable}{0}
\renewcommand{\thestable}{S\arabic{stable}}
\setcounter{sfigure}{0}
\renewcommand{\thesfigure}{S\arabic{sfigure}}
Consider  a setting in which inputs $u$ are drawn from some distribution $\mathcal{D}_u$, i.e., $u\sim\mathcal{D}_u$. Consider also outputs $y$ that are drawn from some conditional  distribution $\mathcal{D}_{y|u}$, i.e., a distribution that depends on the  input $u$. We can more compactly write $(y,u)\sim \mathcal{D}:=\mathcal{D}_{y|u}\times \mathcal{D}_u$. As a consequence,  the variance of the output $y$ may vary greatly as a function of $u$, a phenomenon referred to as heteroskedasticity. To capture such effects, we desire  prediction regions $C_u$ that depend on the sampled input $u$, i.e., we desire $\text{Prob}(y\in C_u|u)\ge 1-\delta$ where $\text{Prob}(\cdot|u)$ is a conditional probability. However, it is known that such guarantees are impossible to obtain without further assumptions, such as a finite input space \cite{foygel2021limits}. The good news is that there are efficient algorithms that provide marginal probability guarantees while capturing effects of heteroskedasticity. \textbf{Conformalized quantile regression} (CQR) is one such algorithm that was originally presented in \cite{romano2019conformalized}. CQR builds on the success of quantile regressors \cite{koenker2005quantile,hao2007quantile}, here estimators  $\hat{q}_{1-\delta}$ of the $1-\delta$ quantile of the random variable $y$. CQR follows a two-step procedure: (1) finding estimates $\hat{q}_{\delta/2}(u)$ and $\hat{q}_{1-\delta/2}(u)$ for the corresponding quantiles of $y$, and (2) computing the nonconformity score \begin{align*}
    R^{(i)}:=\max\big(\hat{q}_{\delta/2}(u^{(i)})-y^{(i)}, y^{(i)}-\hat{q}_{1-\delta/2}(u^{(i)})\big)
\end{align*}
over a calibration dataset $(y^{(i)},u^{(i)})\sim \mathcal{D}$ for $i\in\{1,\hdots,K\}$. Then, for a test point $(y^{(0)},u^{(0)})\sim\mathcal{D}$, Lemma \ref{lem:1} guarantees  
\begin{align}\label{cqr}
    \text{Prob}(y^{(0)}\in [\hat{q}_{\delta/2}(u^{(i)})-C,\hat{q}_{1-\delta/2}(u^{(0)})+C])\ge 1-\delta
\end{align}
where $C:=\text{Quantile}_{1-\delta}( R^{(1)}, \hdots, R^{(K)}, \infty )$. Note specifically that the prediction region in \eqref{cqr} depends on the input. 

One other method is \textbf{locally adaptive conformal prediction} in which we first train a predictor $\mu$ that predicts an output $y$ from an input $u$. We then define the nonconformity score $R^{(i)}:= \|y^{(i)}-\mu(u^{(i)})\|/\sigma(u^{(i)})$ where $\sigma(u^{(i)})>0$ measures the variability in $\|y^{(i)}-\mu(u^{(i)})\|$, e.g., a neural network that was trained to estimate $\|y^{(i)}-\mu(u^{(i)})\|$. Lemma \ref{lem:1} directly gives us
\begin{align}\label{lacp}
    \text{Prob}\big(\|y^{(0)}-\mu(u^{(0)})\|\le C\sigma(u^{(i)})\big)\ge 1-\delta
\end{align}
where again $C:=\text{Quantile}_{1-\delta}( R^{(1)}, \hdots, R^{(K)}, \infty )$. If $\sigma(u^{(i)})$ is a good approximator of the prediction error,  we expect that the term $C\sigma(u^{(i)})$ in \eqref{lacp} captures effects of heteroskedasticity. 
\end{sidebar}

\begin{sidebar}{Conformal Prediction under Distribution Shift}
\section[Conformal Prediction under Distribution Shift]{}\phantomsection
   \label{sidebar-proof-CP}
\setcounter{sequation}{0}
\renewcommand{\thesequation}{S\arabic{sequation}}
\setcounter{stable}{0}
\renewcommand{\thestable}{S\arabic{stable}}
\setcounter{sfigure}{0}
\renewcommand{\thesfigure}{S\arabic{sfigure}}
The conformal prediction guarantees in Lemma \ref{lem:1} rely on the assumption that the random variables $R^{(0)},\hdots, R^{(K)}$ are independent and identically distributed (or exchangeable). In practice, this assumption may not always hold which is one of the major challenges. For instance, in autonomy applications,  calibration data may only be available in the form of simulation data. While this data may come from high-fidelity photorealistic simulators in many applications, e.g., in autonomous driving and robotics, reality will always be different from these simulators. Another example is that the system may encounter different scenarios when deployed, e.g., traffic scenarios or weather conditions. 

One way to address this challenge is by using \textbf{robust conformal prediction} \cite{cauchois2024robust}. There, one assumes that $R^{(0)},\hdots,R^{(K)}$ are again independent, but not identically distributed in the sense that $R^{(0)}\sim\mathcal{R}_0$ while $R^{(1)},\hdots,R^{(K)}\sim\mathcal{R}$ where $\mathcal{R}_0$ is a test distribution. Under the assumption that calibration and test distributions $\mathcal{R}$ and $\mathcal{R}_0$ are ``close'', the calibration data can still be used to bound $R^{(0)}$. Specifically, assume that $D_f(\mathcal{R}_0,\mathcal{R})\le \epsilon$ where $D_f(\cdot)$ is any f-divergence and $\epsilon\ge 0$. It then holds that
    \begin{align*}
        \text{Prob}(R^{(0)}\le \tilde{C})\ge 1-\delta
    \end{align*}
    where the constant $\tilde{C}$ is now instead computed as
    \begin{align}\label{eq:C_tilde}
        \tilde{C}:= \text{Quantile}_{1-\tilde{\delta}}(R^{(1)},\hdots,R^{(K)})
    \end{align}
    where the tightened confidence level $\tilde{\delta}$ is defined as
    \begin{align}\label{eq:tilde_delta}
        \tilde{\delta}&:=1-g^{-1}(1-\delta_n)
    \end{align}
    and obtained by solving the set of convex optimization problems
    \begin{align*}
        \delta_n&:=1-g\big((1+1/K)g^{-1}(1-\delta)\big),\\
        g(\beta)&:=\inf \{z\in[0,1]|\beta f(\frac{z}{\beta})+ (1-\beta)f\Big(\frac{1 - z}{1 - \beta}\Big) \le \epsilon\},\\
        g^{-1}(\tau)&:=\sup \{\beta\in[0,1]|g(\beta)\le \tau\}.
    \end{align*}
The  reader is referred to \cite{cauchois2024robust} for more information. Robust conformal prediction provides valid guarantees without being too conservative when the distribution shift is small. However, large distribution shifts can result in conservative prediction regions $\tilde{C}$, potentially even with  $\tilde{C}=\infty$, e.g., when $1-\tilde{\delta}=1$. The authors in \cite{aolaritei2025conformal} presented a form of robust conformal prediction that uses the L\'evy-Prokhorov metric instead  of an f-divergence, including the $\infty$-Wasserstein and the total variation distance as a special case. Recent work in \cite{xu2025wasserstein} also considers distribution shifts measured in the $1$-Wasserstein distance. Lastly, we mention work on adversarially robust conformal prediction \cite{gendler2021adversarially,ghosh2023probabilistically,jeary2024verifiably}.

If we instead focus on the input-output setting in Sidebar \textbf{Heteroskedasticity and Conformal Prediction} but with covariate distribution shifts, then we can obtain non-conservative guarantees via \textbf{weighted conformal prediction} \cite{tibshirani2019conformal,jonkers2024conformal}. Therefore, consider that $(y^{(i)},u^{(i)})$ for $i\in\{1,\hdots,K\}$ are drawn from some calibration distribution $\mathcal{D}:=\mathcal{D}_{y|u}\times\mathcal{D}_u$, while $(y^{(0)},u^{(0)})$ is drawn from some test distribution $\mathcal{D}_0:=\mathcal{D}_{y|u}\times\tilde{\mathcal{D}}_u$. Note that the conditional distribution  $\mathcal{D}_{y|u}$ remains the same while only the covariate distribution can shift as $\tilde{\mathcal{D}}_u\neq \mathcal{D}_u$. The idea is that one can use a weighted version of conformal prediction when the covariate likelihood ratios $w(u) := \text{d} p(u) / \text{d} \tilde{p}(u)$ are known, where $p(u)$ and $\tilde{p}(u)$ are the probability density functions arising from $\mathcal{D}_u$ and $\tilde{\mathcal{D}}_u$, respectively. By defining the weight function $\pi(u^{(i)}) := \frac{w(u^{(i)})}{\sum_{j=1}^{n} w(u^{(j)}) + w(u^{(i)})}$, we compute the weighted quantile for the nonconformity score $R(u^{(i)},y^{(i)})$ as
\begin{align*}
    C_w:= \text{Quantile}_{1-\delta}\big(\pi(u^{(1)})R(u^{(1)},y^{(1)}),\hdots,\pi(u^{(K)})R(u^{(K)},y^{(K)})\big)
\end{align*}
from where it follows that $\text{Prob}(R(u^{(0)},y^{(0)})\le C_w)\ge 1-\delta$. A generalization of weighted conformal prediction appeared in \cite{prinsterconformal}, enabling conformal prediction  for sequential feedback-loop covariate shifts as  common in autonomous systems. 

An important form of a distribution shift can be observed for time series in which the distribution of the nonconformity score $R^{(t)}$ changes at each time $t\in\{1,2,\hdots\}$, i.e., where $R^{(t)}\sim\mathcal{D}_t$ for some arbitrary distribution $\mathcal{D}_t$. In such cases, we can use a version of \textbf{adaptive conformal prediction} for which we present a basic discussion next, see \cite{zaffran2022adaptive,gibbs2021adaptive,gibbs2024conformal,angelopoulos2024conformal,feldman2022achieving,zhou2024conformalized} for  details. 
The idea here is  to obtain a prediction region $C^{(t+1)}$ adaptively at time $t$  so that $\sum_{t=0}^{T-1}\textrm{Prob}(R^{(t+1)}\le C^{(t+1)})/T$ converges asymptotically to $1-\delta$ as $T\to \infty$. In fact, this region is now computed as
\begin{align*}
    C^{(t+1)}:= \text{Quantile}_{1-\delta_{t+1}}(R^{(1)},\hdots,R^{(t)})
\end{align*}
where the variable $\delta_{t+1}$ is not necessarily equal to $\delta$ and instead adapted online. One such choice (following \cite{gibbs2021adaptive}) can be
\begin{align*}
    \delta_{t+1}:=\delta_{t}+\gamma(\delta-e_{t})
    \; \text{ with }\; e_{t}:=\begin{cases}
    0 &\text{if }\, R^{(t)}\le C^{(t)}\\
    1 &\text{otherwise}
    \end{cases}
\end{align*}
where $\gamma>0$ is a learning rate. Note that adaptive conformal prediction is  least restrictive since no assumptions were made on the distributions $\mathcal{D}_t$. As such, we can capture large distribution shifts and rapidly changing distributions. However, this comes at the cost of weaker asymptotic guarantees. Finally, we note that we later obtain prediction regions for time series by instead assuming to have access to trajectory calibration data.
\end{sidebar}

\subsection{Other Uncertainty Quantification Techniques} 

\textcolor{black}{There is a plethora of statistical techniques that have been used for formal verification and control. In this section, we aim to provide some context and show how conformal prediction relates to these techniques. This being said, exploring connections between these techniques and providing guidelines for when to use which technique is still an active research area, see e.g., \cite{o2025bridging}. To begin with, we can divide these techniques into two broad categories, i.e., techniques for (1) uncertainty quantification, e.g., for quantifying the approximiation error of a neural network, and (2) solving optimization problems under uncertainty, e.g., for solving chance constrained optimization problems.}

\begin{sidebar}{Calibration Conditional Conformal Prediction and Hoeffding's Inequality}
\section[Calibration Conditional Conformal Prediction and Hoeffding's Inequality]{}\phantomsection
   \label{sidebar-CP-Hoeffding}
\setcounter{sequation}{0}
\renewcommand{\thesequation}{S\arabic{sequation}}
\setcounter{stable}{0}
\renewcommand{\thestable}{S\arabic{stable}}
\setcounter{sfigure}{0}
\renewcommand{\thesfigure}{S\arabic{sfigure}}
 Let $S^{(1)},\hdots,S^{(K)}$ be independent and identically distributed  random variables that have bounded support. For simplicity, we here assume that $0\le S^{(1)},\hdots,S^{(K)}\le 1$. For $X:=S^{(1)}+\hdots+S^{(K)}$, Hoeffding's Inequality tells us that  
\begin{align*}
    \text{Prob}_K(X-\mathbb{E}(X)\le t)\ge 1-\exp(-2t^2/K)
\end{align*}
where $\mathbb{E}(X)$ is the expected value of the random variable $X$. Note here that the probability measure $\text{Prob}_K(\cdot)$ is again taken over the randomness of $S^{(1)},\hdots,S^{(K)}$. To explore connections with conformal prediction, let us again consider a test datapoint $S^{(0)}$ and assume that  $S^{(0)},S^{(1)},\hdots,S^{(K)}$ are independent and identically distributed. Note also that $\mathbb{E}(X)=\sum_{i=1}^K\mathbb{E}(S^{(i)})=K\mathbb{E}(S^{(0)})$. If we now set $\beta:=\exp(-2t^2/K)$,  then we obtain 
\begin{align*}
    \text{Prob}_K\Big(X-K\mathbb{E}(S^{(0)})\le \sqrt{\frac{\ln(1/\beta)K}{2}}\Big)\ge 1-\beta.
\end{align*}
If the random variable $S^{(i)}$ is  chosen to be the indicator function of an event related to the nonconformity score $R^{(i)}$, we can draw a connection with calibration conditional conformal prediction in Lemma \ref{lem:2} and equation \eqref{eq:CP_goal_cond}. Particularly, for the  choice of $S^{(i)}:=\mathbb{1}(R^{(i)}\le C')$ where $C'$ is some user-defined constant,  we obtain $\mathbb{E}(S^{(0)})=\text{Prob}(R^{(0)}\le C')$\footnote{Note that the expected value of an indicator function of an event is equivalent to the probability of that event.} so that
\begin{align}\label{eq:compp}
    \text{Prob}_K\Big(\text{Prob}(R^{(0)}\le C')\ge X/K-\sqrt{\frac{\ln(1/\beta)}{2K}}\Big)\ge 1-\beta.
\end{align}

\textcolor{black}{This choice of  $S^{(i)}$ which leads to the guarantee in \eqref{eq:compp} now reveals a connection with equation \eqref{eq:CP_goal_cond}. This is not surprising since, as we mentioned before, Lemma \ref{lem:2} was derived  using Hoeffding's inequality. To compare \eqref{eq:compp} with \eqref{eq:CP_goal_cond}, fix $\beta$ and $C'$ in \eqref{eq:compp} and set $1-\delta:=X/K-\sqrt{\frac{\ln(1/\beta)}{2K}}$ so that  $\delta=1-X/K+\sqrt{\frac{\ln(1/\beta)}{2K}}$. By plugging this $\delta$ into equation  \eqref{eq:CP_goal_cond}, we obtain the bound $\bar{C}:=\text{Quantile}_{X/K}( R^{(1)}, \hdots, R^{(K)}, \infty )$. For \eqref{eq:compp} and \eqref{eq:CP_goal_cond} to give similar guarantees, we expect that $C' \approx \bar{C}$.  We recall that $\bar{C}=R^{(p)}$  where $p:=\lceil (K+1)X/K\rceil$ and $R^{(K+1)}:=\infty$.  For simplicity, assume that $R^{(1)}, \hdots, R^{(K)}$ are not the same. Then, close inspection reveals that $R^{(p-1)}\le C' \le  \bar{C}=R^{(p)}$. }

\textcolor{black}{Finally, we remark that this comparison concerns the first variant of calibration conditional conformal prediction in Lemma~\ref{lem:2}. A comparison with the variant in Lemma~\ref{lem:2______} is not immediate.}


\end{sidebar}

\begin{sidebar}{Calibration Conditional Conformal Prediction and DKW Inequality}
\section[Calibration Conditional Conformal Prediction and Hoeffding's Inequality]{}\phantomsection
   \label{sidebar-CP-Hoeffding}
\setcounter{sequation}{0}
\renewcommand{\thesequation}{S\arabic{sequation}}
\setcounter{stable}{0}
\renewcommand{\thestable}{S\arabic{stable}}
\setcounter{sfigure}{0}
\renewcommand{\thesfigure}{S\arabic{sfigure}}
\textcolor{black}{Let  $R^{(0)},R^{(1)},\hdots,R^{(K)}$ again be independent and identically distributed  random variables. The Dvoretzky–Kiefer–Wolfowitz (DKW) inequality \cite{dvoretzky1956asymptotic,massart1990tight} guarantees that 
\begin{align*}
    \text{Prob}_K\Big(\sup_{r\in\mathbb{R}}F_n(r)-F(r)\le t\Big)\ge 1-\exp(-2Kt^2)
\end{align*}
if $\exp(-2Kt^2)\le 0.5$. Here, $F(r):=\text{Prob}(R^{(0)}\le r)$ is the cumulative distribution function of $R^{(0)}$, while $F_n(r):=\sum_{i=1}^K\mathbb{1}(R^{(i)}\le r)/K$ is the corresponding empirical cumulative distribution function. Setting $\beta:=\exp(-2Kt^2)$ and fixing $r:=C'$, we hence get 
\begin{align*}
    \text{Prob}_K\Big(\text{Prob}(R^{(0)}\le C')\ge F_n(C')- \sqrt{\frac{\ln(1/\beta)}{2K}}\Big)\ge 1-\beta.
\end{align*}
Noting that $F_n(C') = X/K$, we can  make similar arguments as in Sidebar \textbf{Calibration Conditional Conformal Prediction and Hoeffding’s Inequality} to establish a connection between calibration conditional conformal prediction and the DKW inequality.}
\end{sidebar}

\textcolor{black}{\textbf{Techniques for uncertainty quantification. }
Pursuing the same goal as in conformal prediction, concentration inequalities provide bounds for random variables that hold with high probability \cite{vershynin2018high,boucheron2003concentration}. Well known bounds are obtained by Markov's and Chebyshev's inequalities. These, however, require knowledge of the random variable's first and second order moments, respectively. }

\textcolor{black}{Particularly popular are concentration inequalities for sums of independent and identically distributed random variables which do not require knowledge of first and second order moments. The motivation here is that sums of random variables approximate the mean of the random variable if properly normalized, i.e., for $\sum_{i=1}^K S^{(i)}/K$ where $S^{(1)},\hdots,S^{(K)}$ are independent and identically distributed bounded random variables. The Chernoff–Hoeffding inequality provides bounds for sums of Bernoulli random variables, while Hoeffding's inequality does so for sums of bounded random variables. We briefly discuss Hoeffding's inequality and show  how it relates to the first variant of calibration conditional conformal prediction from Lemma \ref{lem:2} in Sidebar \textbf{Calibration Conditional Conformal Prediction and Hoeffding's Inequality}. Closely related is also the Dvoretzky–Kiefer–Wolfowitz (DKW) inequality that can provide a bound on the cumulative distribution function of a random variable in terms of an empirically estimated cumulative distribution function, see Sidebar \textbf{Calibration Conditional Conformal Prediction and DKW Inequality}. We note that these concentration inequalities provide calibration conditional  guarantees, while split and full conformal prediction provide marginal  guarantees.} 

\textcolor{black}{Lastly,  the  Clopper-Pearson bound (discussed in more detail later) can be used to construct confidence bounds for a sequence of Bernoulli random variables. However, we are primarily interested in real-valued random variables that capture quantitative information, such as robustness. }

\textcolor{black}{\textbf{Techniques for solving optimization problems under uncertainty. } Scenario optimization was initially proposed to approximately solve semi-infinite optimization problems, such as those arising in robust control \cite{campi2009scenario,calafiore2005uncertain,calafiore2006scenario}. Effectively, scenario optimization solves chance constrained optimization problems (CCOPs) by (1) collecting a finite number of samples from the uncertain parameters of the CCOP, (2) solving an approximate deterministic problem using these samples, and (3) providing statistical guarantees for constraint satisfaction of the original CCOP in the form of calibration conditional coverage guarantees. Scenario optimization mainly applies to CCOPs with convex cost and constraint functions, but has recently been extended to the nonconvex setting \cite{garatti2024non}.  Very recently, the authors of \cite{lin2024verification,o2025bridging} revealed close connections between scenario optimization and conformal prediction. Specifically, they show how to set up a scenario optimization problem that recovers marginal conformal prediction guarantees. }

\textcolor{black}{Another technique for solving CCOPs is distributionally robust optimization \cite{rahimian2022frameworks,hanasusanto2015distributionally}. Here, the main idea is to (1) define an ambiguity set (i.e., a set of distributions) that contains the underlying distribution of the CCOP with a confidence no less than $1-\beta$, and (2) solve a robust version of the CCOP using this ambiguity set. The ambiguity set is usually selected such that solving the robust CCOP is easier than solving the original problem. For instance, one popular way is to draw samples from the uncertain parameters of the CCOP and construct Wasserstein distance sets around them \cite{mohajerin2018data}. Similar to scenario optimization, calibration conditional coverage guarantees can be obtained for the constraint satisfaction of the underlying CCOP. To the best of our knowledge, there are currently no known connections  between distributionally robust optimization and scenario optimization or conformal prediction. However, the work in \cite{patel2024non} presents initial results for using conformal prediction to obtain ambiguity sets that can be used within the distributionally robust optimization framework.}

\textcolor{black}{Recent approaches  explored the use of conformal prediction for solving CCOPs \cite{zhao2024conformal,vlahakis2024conformal,tumu2023multi}. The idea here is to (1) replace the chance constraint of the CCOP by a deterministic constraint that involves the empirical quantile over samples from the uncertain parameters of the CCOP, (2) rewrite this deterministic constraint in a more tractable way (e.g., via complementarity or mixed integer linear  constraints), and (3) solve this reformulated problem and conformalize the solution using a second set of samples from the uncertain parameters of the CCOP. There are notable differences to scenario optimization and distributionally robust optimization as this approach (1) involves a second set of samples, and (2) cannot a-priori guarantee constraint satisfaction of the underlying CCOP to a desired confidence level. Nonetheless, such approaches can provide a-posteriori guarantees on the constraint satisfaction in the convex as well as in the nonconvex setting. Furthermore, empirical evidence suggests that this approach is more sample efficient than scenario optimization where the bounds on constraint violation scale with the dimension of the decision variable \cite{vlahakis2024conformal}. More detailed comparison should be conducted in future work. }

\textcolor{black}{Finally, we mention PAC-Bayes theory which is a technique for solving supervised learning problems, such as classification or regression problems \cite{maurer2004note,mcallester1998some,shawe1997pac,alquier2024user}. PAC Bayes theory considers randomized predictors via distributions over parameters from a family of parameterized predictors. PAC-Bayes theory then derives calibration conditional coverage guarantees on the error of the predictor, e.g., by using Hoeffding's inequality. Importantly, starting from a given prior distribution of paramters, PAC-Bayes theory provides error bounds that can be optimized over. The authors in \cite{sharma2024pac} present PAC-Bayes generalization bounds for conformal predictors that provide efficient prediction regions.  In the remainder of this survey, we illustrate the benefits in using conformal prediction as a simple, general, and efficient technique for uncertainty quantification applicable to formal verification and control.  }

\section{LEC Verification with Conformal Prediction}

In this section, we  show how to use  conformal prediction for verifying the correctness of LECs, e.g., for computing input-output properties of neural networks. The presented techniques  are a straightforward application of conformal prediction, and our main goal is to ease the reader into the topic and to illustrate conformal prediction's simplicity and efficiency. The main technical sections, presented hereafter, will then present techniques for control design as well as offline and online verification of LEASs. To set the stage, we use $\mu:\mathbb{R}^{n_u}\to\mathbb{R}^{n_y}$ to denote an LEC, e.g., a feedforward neural network. Note that this is the most general description of an LEC possible.

\subsection{Challenges in LEC Verification}

The verification of input-output properties of LECs has been well studied in the computer science community over the last decade. The goal  is to verify (or disprove) that the output of an already trained LEC  satisfies an output property $\phi_\text{out}$ for all inputs that satisfy an input property $\phi_\text{in}$. Formally, we want to verify that $\mu(u)\models \phi_\text{out}$ holds for all inputs $u\in \mathbb{R}^{n_u}$ that satisfy $u\models \phi_\text{in}$, where $\models$ is the satisfaction operator, e.g., $u\models \phi_\text{in}$ means ``the input $u$ satisfies the property $\phi_\text{in}$''.\footnote{One may be interested in  more general properties $\phi$ that are  defined over inputs and outputs of $\mu$ simultaneously, i.e., one may be interested in verifying $\forall u\in\mathbb{R}^{n_u}, (u,\mu(u))\models \phi$. However, the if-then form in equation \eqref{eq:LECS_verifiy} is expressive enough in practice and commonly used within the community. } Equivalently, we can write
\begin{align}\label{eq:LECS_verifiy}
    \forall u\in\mathbb{R}^{n_u}, u\models \phi_\text{in}  \implies \mu(u)\models \phi_\text{out}.
\end{align}

As a concrete example, consider an input set $\mathcal{C}_\text{in}\subseteq\mathbb{R}^{n_u}$ and an output set $\mathcal{C}_\text{out}\subseteq\mathbb{R}^{n_y}$ for which we want to verify that  $\mu(\mathcal{C}_\text{in})\subseteq \mathcal{C}_\text{out}$ with $\mu(\mathcal{C}_\text{in}):=\{y\in \mathbb{R}^{n_y} |\exists u\in\mathcal{C}_\text{in}, \mu(u)=y\}$. This specific example points at an important problem instance regarding the reachability of $\mu$.
 
\textbf{Computational bottlenecks in model-based approaches.} To address this goal, model-based  approaches (often also referred to as optimization-based approaches) were proposed that use a mathematical description of the LEC, usually with a focus on feedforward neural networks, see \cite{liu2021algorithms,kwiatkowska2023trust} for an overview. However, these model-based verification techniques are challenging to solve as the underlying verification problem is NP-hard. A sound verification algorithm provides a positive answer only if equation \eqref{eq:LECS_verifiy} is true, while a complete verification algorithm provides a positive answer if and only if \eqref{eq:LECS_verifiy} is true. Popular complete verification algorithms are based on mixed integer linear programming \cite{dutta2018output,fischetti2018deep,tjeng2017evaluating,lomuscio2017approach} or satisfiable modulo theory solvers \cite{katz2017reluplex,ehlers2017formal,katz2019marabou}.  Computationally more efficient sound, but non-complete verification algorithms were also proposed, e.g., using semidefinite programming \cite{fazlyab2020safety,fazlyab2019efficient}, bound propagation \cite{wang2021beta,zhang2018efficient}, and abstract interpretation techniques \cite{singh2019abstract,gehr2018ai2,singh2018fast}. While complete verification algorithms are subject to scalability challenges in practice, non-complete methods are usually conservative due to the use of over-approximations and hence restrictive. We mention already now that these challenges of model-based approaches will only amplify when attempting to verify LEASs, which  use complex LECs.

\textbf{Efficient statistical verification of LECs. }In what follows, we apply conformal prediction to navigate the  trade-off between scalability and conservatism while additionally being able to deal with uncertainty, e.g., noise affecting the input of an LEC. We will obtain practical algorithms that can compute input-output properties in real-time and with high confidence. In doing so, we give up on soundness of our verification algorithms and  instead obtain probabilistically sound verification algorithms. A probabilistically sound verification algorithm for \eqref{eq:LECS_verifiy} provides a positive answer only if equation \eqref{eq:LECS_verifiy} is true with a probability no less than $1-\delta$. We are thus interested in
\begin{align}\label{eq:LECS_verifiy_}
    u\sim\mathcal{D}_{\phi_\text{in}}  \implies \text{Prob}(\mu(u)\models \phi_\text{out})\ge 1-\delta
\end{align} 
where $\mathcal{D}_{\phi_\text{in}}$ is a distribution with support over the input set $\mathcal{C}_\text{in}:=\{u\in\mathbb{R}^{n_u}|u\models\phi_\text{in}\}$. Indeed, the data in verification problems is not always deterministic, e.g., in image classification or state estimation where stochasticity has to be taken into account. Verification problems are thus not always expressed in the form of \eqref{eq:LECS_verifiy}  and one instead formulates them in the form of \eqref{eq:LECS_verifiy_}.

\begin{remark}\label{rem1}
    So far, our exposition focused solely on computing input-output properties of an LEC $\mu$. However, LECs are usually evaluated on input-output data, e.g., to predict states from measurements in state estimation or to predict labels from images in image classification. We note that equation \eqref{eq:LECS_verifiy_} can easily be modified to capture requirements in such settings, i.e., let
    \begin{align*}
    (u,y)\sim\mathcal{D}  \implies \text{Prob}((\mu(u),y)\models \phi)\ge 1-\delta
\end{align*} 
where inputs and outputs $(u,y)\sim\mathcal{D}$ are drawn from some distribution $\mathcal{D}$, which usually arises from an underlying data-generating distribution, and the output property $\phi$, which is now defined over the output $y$ and the output of the LEC $\mu(u)$. For instance, an output property $\phi:=\|\mu(u)-y\|\le \epsilon$ expresses that the prediction error $\|\mu(u)-y\|$ is not larger than $\epsilon>0$.
\end{remark}

In verification problems that are not deterministic, e.g., when the underlying data is stochastic as in Remark \ref{rem1}, the distribution $\mathcal{D}_{\phi_\text{in}}$  arises naturally as part of the problem formulation. However, as mentioned before, we can also obtain probabilistically sound verification algorithms for  \eqref{eq:LECS_verifiy} by instead solving  \eqref{eq:LECS_verifiy_} for a carefully constructed distribution $\mathcal{D}_{\phi_\text{in}}$, e.g., by uniform sampling over $\mathcal{C}_\text{in}$. When dealing with verification problems of the form \eqref{eq:LECS_verifiy_}, we hence advise the reader to  carefully investigate when $\mathcal{D}_{\phi_\text{in}}$ arises naturally from the problem formulation and when $\mathcal{D}_{\phi_\text{in}}$ is  constructed. We will encounter both cases, sometimes mixed, in this survey.  While the verification problems in \eqref{eq:LECS_verifiy} and \eqref{eq:LECS_verifiy_} are stated for LECs, we will later extend these to apply to LEASs.

Lastly, we emphasize two distinct advantages of such statistical approaches over model-based verification techniques. First, model-based verification techniques face various challenges in the presence of uncertainty (inducing more conservatism and/or computational complexity), while statistical techniques elegantly reason about stochastic uncertainty. Second, model-based techniques are model dependent and often tailored to specific types of neural networks, while statistical techniques apply to any contemporary black box LEC. We argue that this is an important property when using algorithms from a fast-paced field such as machine learning.

\subsection{Probabilistically Sound LEC Verification}
We now illustrate how to obtain probabilistically sound verification algorithms for input-output properties of LECs using conformal prediction. The key idea will be to define nonconformity scores that are in accordance with the input-output properties in hand. Throughout this section, we make the following assumption.
\begin{assumption}\label{ass1}
    We have independently sampled a calibration dataset of $K$ inputs from the distribution $\mathcal{D}_{\phi_\text{in}}$, i.e., we have access to samples $u^{(i)}\sim \mathcal{D}_{\phi_\text{in}}$ for $i\in\{1,\hdots,K\}$.
\end{assumption}

We proceed by presenting solutions to problems of reachability and general verification of LECs. Specifically, in the former case the set $\mathcal{C}_\text{out}:=\{y\in\mathbb{R}^{n_y}|y\models\phi_\text{out}\}$ is explicitly given (e.g., a polytope in $\mathbb{R}^{n_y}$), while in the latter case only knowledge of $\phi_\text{out}$ is available. 

\textbf{Reachability of LECs. } Assume that we are given an input set $\mathcal{C}_\text{in}\subseteq\mathbb{R}^{n_u}$ and an output set $\mathcal{C}_\text{out}\subseteq\mathbb{R}^{n_y}$. Then, we can directly define the nonconformity score
\begin{align}\label{R_reach_LEC}
R^{(i)}:=\text{dist}\big(\mu(u^{(i)}),\mathcal{C}_\text{out}\big)
\end{align}
where we recall that $u^{(i)}\sim \mathcal{D}_{\phi_\text{in}}$ for $i\in\{0,1,\hdots,K\}$, and where $\text{dist}(\mu(u^{(i)}),\mathcal{C}_\text{out})$ measures the distance between the output $\mu(u^{(i)})$ and the output set $\mathcal{C}_\text{out}$.\footnote{A common choice to measure the distance between a point $y$ and a set $\mathcal{C}$ is the signed distance  $\text{dist}(y,\mathcal{C}):=\inf_{y'\in\mathcal{C}} \|y-y'\|$ if $y\not\in \mathcal{C}$ and $\text{dist}(y,\mathcal{C}):=-\inf_{y'\in \mathbb{R}^{n_y}\setminus \mathcal{C}} \|y-y'\|$ otherwise.} As  before, the distribution $\mathcal{D}_{\phi_\text{in}}$ is assumed to have support over $\mathcal{C}_\text{in}$, and $\mathcal{D}_{\phi_\text{in}}$ can either arise as part of the problem formulation or by sampling the input set $\mathcal{C}_\text{in}$. By applying Lemma \ref{lem:1}, we directly obtain the following result.
\begin{corollary}\label{cor1}
    Given an output set  $\mathcal{C}_\text{out}\subseteq\mathbb{R}^{n_y}$, the distribution $\mathcal{D}_{\phi_\text{in}}$ with support over an input set $\mathcal{C}_\text{in}\subseteq\mathbb{R}^{n_u}$, and a test input $u^{(0)}\sim\mathcal{D}_{\phi_\text{in}}$, we have that
\begin{align*}
    \text{Prob}(\text{dist}\big(\mu(u^{(0)}),\mathcal{C}_\text{out}\big)\le C)\ge 1-\delta,
\end{align*}
where  $C:=\text{Quantile}_{1-\delta}( R^{(1)}, \hdots, R^{(K)}, \infty )$ with $R^{(i)}$ being defined in equation \eqref{R_reach_LEC}. 
\end{corollary}

 By the definition of the signed distance function, we obtain a positive verification answer if $C< 0$.\footnote{Corollary \ref{cor1} implies  $\text{Prob}(C<0 \implies \mu(u^{(0)})\in\mathcal{C}_\text{out})\ge 1-\delta$.}
In this case, the absolute value of $C$ indicates how robustly the reachability specification is satisfied.  Note that this is a sufficient condition, which means that our algorithm is inconclusive if $C\ge 0$. In this case, we can instead define the nonconformity score $ R^{(i)}:=-\text{dist}\big(\mu(u^{(i)}),\mathcal{C}_\text{out}\big)$ with which we can compute $C$ such that $\text{Prob}(-\text{dist}\big(\mu(u^{(0)}),\mathcal{C}_\text{out}\big)\le C)\ge 1-\delta$. We now obtain a negative verification answer if $-C>0$.

We note that the distance $\text{dist}\big(\mu(u^{(i)}),\mathcal{C}_\text{out}\big)$ may not  be easy to compute if the set $\mathcal{C}_\text{out}$ has non-trivial geometric structure. If the output set is instead parameterized as $\mathcal{C}_\text{out}:=\{y\in\mathbb{R}^{n_y}| h_\text{out}(y)\le 0\}$ for some function $h_\text{out}:\mathbb{R}^{n_y}\to\mathbb{R}$, we can select the nonconformity score $R^{(i)}:=h_\text{out}(\mu(u^{(i)}))$ and proceed in the same way as above. Let us illustrate Corollary \ref{cor1} in a simple example.

\begin{example}\label{ex1}
    Consider a robot described by the initial state $u:=(p_{0}^x, p_{0}^y, \theta_0, v_0)$ where $p_{0}^x$ and $p_{0}^y$ denote the two-dimensional position, $\theta_0$ denotes the angle, and $v_0$ denotes the translational velocity. The unknown model of the robot follows discrete-time unicycle dynamics with constant translational and rotational velocity and additive noise on the position following the normal distribution $\mathcal{N}(0, 0.01^2)^2$. We want to predict $p_{10}^x$ and $p_{10}^y$ from the initial state $u$  where $p_{10}^x$ and $p_{10}^y$ are the  robot's position ten time steps ahead. We thus train a simple feedforward neural network $\mu$ to obtain predictions $y := \mu(u)\approx (p_{10}^x,p_{10}^y)$. How can we now check if the model $\mu$ gives reasonable predictions? We first fix an input distribution $\mathcal{D}_{\phi_\text{in}} \coloneqq \mathcal{U}([0,1\textcolor{black}{)})^2 \times \textcolor{black}{\mathcal{N}(0, 0.1^2)} \times \mathcal{TN}(1, 0.1^2)^2$ where $\mathcal{U}$ is the uniform distribution, $\mathcal{N}$ is the normal distribution, and $\mathcal{TN}$ is the normal distribution truncated at $[0,2]$. Using  prior physical knowledge of the robot, we postulate that the safe set $\mathcal{C}_\text{out}:=\{p:=(p^x,p^y)\in\mathbb{R}^2|h_\text{out}(p)\le 0\}$ with $h_\text{out}(p) \coloneqq (p^x-\textcolor{black}{13.5})^2 + (p^y-\textcolor{black}{0.5})^2 - \textcolor{black}{3}^2$ contains $p_{10}^x$ and $p_{10}^x$, i.e., $(p_{10}^x,p_{10}^y)\in \mathcal{C}_\text{out}$. We set the failure probability to $\delta \coloneqq 0.05$ and use the nonconformity score $R^{(i)}:=h_\text{out}(\mu(u^{(i)}))$ as discussed before where $u^{(i)}\sim\mathcal{D}_{\phi_\text{in}}$. Following Corollary \ref{cor1}, we know that $\text{Prob}(h_\text{out}(\mu(u^{(0)})) \le C)\ge 1-\delta$. We conduct $N \coloneqq 500$ experiments for calibration set sizes of $K \in \{100, 500, 1000\}$. Again, we first  verify   statistical validity empirically  and compute the empirical coverage $EC$ according to equation \eqref{eq:emp_coverage} as \textcolor{black}{$0.962$}, \textcolor{black}{$0.964$}, and \textcolor{black}{$0.96$}, respectively. We also plot the conditional empirical coverage $CEC_n$ according to equation \eqref{eq:cond_emp_coverage} with $J := 500$ in Figure \ref{fig:example1_coverage_cp}. 
    To check the quality of the learned model $\mu$, we compute $\sum_{n=1}^NC_n/N$ which is the average of the bound $C_n$ over all $N$ experiments, where we recall from equations \eqref{eq:emp_coverage} and \eqref{eq:cond_emp_coverage} that $C_n:=C(R^{(1)}_{n},\hdots,R^{(K)}_{n})$. For $K \in \{100, 500, 1000\}$, we obtain the values \textcolor{black}{$-0.51$}, \textcolor{black}{$-0.75$}, and $-0.81$, respectively. We can thus conclude that the model $\mu$ reasonably captures the reachable set of the  unicycle dynamics, which was postulated as $\mathcal{C}_\text{out}$, at the confidence level of $1-\delta$. In Figure \ref{fig:example1_coverage_c}, we also plot the histogram of $C_n$ over all $N$ experiments. 
    Finally, we illustrate $1000$ test datapoints in Figure~\ref{fig:example1_ill}.
    
    \begin{figure*}
    \centering
    \begin{subfigure}[t]{0.31\textwidth} 
        \includegraphics[width=\textwidth]{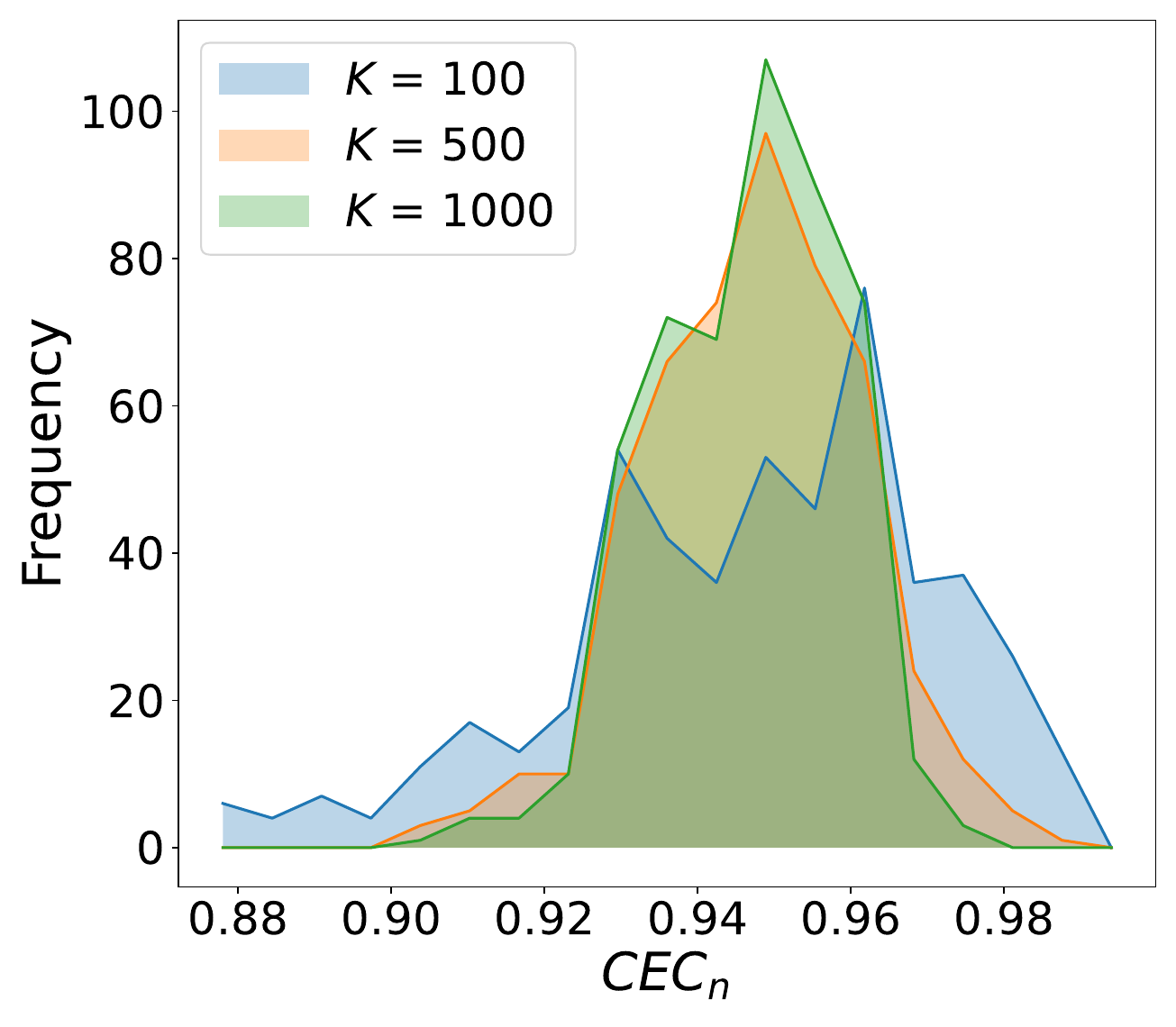}
        \caption{Histogram of the conditional empirical coverage $CEC_n$ over all $N$ experiments.}
        \label{fig:example1_coverage_cp}
    \end{subfigure}
    \hspace{2mm}
    \begin{subfigure}[t]{0.31\textwidth} 
        \includegraphics[width=\textwidth]{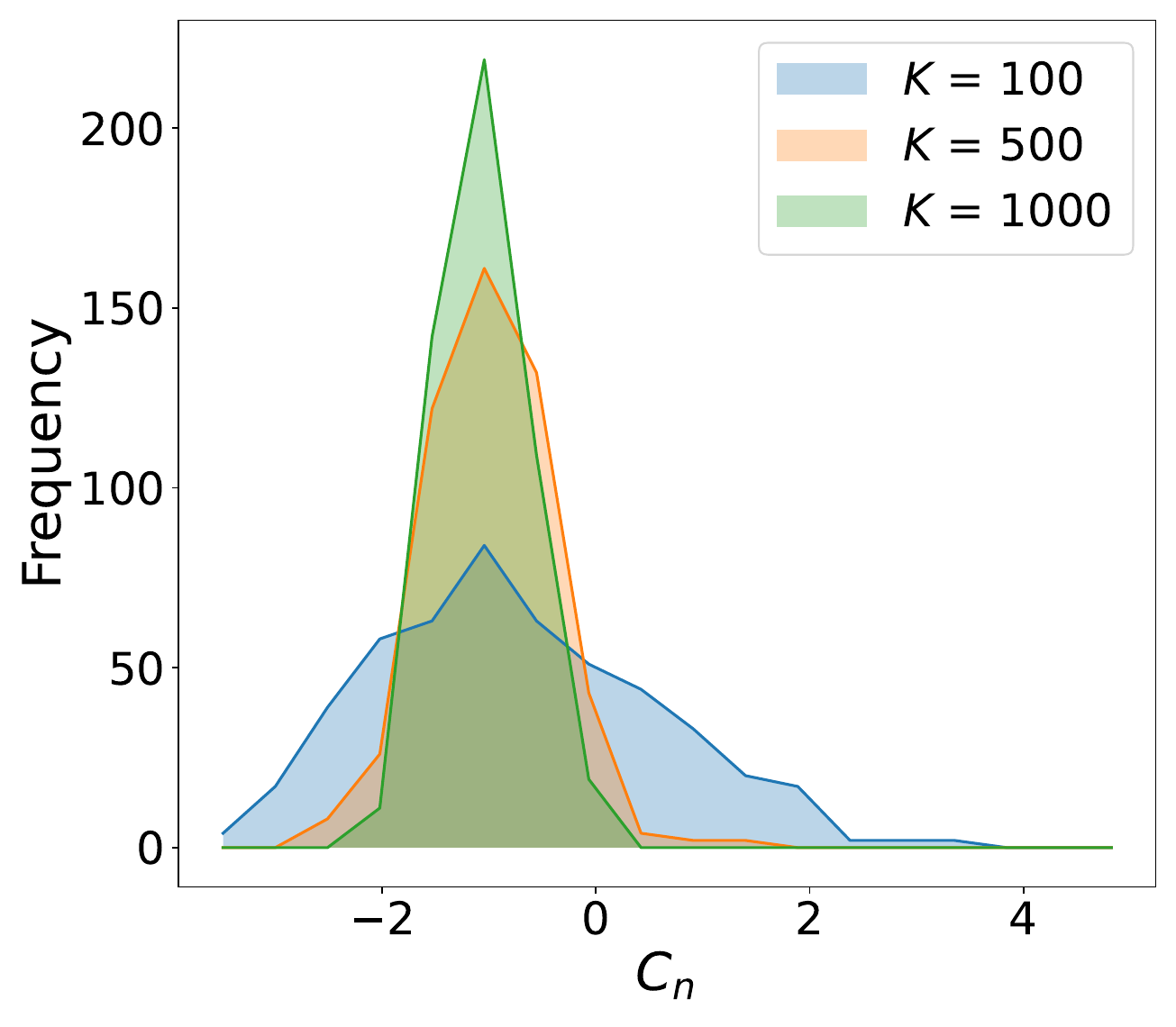}
        \caption{Histogram of the bound $C_n$ over all $N$ experiments.}
        \label{fig:example1_coverage_c}
    \end{subfigure}
    \hspace{2mm}
    \begin{subfigure}[t]{0.31\textwidth}
        \includegraphics[width=\textwidth]{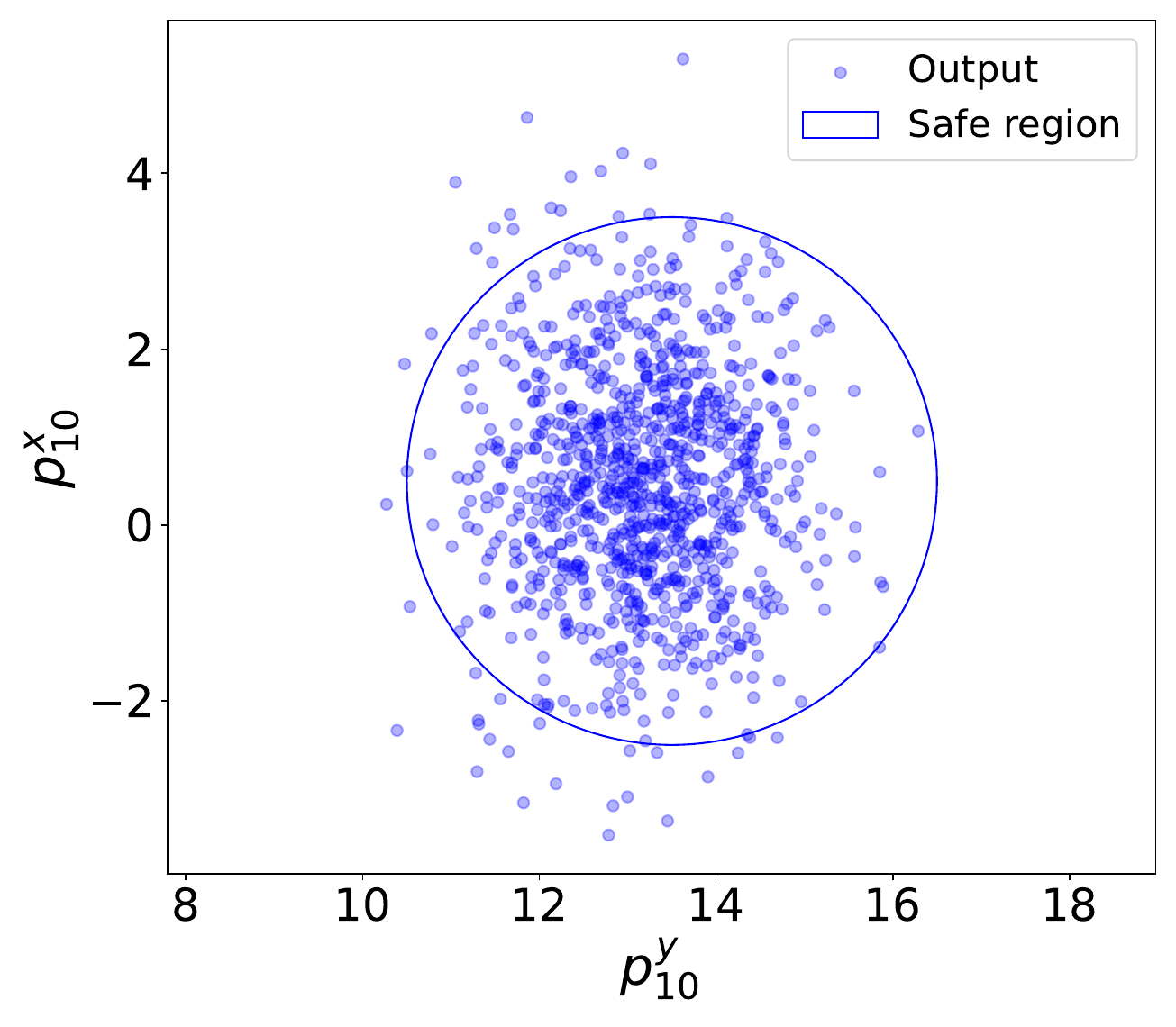}
        \caption{Illustration of $1000$ test datapoints.}
        \label{fig:example1_ill}
    \end{subfigure}
    \caption{Empirical validation for Example \ref{ex1}.}
    \label{fig:case_1_cp}
\end{figure*}
    
\end{example}

\begin{sidebar}{LEC Specifications in Predicate Logic}
\section[LEC Specifications in Predicate Logic]{}\phantomsection
   \label{sidebar-proof-CP}
\setcounter{sequation}{0}
\renewcommand{\thesequation}{S\arabic{sequation}}
\setcounter{stable}{0}
\renewcommand{\thestable}{S\arabic{stable}}
\setcounter{sfigure}{0}
\renewcommand{\thesfigure}{S\arabic{sfigure}}
We begin by defining the syntax of predicate logic  over $\mu$ as
\begin{align}\label{Boolean_fragment}
 \phi ::= \text{True} \ |\ h \ |\ \neg \phi'\ |\ \phi' \wedge \phi'' 
\end{align}
where $h:\mathbb{R}^{n_y}\to\mathbb{R}$ is a constraint function to impose constraints of the form $h(\mu(u))\ge 0$, $\neg \phi'$ denotes the negation of a predicate logic specification $\phi'$, and $\phi' \wedge \phi''$ denotes the conjunction of the predicate logic specifications $\phi'$ and $\phi''$. This syntax defines a set of rules according to which predicate logic specifications can be constructed. The notation $::=$ means that the left-hand side in \eqref{Boolean_fragment}, where $\phi$ is a free variable, is assigned to be one of the expressions from the right-hand side (separated by vertical bars). In other words, the expressions $\text{True}$ and $h$ are predicate logic specifications that we can denote by $\phi'$ or $\phi''$. We  then recursively construct new predicate logic specification using negations and conjunctions. We can further define  disjunction, implication, and equivalence  operators as
\begin{align*}
    \phi'\vee \phi'' &:= \neg (\neg \phi' \wedge \neg \phi'')&\text{ (disjunction operator)},\\
    \phi'\Rightarrow \phi'' &:= \neg \phi' \vee \phi''&\text{ (implication operator)},\\
    \phi'\Leftrightarrow \phi'' &:= (\phi'\Rightarrow\phi'')\wedge(\phi''\Rightarrow\phi')&\text{ (equivalence operator)}.
\end{align*}
 To formally determine if a predicate logic specification as defined in \eqref{Boolean_fragment} is satisfied, we define the Boolean  semantics as
 \begin{align*}
    \mu(u) \models
	 \text{True} &\;\;\;\text{ iff } \;\;\;	\text{holds by definition}, \\
    \mu(u)  \models
	 h  &\;\;\;\text{ iff } \;\;\;	h(\mu(u))\ge 0 , \\
	\mu(u)  \models \neg\phi' &\;\;\;\text{ iff } \;\;\;\mu(u)  \not\models \phi', \\
	 \mu(u)  \models \phi' \wedge \phi'' &\;\;\;\text{ iff } \;\;\;\mu(u) \models \phi' \text{ and } \mu(u) \models \phi''.
\end{align*}
Finally, we can define quantitative semantics $\rho^\phi:\mathbb{R}^{n_y}\to\mathbb{R}$ that will reduce checking $\mu(u) \models \phi$ to instead checking positivity of $\rho^\phi(\mu(u))$.\footnote{In the main text of this survey article, we refer to $\rho^\phi$ also as a performance function.}  Formally, we define the quantitative semantics as 
\begin{align*}
	\rho^{\text{True}}(\mu(u))& := \infty,\\
	\rho^{h}(\mu(u))& := h(\mu(u)),\\
	\rho^{\neg\phi'}(\mu(u)) &:= 	-\rho^{\phi'}(\mu(u)),\\
	\rho^{\phi' \wedge \phi''}(\mu(u)) &:= 	\min\big(\rho^{\phi'}(\mu(u)),\rho^{\phi''}(\mu(u))\big).
\end{align*}
It is known that these semantics are sound \cite{donze2010robust,fainekos2009robustness}, i.e., that it holds that $\mu(u)\models\phi$ holds if $\rho^\phi(\mu(u))>0$. Predicate logic additionally allows to include existential ($\exists$) and universal ($\forall$) quantifiers over the input $u$, which we mention here but omit from our presentation for the sake of simplicity.

To give more intuition, we consider an academic example via the specification $\phi:=(h_1\vee h_2)\wedge \neg h_3$. The conjunction $\wedge$ is the parent operator that connects to the expressions $h_1\vee h_2$ and $\neg h_3$. Following the semantics, we know   
\begin{align*}
    \mu(u) \models \phi \;\;\;\;\text{iff}\;\;\;\; \mu(u) \models h_1\vee h_2 \;\text{ and }\; \mu(u) \models \neg h_3.
\end{align*} 
For $h_1\vee h_2$ and $\mu(u) \models \neg h_3$, we also know
\begin{align*}
    \mu(u) \models h_1\vee h_2 \;\;\;\; &\text{iff} \;\;\;\; \mu(u) \models h_1 \;\text{ or }\; \mu(u) \models h_2\\
    \mu(u) \models \neg h_3 \;\;\;\; &\text{iff}\;\;\;\; \mu(u) \not\models h_3.
\end{align*} 
From this, we conclude that $\mu(u) \models \phi$ if and only if
\begin{itemize}
    \item $h_1(\mu(u))\ge 0$ or $h_2(\mu(u))\ge 0$, and
    \item $h(\mu(u))< 0$.
\end{itemize} 
\end{sidebar}

\textbf{General Verification of LECs.} We can verify other input-output properties as long as we can encode them as real-valued functions, e.g., any performance function that is defined over the output of the LEC. In the remainder, we focus on the case when $\phi_\text{out}$ is expressed in predicate logic over outputs of the LEC $\mu$ due to its generality. For a survey on formal specifications for the verification of learning-enabled components, we refer the reader to \cite{seshia2018formal}. To give a motivating example, consider the specification 
\begin{align*}
    \phi_\text{out}:=&\big((\mu_1(u)\ge 0) \implies (\mu_2(u)<0)\big)\\
    &\wedge \big((\mu_2(u)\ge 0) \implies (\mu_1(u)<0)\big)
\end{align*} that expresses mutual exclusivity of the two outputs $\mu_1$ and $\mu_2$ of  $\mu$. As we do not want to distract the reader from the main developments of this survey, we refer the reader to Sidebar \textbf{LEC Specifications in Predicate Logic} for more details on how predicate logic specifications over $\mu$ can  formally be defined.  Given a predicate logic specification $\phi_\text{out}$ over $\mu$, this sidebar also instructs how one can construct a real-valued performance function $\rho^{\phi_\text{out}}:\mathbb{R}^{n_y}\to\mathbb{R}$ that enjoys the following soundness property
\begin{align*}
    \rho^{\phi_\text{out}}(\mu(u))>0 \; \implies \; \mu(u)\models\phi_\text{out}.
\end{align*}
We note that larger values of $\rho^{\phi_\text{out}}(\mu(u))$ are beneficial as they indicate robustness against perturbation in $u$, see \cite{donze2010robust,fainekos2009robustness}. For the previous mutual exclusivity property, one obtains
\begin{align*}
    \rho^{\phi_\text{out}}(\mu(u)):=\min\big(\max&(-\mu_1(u),-\mu_2(u)),\\
    &\max(-\mu_2(u),-\mu_1(u))\big),
\end{align*}
and it can easily be verified that the soundness property holds, which we leave as an exercise for the reader.

To now verify LEC specifications in predicate logic, we can define the nonconformity score
\begin{align}\label{R_ver_LEC}
    R^{(i)}:=- \rho^{\phi_\text{out}}(\mu(u^{(i)}))
\end{align}
from which we get the following result using Lemma \ref{lem:1}.
\begin{corollary}\label{cor2}
    Given an output property  $\phi_\text{out}$ with sound performance function $\rho^{\phi_\text{out}}:\mathbb{R}^{n_y}\to\mathbb{R}$, the distribution $\mathcal{D}_{\phi_\text{in}}$ with support over the set $\mathcal{C}_\text{in}:=\{u\in\mathbb{R}^{n_u}|u\models \phi_\text{in} \} \subseteq\mathbb{R}^{n_u}$ defined by an input property $\phi_\text{in}$, and a test input $u^{(0)}\sim\mathcal{D}_{\phi_\text{in}}$, we have
\begin{align*}
    \text{Prob}(-C\le \rho^{\phi_\text{out}}(\mu(u^{(0)})))\ge 1-\delta.
\end{align*}
where  $C:=\text{Quantile}_{1-\delta}( R^{(1)}, \hdots, R^{(K)}, \infty )$ with $R^{(i)}$ being defined in equation \eqref{R_ver_LEC}. 
\end{corollary}

Lemma \ref{lem:1} gives $C$ such that $\text{Prob}(-\rho^{\phi_\text{out}}(\mu(u^{(0)}))\le C)\ge 1-\delta$ from which the above result is derived by simple manipulation. We then conclude that the algorithm produces a positive verification result if  $C<0$.\footnote{Corollary \ref{cor2} implies 
$\text{Prob}(C<0 \implies \mu(u^{(0)})\models \phi_\text{out})\ge 1-\delta$.} 

\textbf{Limitations of statistical  techniques. }  Corollaries \ref{cor1} and \ref{cor2} are powerful and enable verification of arbitrary LECs. However, we want to remind the reader that the coverage guarantees obtained via conformal prediction are marginal in the sense that the probability measure $\text{Prob}(\cdot)$ in Corollaries \ref{cor1} and \ref{cor2} is defined jointly over the randomness in the test and calibration datapoints $u^{(i)}\sim \mathcal{D}_{\phi_\text{in}}$ for $i\in\{0,\hdots,K\}$. The ideal result, in contrast, would be that the probability measure $\text{Prob}(\cdot)$ is only defined over the randomness in the test datapoint $u^{(0)}\sim \mathcal{D}_{\phi_\text{in}}$. This limitation is unavoidable with statistical techniques, e.g., it is also present with calibration conditional coverage guarantees obtained via calibration conditional conformal prediction (see Section \textbf{Different Variants of Conformal Prediction}) or scenario optimization.

\textbf{More properties.} Other examples of input-output properties are checking probabilistic prediction sets for classification tasks (see \cite{angelopoulos2021gentle}) or estimating function properties of $\mu$, such as Lipschitz constants (see \cite{qin2023conformance}), and extreme values, among others. To provide one last example, consider that we want to estimate the extreme value of $\mu$ when $n_y:=1$, i.e., we want to estimate $\mu^*:=\sup_{u\in\mathcal{C}_\text{in}} \mu(u)$. By defining the nonconformity score $R^{(i)}:=\mu(u^{(i)})$, we can obtain $C$ such that $\text{Prob}(\mu(u^{(0)})\le C)\ge 1-\delta$. So far, our mental model was to  fix a failure probability $\delta\in(0,1)$ and apply conformal prediction. However, this approach would not make much sense here as we are interested in the largest value of $\mu(u^{(i)})$. We thus reverse our approach, and select the smallest $\delta\in(0,1)$ that satisfies $\lceil (K+1)(1-\delta)\rceil\le K$. This choice, as discussed before, guarantees that $R^{(i)}\neq \infty$ while finding $C$ that is equal to the largest value of $\mu(u^{(i)})$. \textcolor{black}{Our intention with this paragraph was to illustrate that conformal prediction can be a versatile tool. To estimate extreme values, we note that there exist specialized techniques such as extreme value theory, see e.g., \cite{haan2006extreme}.}

\begin{sidebar}{Conformalizing the Kalman Filter}
\section[Conformalizing the Kalman Filter]{}\phantomsection
   \label{sidebar-proof-CP}
\setcounter{sequation}{0}
\renewcommand{\thesequation}{S\arabic{sequation}}
\setcounter{stable}{0}
\renewcommand{\thestable}{S\arabic{stable}}
\setcounter{sfigure}{0}
\renewcommand{\thesfigure}{S\arabic{sfigure}}
Let $z_{t+1}=f(z_t,v_t)$ be a dynamical system  where $z_t\in\mathbb{R}^{n_z}$ and $v_t\in\mathbb{R}^{n_v}$ are the state and the disturbance at time $t$, respectively. Let $y_t=p(z_t,w_t)\in\mathbb{R}^{n_y}$ be measurements where $w_t\in\mathbb{R}^{n_w}$ is sensor noise. The functions $f:\mathbb{R}^{n_z}\times\mathbb{R}^{n_u}\times \mathbb{R}^{n_v}\to\mathbb{R}^{n_z}$ and $p:\mathbb{R}^{n_z}\times \mathbb{R}^{n_w}\to\mathbb{R}^{n_y}$ model system dynamics and sensors. We assume that the input $u:=(z_0, v_t, w_t)$ of the dynamical system follows a distribution $\mathcal{D}$ from which a trajectory $z:=(z_0,z_1,\hdots)$ is generated. For calibration purposes, we assume that we have available $K$ inputs $u^{(i)}$ sampled from $\mathcal{D}$ along with the corresponding trajectories $z^{(i)}$. We note that knowledge of calibration trajectories $z^{(i)}$ is a strong assumption at this point, which we will comment on further in Section \textbf{State Estimation and Perception}. 

State estimators use past and present measurements $\{y_s\}_{s=0}^t$ to compute an estimate $\hat{z}_t$ of the state $z_t$ at time $t$. Popular techniques are based on Kalman or particle filtering, see \cite{simon2006optimal} for an overview. However, recursive state estimators such as the Kalman filter are only valid under restrictive assumptions on the system, while the particle filter is computationally prohibitive in practice. Specifically, the Kalman filter is an optimal state estimator that minimizes the expected mean square estimation error if: (1) the functions $f$ and $p$  are linear, and (2) the random variables $v_t$ and $w_t$ are Gaussian. In practice, these assumptions rarely hold, motivating us to conformalize the Kalman filter. Consider therefore the nonconformity score
\begin{align*}
    R^{(i)}:=\max(\|\hat{z}^{(i)}_1-z^{(i)}_1\|,\hdots,\|\hat{z}^{(i)}_T-z^{(i)}_T\|)
\end{align*}
which we evaluate over the calibration dataset, i.e., over ground truth information  $z^{(i)}$ for $i\in \{1,\hdots,K\}$. Having defined this nonconformity score, it is easy to see that 
\begin{align*}
    \text{Prob}(\|\hat{z}^{(0)}_t-z^{(0)}_t\|\le C, \forall t\in \{1,\hdots,T\})\ge 1- \delta
\end{align*}
by the choice of $C:=\text{Quantile}_{1-\delta}( R^{(1)}, \hdots, R^{(K)}, \infty )$. We specifically note that the use of the max-operator in the definition of the nonconformity score enables us to get simultaneous coverage of $\|\hat{z}^{(0)}_t-z^{(0)}_t\|$ for all times $t\in \{1,\hdots,T\}$ with high probability. However, using the maximum in this way may let time(s) where the estimation error is large dominate. We will later discuss in Section \textbf{Statistical Abstractions of Dynamic
Environments} how to address this subtlety and present a normalized version of this nonconformity score. 
\end{sidebar}

\textbf{State estimation.} Before concluding this section,  Sidebar \textbf{Conformalizing the Kalman Filter} shows how state estimators, such as the Kalman Filter, can be verified. We note that the algorithm proposed in the sidebar indeed applies to arbitrary state estimators. Our goal with this example is to show that conformal prediction can complement well established algorithms within control and estimation theory, such as the Kalman filter which may not provide guarantees if certain assumptions break.

\section{Control Synthesis for LEAS with Conformal Prediction}

On an abstract level, LEASs are dynamical control systems that use learning-enabled controllers, either learned in an end-to-end fashion or as a composition of object detection, state estimation, trajectory prediction, decision making, and low-level feedback control layers. We can model  LEASs as discrete-time stochastic dynamical systems
\begin{subequations}\label{eq:LEAS_model}
\begin{align}
        z_{t+1}&=f(z_t,u_t,v_t)\\
    y_t&=p(z_t,w_t)
\end{align}
\end{subequations}
where we follow the convention that $z_t\in\mathbb{R}^{n_z}$, $u_t\in\mathbb{R}^{n_u}$, and $y_t\in\mathbb{R}^{n_y}$ describe the system's state, control input, and measurement at time $t$, while $v_t\in\mathbb{R}^{n_v}$ and $w_t\in\mathbb{R}^{n_w}$ describe generic disturbances and sensor noise, respectively. We assume that $v_t$ and $w_t$ follow distributions $\mathcal{D}_v$ and $\mathcal{D}_w$, respectively. Assume also that we sampled the initial condition $z_0$ from a distribution $\mathcal{D}_z$. The functions $f:\mathbb{R}^{n_z}\times\mathbb{R}^{n_u}\times \mathbb{R}^{n_v}\to\mathbb{R}^{n_z}$ and $p:\mathbb{R}^{n_z}\times \mathbb{R}^{n_w}\to\mathbb{R}^{n_y}$ model the dynamics and the sensors of the system, respectively. We note that the model in \eqref{eq:LEAS_model} is restricted  to  discrete-time models, but otherwise fairly general. 

The control input $u_t$ is assumed to be a function of past measurements $\{y_s\}_{s=0}^t$, i.e., 
\begin{align}\label{eq:LEAS_control}
    u_t:=\pi(\{y_s\}_{s=0}^t),
\end{align} 
which can reduce to a state feedback controller $\pi(z_t)$ in the simplest case. We recall that the system in \eqref{eq:LEAS_model} is learning-enabled when $\pi$ contains LECs, e.g., a reinforcement learning controller or  a combination of a model predictive controller with learning-enabled state predictors as we mostly consider in this section.

\textbf{Objective. }Our next goal is to design a learning-enabled controller $\pi$ that satisfies a given property with high confidence. This is in contrast to later sections on online and offline verification of LEAS with conformal prediction where $\pi$ can be an arbitrary controller of the form \eqref{eq:LEAS_control} that may not  be guaranteed to satisfy the required property, hence motivating us to verify \eqref{eq:LEAS_model} under $\pi$.

Autonomous systems often operate in dynamic and uncertain environments. The state $z_t$ of the system in \eqref{eq:LEAS_model} hence consists of physical states $x_t\in\mathbb{R}^{n_x}$ and environment states $e_t\in\mathbb{R}^{n_e}$, e.g., describing uncontrollable agents such as pedestrians or objects of interest. To make this distinction explicit, let us write the LEAS in \eqref{eq:LEAS_model} as
\begin{align*}
    \begin{bmatrix}x_{t+1} \\ e_{t+1}\end{bmatrix}&=\begin{bmatrix}f_x(x_t,u_t) \\ f_e(e_t,v_t) \end{bmatrix}\label{eq:dynamics}\\
    y_t&=(x_t,e_t)
\end{align*}
where we note that we, at least for now, decoupled the $x$ and $e$ sub-systems. This assumption will help us to illustrate the main ideas while being able to provide strong guarantees. In Section \textbf{Safe Control in Dynamic Environments}, we will explain how this assumption can be relaxed. Further note that only the $x$ system is controllable by the input $u_t$, while only the $e$ sub-system is affected by the disturbance $v_t$.  We assume to know the function $f_x:\mathbb{R}^{n_x}\times\mathbb{R}^{n_u}\to\mathbb{R}^{n_x}$, i.e., the system dynamics, while the function $f_e:\mathbb{R}^{n_e}\times\mathbb{R}^{n_v}\to\mathbb{R}^{n_e}$ that describes the environment is not known. In fact, we do not need to assume Markovian dynamics for the $e$ sub-system and can instead assume that the sequence $(e_0,\hdots,e_T)$ follows an unknown discrete-time random process, i.e.,  $(e_0,\hdots,e_T)\sim \mathcal{D}_e$. Therefore, we consider the system
\begin{subequations}\label{eq:LEAS_model_}
\begin{align}
x_{t+1}&=f_x(x_t,u_t) \\ 
(e_0,\hdots,e_T)&\sim \mathcal{D}_e\label{eq:LEAS_model_eee}\\
    y_t&=(x_t,e_t),
\end{align}
\end{subequations}
and assume to have access to trajectories of finite length $T$ from $\mathcal{D}_e$ where $T>0$ is some finite task horizon.
\begin{assumption}\label{ass3}
    We have independently sampled a calibration dataset of $K$ trajectories from the environment  in equation \eqref{eq:LEAS_model_eee}, i.e., we have access to samples $e^{(i)}:=(e_0^{(i)},\hdots, e_T^{(i)}) \sim \mathcal{D}_e$ for $i\in\{1,\hdots,K\}$.
\end{assumption}
Lastly, note that we here assumed an ideal sensor with $y_t=(x_t, e_t)$, i.e., we have access to $x_t$ and $e_t$ at time $t$. Later in Section \textbf{State Estimation and Perception}, we will relax this assumption and consider dealing with perception and estimation uncertainty. 

We are interested in designing safe controllers, i.e., controllers that achieve the satisfaction of constraints formulated in terms of $x$ and $e$. For the most part of this section, we are interested in solving the following problem. 

\begin{probl}\label{prob1}
    Given the system in \eqref{eq:LEAS_model_}, a safety constraint $c:\mathbb{R}^{n_x}\times\mathbb{R}^{n_e}\to \mathbb{R}$ over states $x_t$ and $e_t$, and a task horizon $T>0$, design control inputs $u_0,\hdots,u_{T-1}$ such that $x_t$ satisfies $c$ for all times with a probability no less than $1-\delta$, i.e., such that
    \begin{align*}
        \text{Prob}(c(x_t,e_t)\ge 0, \forall t\in\{1,\hdots,T\})\ge 1-\delta.
    \end{align*}
\end{probl}

The safety constraint $c$ is enforced point-wise in time.  In Section \textbf{Temporal Logic-Constrained Control}, we will  consider a more general setting  where we enforce trajectory-wise constraints, e.g.,  temporal logic 
specifications $\phi$  defined over the trajectories $x$ and~$e$. We also note that the task horizon $T$ is finite. While this assumption does exclude us from expressing asymptotic properties such as stability and forward invariance, it still allows for expressing a broad range of properties in practice. Similar assumptions can commonly be found in the learning-based control and formal verification literature.

In what follows, we consider a compositional approach to design $\pi$. Indeed, conformal prediction is particularly suited for compositional reasoning and control design, as will become evident in this section. The high-level idea here is to construct efficient statistical abstractions of LECs, e.g., used for state estimation and prediction, which can then be used for control design, recall Figure \ref{fig:idea}. 

\subsection{Statistical Abstractions of Dynamic Environments } 
In order to safely control the system in \eqref{eq:LEAS_model_}, we need to reliably quantify the future behavior of the environment.  

\textbf{Environment prediction. }At each time $t$, we need to predict future states $(e_{t+1},\hdots,e_{T})$  from past and present observations $(e_0,\hdots,e_t)$. We denote predictions of $e_\tau$ for future times $\tau>t$ by $\hat{e}_{\tau|t}$. One can think of the predictor at time $t$ as a function $\mu:\mathbb{R}^{(t+1)n_e}\to\mathbb{R}^{(T-t)n_e}$ such that
\begin{align*}
(\hat{e}_{t+1|t},\hdots,\hat{e}_{T|t})=\mu(e_0,\hdots,e_t).
\end{align*} 
Popular prediction models are recurrent neural networks \cite{rudenko2020human}, long short-term memory networks \cite{graves2013generating,salzmann2020trajectron,hochreiter1997long}, transformers \cite{nayakanti2023wayformer}, or more traditional auto regressive moving average models  \cite{benjamin2003generalized}. These models may achieve varying accuracy on different prediction tasks, which we will account for by quantifying uncertainty of these predictions via  statistical abstractions. Finally, we remark that prediction models have to be trained on datasets that are independent of the calibration dataset from Assumption \ref{ass3}.

The idea for computing  statistical abstractions of the environment is to use  trajectory predictors  and conformal prediction in combination. We hence start by computing predictions of the trajectory predictor over the calibration dataset. Specifically, we compute the predictions 
\begin{align*}
(\hat{e}_{t+1|t}^{(i)},\hdots,\hat{e}_{T|t}^{(i)}):=\mu(e_0^{(i)},\hdots,e_t^{(i)})
\end{align*} 
for all times $t\ge 0$ from the  calibration data $i\in\{1,\hdots,K\}$. 

\textbf{Uncertainty representation via statistical abstractions. } So far, we have been using the terminology of a statistical abstraction rather abstractly. Formally, a statistical abstraction of the environment  is a set $E\subseteq \mathbb{R}^{Tn_e}$ such that
\begin{align*}
    \text{Prob}((e_1,\hdots,e_T)\in E)\ge 1-\delta.
\end{align*}
\textcolor{black}{Ideally, one would like to compute a statistical abstraction $E$ that is minimal according to some quantifiable metric. However, such minimal sets $E$ can generally be difficult to compute and may be nonconvex. We are here motivated from a practical point of view and having in mind that we want to use these sets downstream. }

Once we have obtained such a statistical abstraction, we will be able to use it for control design. For this purpose, we are interested in two slightly different statistical abstractions that enable the design of an open-loop and a closed-loop controller that solve Problem \ref{prob1} with different information available. For open-loop control, we only have information available at time $t=0$, and we will compute a statistical abstraction of the form 
\begin{align}\label{eq:cp_open_}
\begin{split}
         \text{Prob}(||e_\tau - \hat{e}_{\tau|0}|| \le C_{\tau|0}, \forall \tau \in \{1, \dots, T\}) \ge 1-\delta
    \end{split}
    \end{align}
where  $C_{\tau|0}$ indicates the $\tau$-step ahead prediction error for predictions $\hat{e}_{\tau|0}$ made at time $t=0$. For closed-loop control,  we use information available at all times $t\in\{0,\hdots,T-1\}$ to make one-step ahead predictions $\tau= t+1$. Specifically, we will compute statistical abstractions of the form
\begin{align}\label{eq:cp_closed_}
\begin{split}
         \text{Prob}(||e_{t+1} &- \hat{e}_{t+1|t}|| \le C_{t+1|t}, \forall t \in \{0, \dots, T-1\}) \ge 1-\delta  
    \end{split}
\end{align}
where $C_{t+1|t}$ indicates the one-step ahead prediction error for predictions $\hat{e}_{t+1|t}$ made at time $t$. 

We note that there are different ways in which we can compute abstractions of the form \eqref{eq:cp_open_} and \eqref{eq:cp_closed_}.  We will now present two such approaches: one naive approach that is intuitive but potentially conservative, as  presented in similar form in \cite{stankeviciute2021conformal,lindemann2022safe}, and another non-conservative approach, as originally presented in \cite{cleaveland2023conformal}. 

\textbf{Naive approach. } In this approach, we naively compute a probabilistic prediction region for each error $\|e_\tau-\hat{e}_{\tau|0}\|$ (or $\|e_{t+1}-\hat{e}_{t+1|t}\|$) and combine these  into guarantees of the form \eqref{eq:cp_open_} (or \eqref{eq:cp_closed_}) using Boole's inequality.\footnote{For a countable set of events $A_1, A_2,\hdots$, Boole's inequality states that the probability that at least one of the events is true is upper bounded by the sum of the probabilities of all individual events being true, i.e., that $\text{Prob}(\cup_{i=1}^\infty A_i)\le \sum_{i=1}^\infty \text{Prob}(A_i)$.} We define 
\begin{align}\label{eq:non_naiv}
    R_{\tau|t}^{(i)}:=\|e_\tau^{(i)}-\hat{e}_{\tau|t}^{(i)}\|
\end{align}
as the nonconformity score, and then compute
\begin{align}\label{eq:C_naiv}
    C_{\tau|t}:=\text{Quantile}_{1-\bar{\delta}}( R^{(1)}_{\tau|t}, \hdots, R^{(K)}_{\tau|t}, \infty )
\end{align}
where $\bar{\delta}:=\delta/T$ is a tightened failure probability that is such that $\bar{\delta}\le \delta$. By using $\bar{\delta}$ instead of $\delta$, we make a correction that will enable us to obtain guarantees of the form \eqref{eq:cp_open_} and \eqref{eq:cp_closed_}. Let us summarize these results

\begin{theorem}\label{thm:1}
    Given a test trajectory $e^{(0)}\sim\mathcal{D}_e$, the statistical abstractions in \eqref{eq:cp_open_} and \eqref{eq:cp_closed_} are valid by the choice of $C_{\tau|t}$ in equation \eqref{eq:C_naiv} with $\bar{\delta}:=\delta/T$ and  the nonconformity score \eqref{eq:non_naiv}.

\textbf{Proof: }
Let us show that the  abstraction in equation \eqref{eq:cp_open_} is valid by the choice of $C_{\tau|t}$, while the proof for the abstraction in \eqref{eq:cp_closed_} follows similarly. From Lemma \ref{lem:1} and the definition of the nonconformity score $R_{\tau|t}^{(i)}$ it follows immediately that $\text{Prob}(\|e_\tau-\hat{e}_{\tau|t}\|\le C_{\tau|0})\ge 1-\bar{\delta}$ holds for each $\tau\in\{1,\hdots,T\}$ individually. We consequently know that $\text{Prob}(\|e_\tau-\hat{e}_{\tau|0}\|> C_{\tau|0})\le \bar{\delta}$. Applying Boole's inequality gives us
\begin{align*}
    \text{Prob}(\exists \tau>0 \text{ s.t. }\|e_\tau-\hat{e}_{\tau|0}\|> C_{\tau|0})\le\sum_{i=t}^T\bar{\delta}= \sum_{i=t}^T\frac{\delta}{T}=\delta
\end{align*}
so that we can finally conclude that
\begin{align*}
    \text{Prob}(\|e_\tau-\hat{e}_{\tau|0}\|\le C_{\tau|0},\; \forall \tau\in\{1,\hdots,T\})\ge 1- \delta
\end{align*}
by which we have shown that \eqref{eq:cp_open_} holds. \qed
\end{theorem}

As a consequence of using the tightened failure probability $\bar{\delta}$ instead of $\delta$, this naive approach requires more data as now  $\lceil (K+1)(1-\bar{\delta})\rceil\le K$ has to hold to obtain nontrivial $C_{\tau|t}$, i.e., $C_{\tau|t}$ that are not $\infty$. Another consequence is that the statistical abstractions will be  conservative  since 
\begin{align*}
    \text{Quantile}_{1-\bar{\delta}}&( R^{(1)}_{\tau|t}, \hdots, R^{(K)}_{\tau|t}, \infty )\\
    &\ge \text{Quantile}_{1-{\delta}}( R^{(1)}_{\tau|t}, \hdots, R^{(K)}_{\tau|t}, \infty ).
\end{align*}
Specifically, we will notice conservatism when the distribution of $R^{(i)}_{\tau|t}$ has a long tail. This motivates our second approach, as presented next. 

\textbf{Single nonconformity score approach. } To avoid using the tightened failure probability $\bar{\delta}$ and combining $T$ conformal prediction instances via Boole's inequality, we define a single nonconformity score that enables us to reason over multiple prediction errors simultaneously. We  obtain the open-loop abstraction in \eqref{eq:cp_open_} by the nonconformity score 
\begin{align}\label{eq:R_max1}
    R^{(i)}_\text{OL}:=\max_{\tau\in \{1,\hdots,T\}} \alpha_{\tau|0}\|e^{(i)}_\tau-\hat{e}^{(i)}_{\tau|0}\|
\end{align}
where the choice of the constants $\alpha_{\tau|0}>0$ will be crucial to obtain tight abstractions, as we will discussed later. We then proceed by computing
\begin{align}\label{eq:C_max1}
    C_\text{OL}:=\text{Quantile}_{1-\bar{\delta}}( R^{(1)}_\text{OL}, \hdots, R^{(K)}_\text{OL}, \infty ).
\end{align}
Almost equivalently, we  obtain the closed-loop abstraction in \eqref{eq:cp_closed_}  by the nonconformity score 
\begin{align}\label{eq:R_max2}
    R^{(i)}_\text{CL}:=\max_{t\in\{0,\hdots,T-1\}} \alpha_{t+1|t}\|e^{(i)}_{t+1}-\hat{e}^{(i)}_{t+1|t}\|
\end{align}
where we again use constants $\alpha_{t+1|t}>0$, and then compute
\begin{align}\label{eq:C_max2}
    C_\text{CL}:=\text{Quantile}_{1-\bar{\delta}}( R^{(1)}_\text{CL}, \hdots, R^{(K)}_\text{CL}, \infty ).
\end{align}

Next, we state the main result upfront, again utilizing Lemma \ref{lem:1}, which is followed by a discussion on the choice of the constants $\alpha_{\tau|0}>0$ and $\alpha_{t+1|t}>0$.

\begin{theorem}\label{thm:2}
    Given a test trajectory $e^{(0)}\sim\mathcal{D}_e$ and positive constants $\alpha_{\tau|0}>0$ and $\alpha_{t+1|t}>0$, the statistical abstractions in \eqref{eq:cp_open_} and \eqref{eq:cp_closed_} are valid by the choices of
    \begin{align*}
        C_{\tau|0}&:={C}_\text{OL}/\alpha_{\tau|0},\\
        C_{t+1|t}&:={C}_\text{CL}/\alpha_{t+1|t},
    \end{align*}
    where ${C}_\text{OL}$ and ${C}_\text{CL}$ are following from equations \eqref{eq:C_max1} and \eqref{eq:C_max2} with the nonconformity scores in equations \eqref{eq:R_max1} and \eqref{eq:R_max2}, respectively.
    
\textbf{Proof: }
Let us  show that the  abstraction in equation \eqref{eq:cp_open_} is valid by the choice of $C_{\tau|t}$, while the proof for the abstraction in \eqref{eq:cp_closed_} again follows similarly. From Lemma \ref{lem:1} and the definition of the nonconformity score $R_\text{OL}^{(i)}$ it immediately follows  that 
\begin{align*}
    \text{Prob}\big(\max_{\tau\in \{1,\hdots,T\}} \alpha_{\tau|0}\|e^{(0)}_\tau-\hat{e}^{(0)}_{\tau|0}\|\le C_{\tau|0}\big)\ge 1-\delta
\end{align*}
which, since $\alpha_{\tau|0}>0$, is equivalent to
\begin{align*}
    \text{Prob}(\|e^{(0)}_\tau-\hat{e}^{(0)}_{\tau|0}\|\le C_{\tau|0}/\alpha_{\tau|0},\forall \tau\in \{1,\hdots,T\})\ge 1-\delta
\end{align*}
by which we have shown that \eqref{eq:cp_open_} holds with $C_{\tau|0}:={C}_\text{OL}/\alpha_{\tau|0}$. \qed
\end{theorem}

While Theorem \ref{thm:2} holds for arbitrary positive constants  $\alpha_{\tau|0}$ and $\alpha_{t+1|t}$, we  obtain informative, i.e., non-conservative,  abstractions only if these constants are computed intelligently. Let us focus  on the nonconformity score $R_\text{OL}^{(i)}$ in \eqref{eq:R_max1} that result in the  abstraction \eqref{eq:cp_open_} to get some intuition. Why do we not simply select $\alpha_{\tau|0}:=1$ for all $\tau\in\{1,\hdots,T\}$? The reason is that the prediction errors $\|e^{(i)}_\tau-\hat{e}^{(i)}_{\tau|0}\|$ have different distributions, e.g., the prediction error for $\tau=T$ typically has the largest $1-\delta$  quantile and would hence dominate the max-operator in  $R_\text{OL}^{(i)}$. In other words, the upper bound $C_{\tau|0}$ from Theorem \ref{thm:2} would be the same for each time $\tau\in\{1,\hdots,T\}$, which could be  tight for $\tau=T$ but may be loose for $\tau<T$.

To overcome this challenge, one simple idea is to select $\alpha_{\tau|0}$ such that it normalizes the prediction error, i.e., such that $\alpha_{\tau|0}\|e^{(i)}_\tau-\hat{e}^{(i)}_{\tau|0}\|$ lies within the interval $[-1,1]$. To achieve this (approximately) in practice, we sample an additional dataset of $M$ trajectories from the environment in \eqref{eq:LEAS_model_eee} so that we have access to  $e^{(i)}:=(e_0^{(i)},\hdots, e_T^{(i)}) \sim \mathcal{D}_e$ for $i\in\{K+1,\hdots,K+M\}$.\footnote{We remark that this dataset could simply  be the dataset that was used to train the predictor $\mu$. However, we emphasize that the dataset cannot be the calibration dataset from Assumption \ref{ass3}, which would result in dependent nonconformity scores.} We can now compute the normalization constants for \eqref{eq:R_max1} and \eqref{eq:R_max2}  as
\begin{subequations}
\begin{align}
    \alpha_{\tau|0}&:=\frac{1}{\max_{i\in \{K+1,\hdots,K+M\}} \|e^{(i)}_{\tau} - \hat{e}^{(i)}_{\tau|0}\|},\label{eq:normalization_const_open}\\
    \alpha_{t+1|t}&:=\frac{1}{\max_{i\in \{K+1,\hdots,K+M\}} \|e^{(i)}_{t+1} - \hat{e}^{(i)}_{t+1|t}\|}.
\end{align}
\end{subequations}

We note that the computation of $\alpha_{\tau|0}$ and $\alpha_{t+1|t}$ is highly efficient, but not necessarily incentivizing any form of optimality. Instead, we can cast the problem of finding the parameters $\alpha_{\tau|0}$ and  $\alpha_{t+1|t}$ as an optimization problem in which we minimize the $1-\delta$ quantile over the dataset $e^{(i)}:=(e_0^{(i)},\hdots, e_T^{(i)}) \sim \mathcal{D}_e$ for $i\in\{K+1,\hdots,K+M\}$, following ideas from \cite{cleaveland2023conformal}. We refer the reader to Sidebar \textbf{Normalization Constants via Mixed Integer Linear Complementarity Programming} for details. Finally, we  provide an example that illustrate the two approaches that we presented to compute statistical abstractions.

\begin{sidebar}{Normalization Constants via Mixed Integer Linear Complementarity Programming}
\section[Normalization Constants via Mixed Integer Linear Complementarity Programming]{}\phantomsection
   \label{sidebar-proof-CP}
\setcounter{sequation}{0}
\renewcommand{\thesequation}{S\arabic{sequation}}
\setcounter{stable}{0}
\renewcommand{\thestable}{S\arabic{stable}}
\setcounter{sfigure}{0}
\renewcommand{\thesfigure}{S\arabic{sfigure}}
Let us focus on the computation of the normalization constants $\alpha_{\tau|0}$, while the constants $\alpha_{t+1|t}$ can be computed similarly. We cast the problem of finding   $\alpha_{\tau|0}$ as the  optimization problem
\begin{subequations}\label{eq:highLevelOptimization_}
\begin{align} 
     &\min_{ \alpha_{1|0},\hdots,\alpha_{T|0}} \text{Quantile}_{1-\delta}( R^{(K+1)}, \hdots, R^{(K+M)},\infty) \label{eq:highLevelOptimization} \\
    \text{s.t. } & R^{(i)} = \max_{\tau\in\{1,\hdots,T\}}\alpha_{\tau|0} \|e_\tau^{(i)}-\hat{e}_{\tau|0}^{(i)}\|, \forall i\in\{K+1,\hdots,K+M\} \label{eq:highLevelOptimizationb}  \\
    & \sum_{\tau=1}^{T} \alpha_{\tau|0} = 1 \label{eq:highLevelOptimizationc}  \\
    &  \alpha_{\tau|0} \ge 0, \forall \tau\in\{1,\hdots,T\} \label{eq:highLevelOptimizationd}.
\end{align}
\end{subequations}
We note that the equality constraint in \eqref{eq:highLevelOptimizationb} computes the nonconformity score $R^{(i)}$ defined in \eqref{eq:R_max1} over the additional dataset $i\in\{K+1,\hdots,K+M\}$, while the cost function in \eqref{eq:highLevelOptimization}  minimizes the $1-\delta$ quantile over $R^{(K+1)}, \hdots, R^{(K+M)},\infty$. Lastly, the inequality constraint in \eqref{eq:highLevelOptimizationd} enforces positive normalization constants, while the equality constraint in \eqref{eq:highLevelOptimizationc} enforces a normalization budget, avoiding that $\alpha_{\tau|0}=0$ for all $\tau\in\{1,\hdots,T\}$ which would be a trivial solution. In fact, this constraint  enforces that $\alpha_{\tau|0}>0$ for all $\tau\in\{1,\hdots,T\}$, see \cite[Theorem 1]{cleaveland2023conformal}. The optimization problem in equation \eqref{eq:highLevelOptimization_} is always feasible since any set of parameters $\alpha_{1|0},\hdots,\alpha_{T|0}$ that satisfy constraints \eqref{eq:highLevelOptimizationc} and \eqref{eq:highLevelOptimizationd} are a feasible solution of \eqref{eq:highLevelOptimization_}.

Given the optimization problem \eqref{eq:highLevelOptimization_}, how can we now  solve this problem in practice? Remembering Sidebar \textbf{Empirical Quantiles as Linear Programs}, we note that $\text{Quantile}_{1-\delta}( R^{(K+1)}, \hdots, R^{(K+M)},\infty)$ in  \eqref{eq:highLevelOptimization} can be written as a linear program following \eqref{eq:quantileAsLP}, hence making the optimization problem \eqref{eq:highLevelOptimization_} a bilevel optimization problem. To address this issue, we  rewrite the inner optimization problem into a set of equivalent linear complementarity constraints using the KKT conditions of the inner optimization problem. Finally, we  use a standard reformulation of the max-operator into a set of equivalent mixed integer constraints \cite{bemporad1999control}, making the optimization problem \eqref{eq:highLevelOptimization_} a mixed integer linear complementarity program. The authors of  \cite{cleaveland2023conformal} also showed that the integer constraints can be removed without losing optimality, see \cite{cleaveland2023conformal} for details. We provide a numerical evaluation of the statistical abstraction that is obtained in this way in Example \ref{ex:abstractions}.

The presented approach follows a two-step procedure requiring two independent datasets on which we solve the optimization problem in \eqref{eq:highLevelOptimization_} and apply Theorem \ref{thm:2}. Recent work in \cite{sharma2024pac} enables the use of PAC-Bayes theory to obtain generalization bounds without splitting datasets. Theoretical connections between the size of prediction regions and generalization properties of the underlying predictor were established in \cite{zecchin2024generalization}.  
\end{sidebar}

\begin{example}\label{ex:abstractions}

\begin{figure*}
    \centering
    \begin{subfigure}[t]{0.31\textwidth}
        \includegraphics[width=\textwidth]{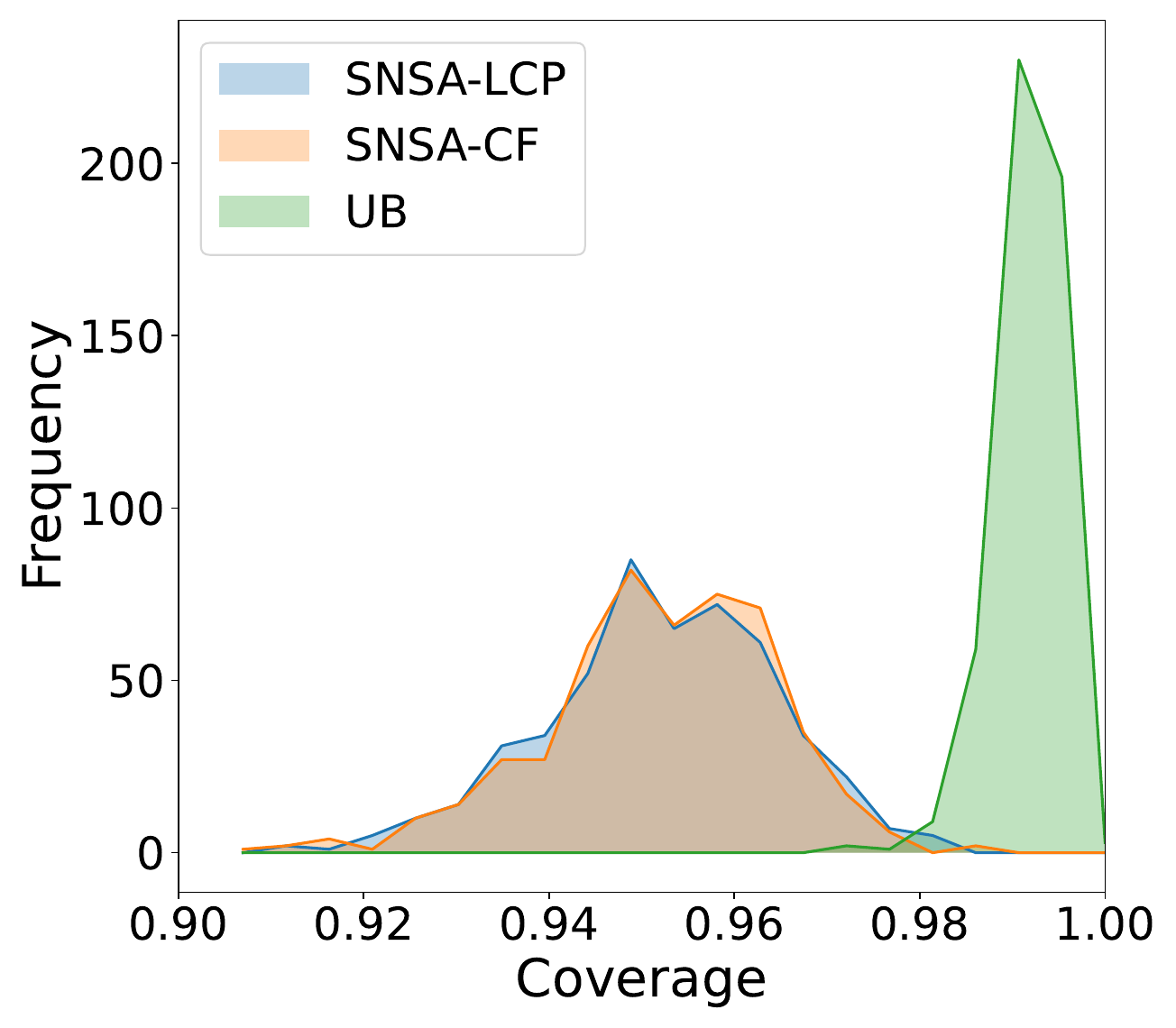}
        \caption{Histogram of $||e_\tau - \hat{e}_{\tau|0}|| \le C_{\tau|0}, \forall \tau \in \{1, \dots, T\}$ over all $N$ experiments.}
        \label{fig:example2_CEC}
    \end{subfigure}
    \hspace{2mm}
    \begin{subfigure}[t]{0.31\textwidth}
        \includegraphics[width=\textwidth]{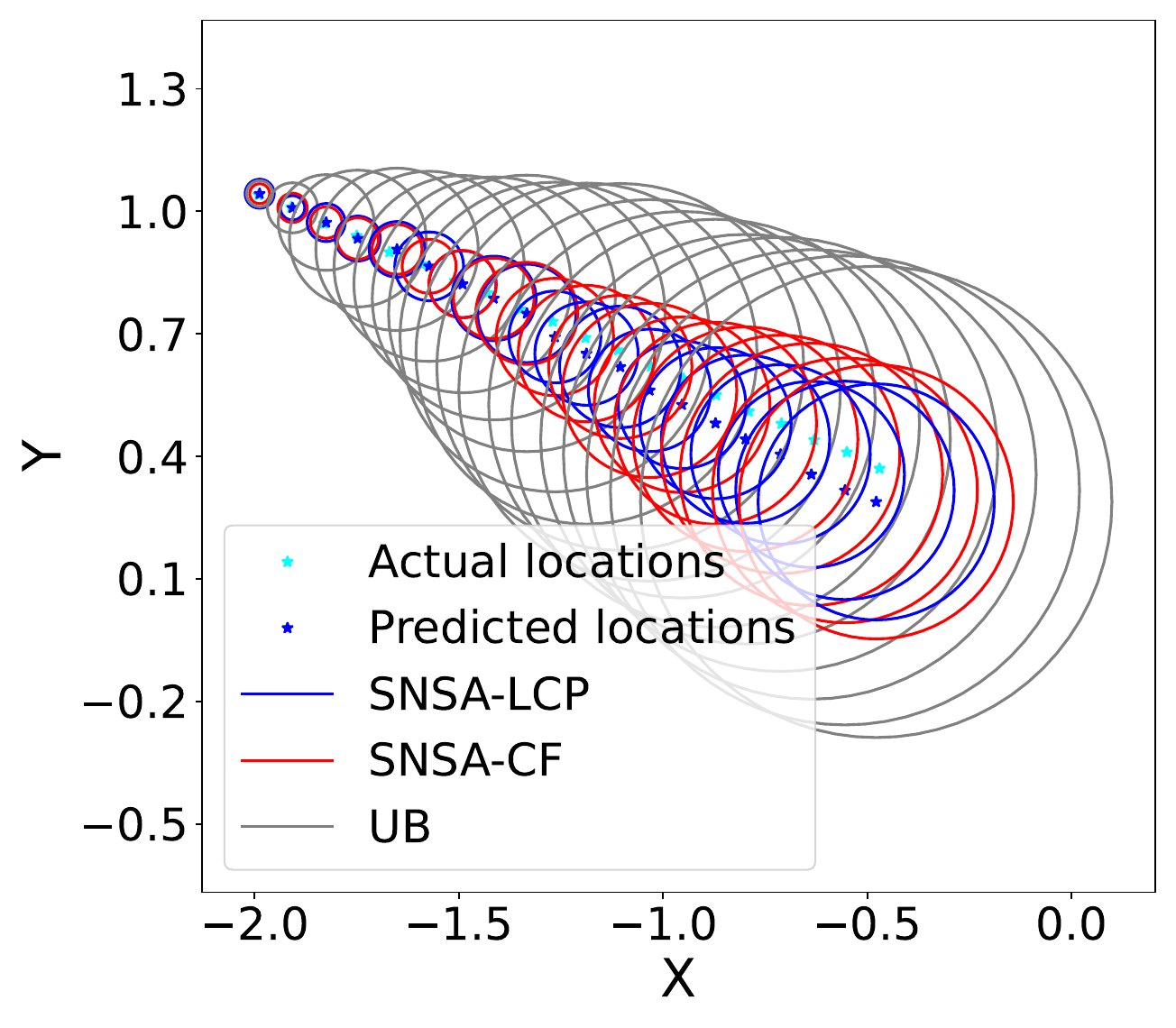}
        \caption{Conformal prediction regions}
        \label{fig:example2_general}
    \end{subfigure}
    \hspace{2mm}
    \begin{subfigure}[t]{0.31\textwidth} 
        \includegraphics[width=\textwidth]{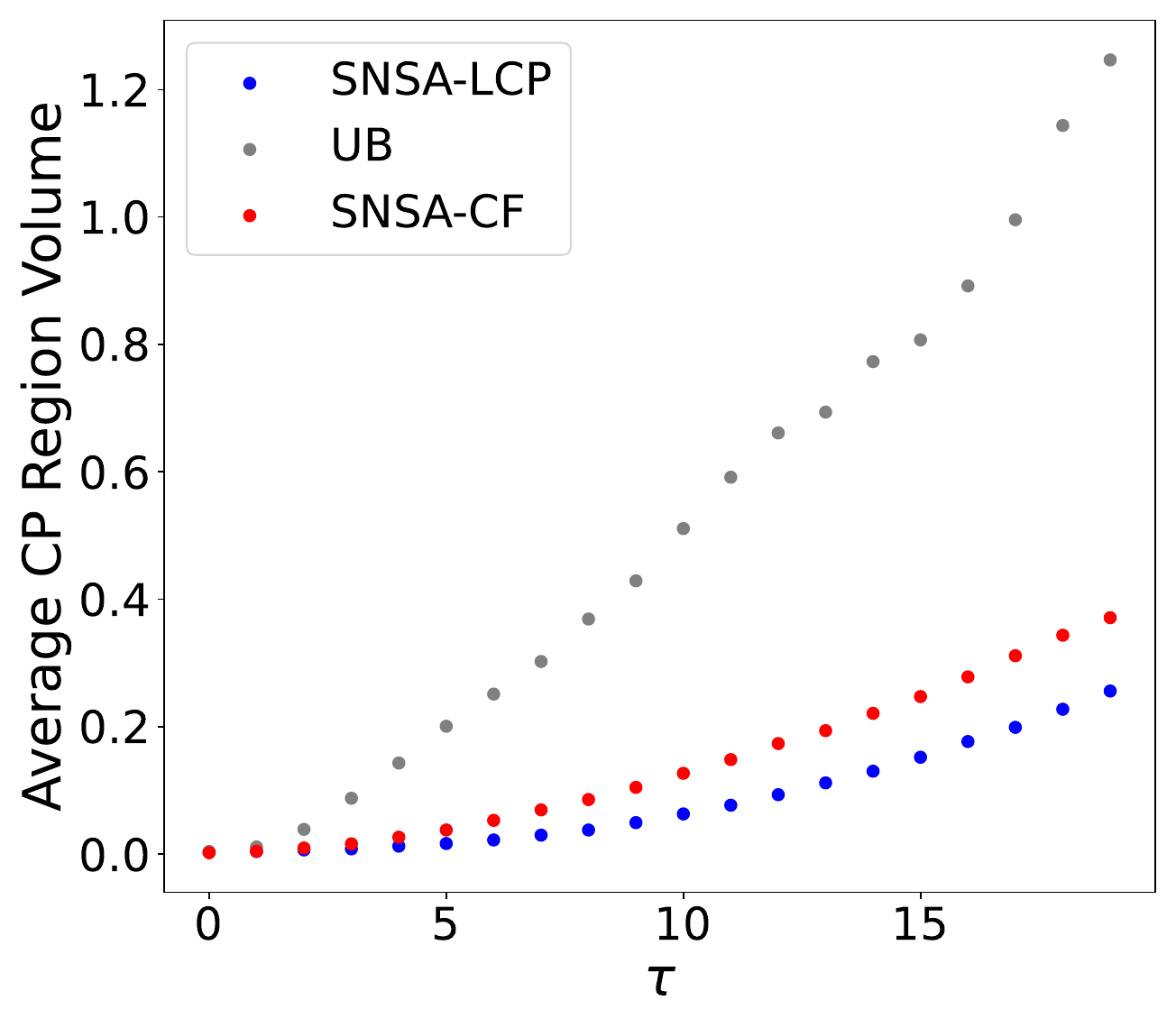}
        \caption{Average area of the prediction regions}
        \label{fig:example2_averagearea}
    \end{subfigure}
    \caption{Experimental Results for Example \ref{ex:abstractions}.}
\end{figure*}

\begin{figure*}
    \centering
    \begin{subfigure}[t]{0.31\textwidth}
        \includegraphics[width=\textwidth]{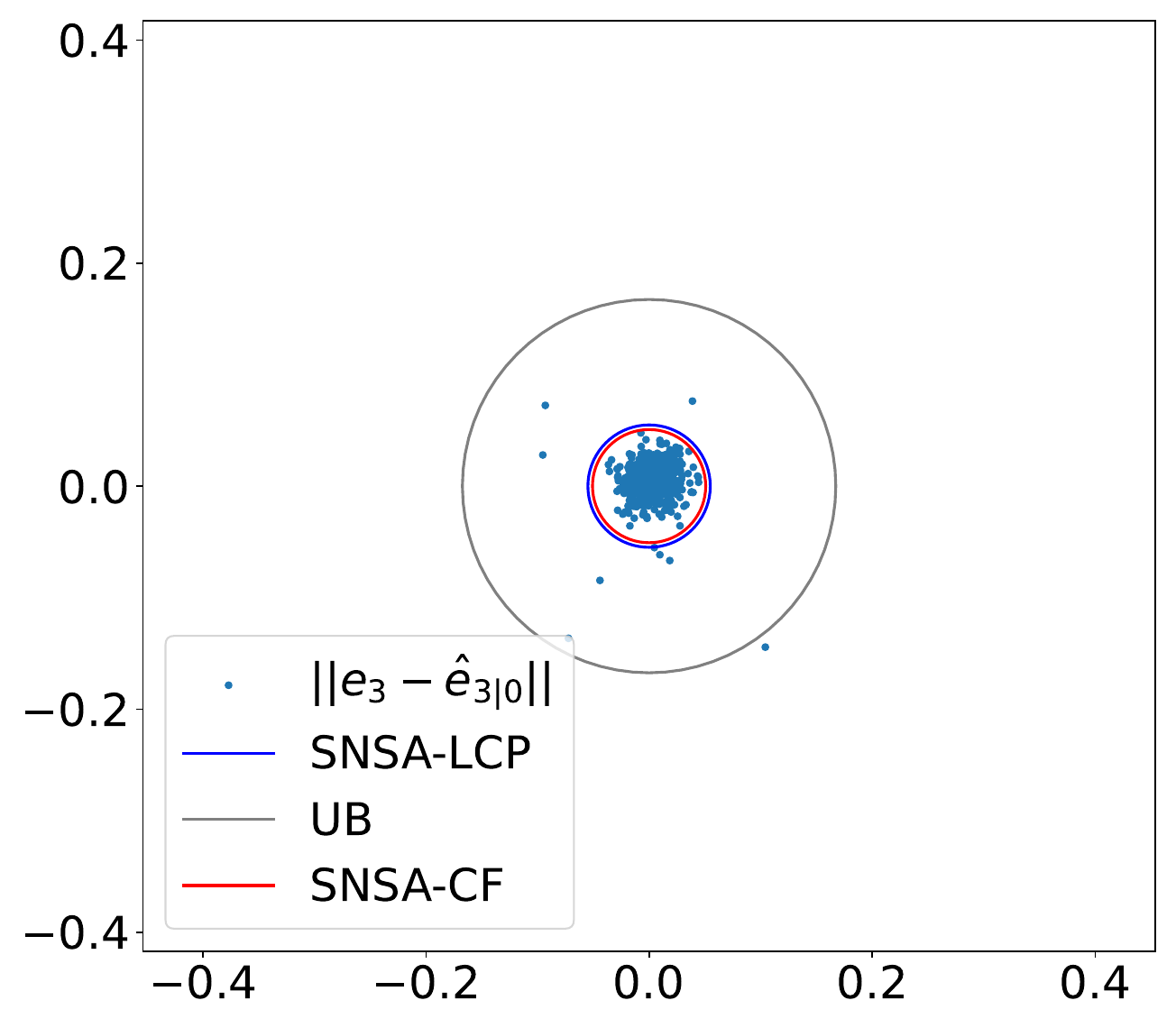}
        \caption{$||e_3 - \hat{e}_{3|0}||$}
    \end{subfigure}
    \hspace{2mm}
    \begin{subfigure}[t]{0.31\textwidth}
        \includegraphics[width=\textwidth]{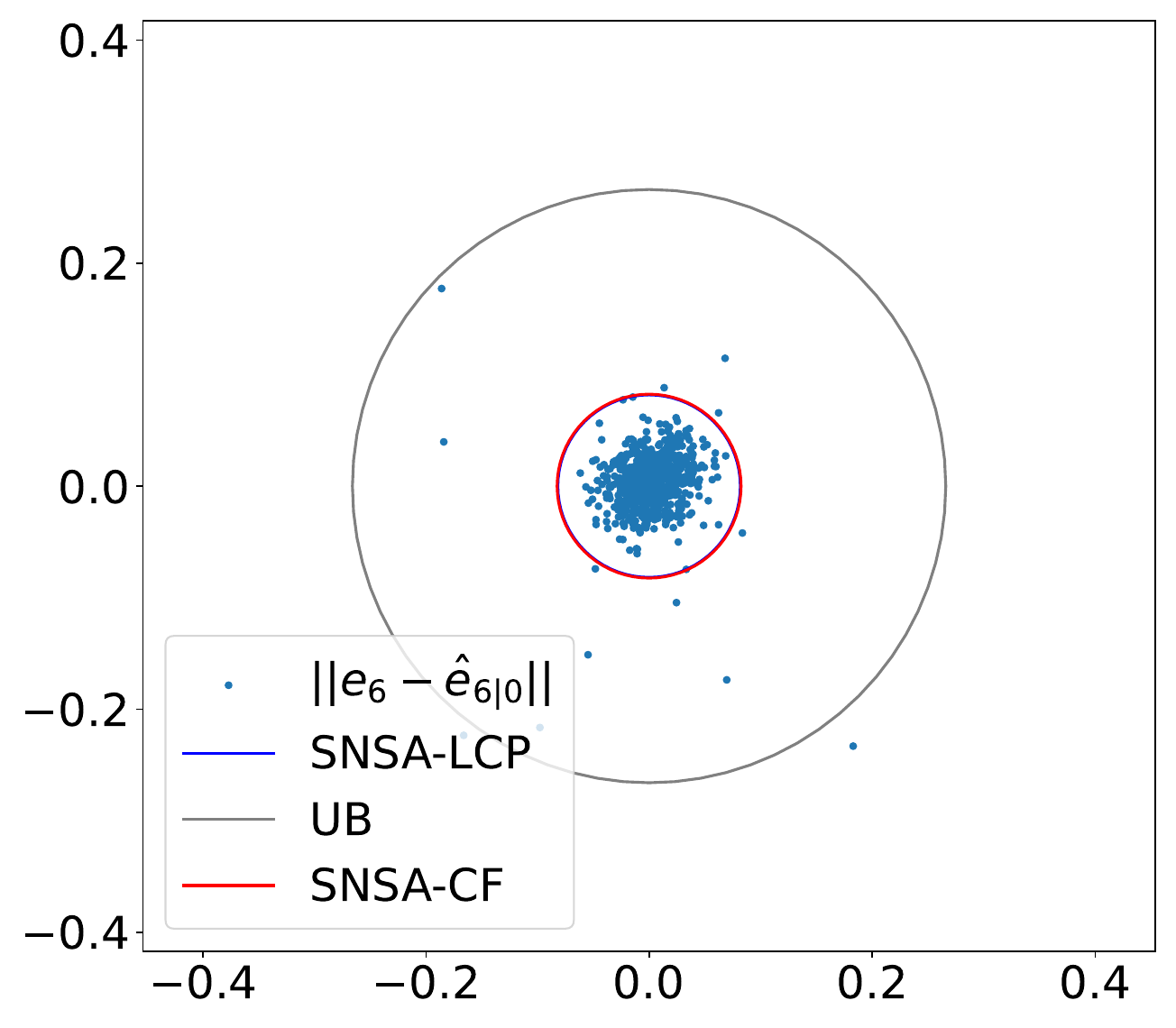}
        \caption{$||e_6 - \hat{e}_{6|0}||$}
    \end{subfigure}
    \hspace{2mm}
    \begin{subfigure}[t]{0.31\textwidth} 
        \includegraphics[width=\textwidth]{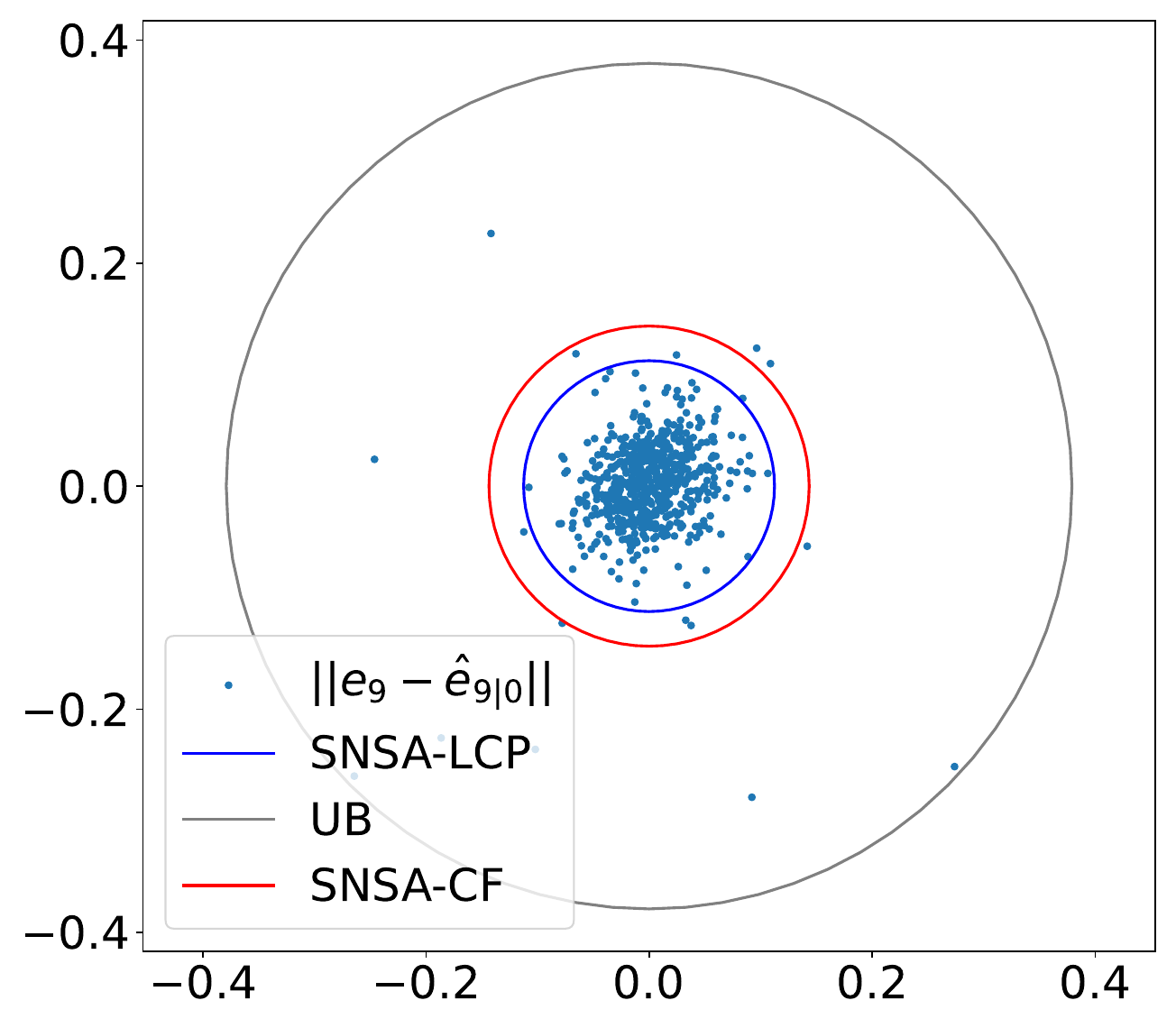}
        \caption{$||e_9 - \hat{e}_{9|0}||$}
    \end{subfigure}
    \caption{Conformal prediction regions and errors for various prediction horizons for Example \ref{ex:abstractions}.}
    \label{fig:example2_validation}
\end{figure*}

We construct open-loop statistical abstractions as in equation \eqref{eq:cp_open_} for pedestrians modeled by a distribution $e\sim \mathcal{D}_e$ that is implicitly defined by the ORCA simulator \cite{van2008reciprocal}. We consider the current time $t:=0$ and a mission horizon of $T=20$ time steps, requiring us to make predictions for times $\tau\in\{1,\hdots,20\}$.\footnote{We assume that $20$ time steps of observations are already available as inputs for the predictor, but we keep $t:=0$ nonetheless for notational consistency. We also discretized the original signal at a rate of $8$ Hz so that we have a total of $5$ seconds of data, with observations and predictions of $2.5$ seconds each.} First, we train a long short-term memory network that takes social interaction between pedestrians into account, see \cite{alahi2016social,kothari2021human} for details. In the remainder, we set the failure probability to $\delta := 0.05$ and compute three different types of statistical abstractions using: (1) the ``naive approach'', (2) the ``single nonconformity score approach'' with normalization constants as in equation \eqref{eq:normalization_const_open}, and (3) the ``single nonconformity score approach'' with normalization constants as in Sidebar \textbf{Normalization Constants via Mixed Integer Linear Complementarity Programming}. We refer to these three methods as UB, SNSA-CF, and SNSA-LCP, respectively.\footnote{UB stands for union bound, while SNSA stands for single nonconformity score approach with extensions CF and LCP for closed-form and linear complementarity programming, respectively.}  

\textbf{Computing normalization constants. } We first compute the normalization constants for SNSA-CF and SNSA-LCP by using $M:=50$ trajectories sampled from $\mathcal{D}_e$. The average computation times over $500$ experiments for SNSA-CF and SNSA-LCP are $0.00016$ and $0.08291$ seconds, respectively.   While the normalization constants for SNSA-LCP provide tighter statistical abstractions (details and plots provided below), they require larger computation times. We also remark that the computation times for $SNSA-LCP$ may increase drastically when increasing $T$ or $M$ (recall the optimization problem in \eqref{eq:highLevelOptimization_}).

\textbf{Statistical validation. } We conduct $N \coloneqq 500$ experiments  for calibration set sizes of $K := 596$ and test set sizes of $J:=500$. We plot the conditional empirical coverage of $||e_\tau - \hat{e}_{\tau|0}|| \le C_{\tau|0}$ (defined in the same way $CEC_n$ is defined in equation \eqref{eq:cond_emp_coverage}) for all three methods in Figure \ref{fig:example2_CEC}.\footnote{The computation of $C_{\tau|0}$ for UB follows equation \eqref{eq:C_naiv}, while the computation of $C_{\tau|0}$ for SNSA-CF and SNSA-LCP follows Theorem \ref{thm:2}.} As expected, it can be seen that UB is overly conservative, while SNSA-CF and SNSA-LCP provide exact coverage around $1-\delta$ confidence. The average conditional empirical coverage of $||e_\tau - \hat{e}_{\tau|0}|| \le C_{\tau|0}$ is $0.994$, $0.946$, and $0.946$ for UB, SNSA-CF, and SNSA-LCP, respectively.

\textbf{Comparing all three methods. } For one experiment, we illustrate the pedestrian trajectory alongside with the predictions and the statistical abstractions in Figure \ref{fig:example2_general}. In Figure \ref{fig:example2_averagearea}, we also plot the average value of $||e_\tau - \hat{e}_{\tau|0}||$ for each time $\tau$, which can here be interpreted as the volume of the abstraction. We see again that UB has the largest coverage due to the induced conservatism, while SNSA-CF and SNSA-LCP provide valid statistical abstractions with smaller volumes. We also note that SNSA-LCP has the smallest volume, which comes at the cost of increased computation times for the normalization constants (as discussed before).  Finally, we show two-dimensional plots of the statistical abstractions along with the test trajectories at times $\tau\in\{3,6,9\}$ in Figure \ref{fig:example2_validation}. We again see that SNSA-CF and SNSA-LCP tightly cover a $1-\delta$ fraction of the test data.
\end{example}

\textbf{Shaping the geometry of statistical abstractions.} In both methods (naive and single nonconformity score), we have fixed the spatial shape of the abstraction  to be $\|e_\tau-\hat{e}_{\tau|t}\|$. In other words,  for \eqref{eq:cp_open_} (and similarly for \eqref{eq:cp_closed_})  the abstraction will be a ball of radius $C_{\tau|0}$ centered around the prediction $\hat{e}_{\tau|0}$. This choice may lead to conservative abstractions when $e_\tau-\hat{e}_{\tau|0}$ is not symmetrically distributed. One such example is when the environment consists of multiple agents, i.e., when $e_t:=(e_{t,1},\hdots,e_{t,N})$ for $N$ different agents, and some agent predictions have more uncertainty than others.  Following a similar idea as in equations \eqref{eq:R_max1} and \eqref{eq:R_max2}, we can define a nonconformity score that uses the maximum prediction error $\|e_{t,n}-\hat{e}_{\tau|t,n}\|$ over all $n\in\{1,\hdots,N\}$ agents \cite{yu2023signal}. We discuss this approach in more detail in the end of the next sub-section. Another idea was presented in \cite{muthali2023multi} where quantile regressors are trained and calibrated for each agent, motivated by success in conformalized quantile regression as discussed in Sidebar \textbf{Heteroskedasticity and Conformal Prediction}.

More generally, and not within a multi-agent context, we can use learned kernel density estimators, i.e., estimators $\hat{p}(e_t)$ of the probability density function of $e_t$ (these estimators may even be conditioned on $e_0,\hdots,e_{t-1}$). We then obtain the statistical abstraction  $\{\bar{e}_t\in\mathbb{R}^{n_e} | -\hat{p}(\bar{e}_t) \le C \}$ where $C$ is  the empirical quantile of the nonconformity score $R^{(i)} := -\hat{p}(e_t^{(i)})$, see e.g., \cite{lei2011efficient,lei2013distribution,lei2014distribution,smith2014anomaly}. Such approaches  capture complex spatial shapes and multi-modality in the distribution of $e_t$. However,  computing the set $\{\bar{e}_t\in\mathbb{R}^{n_e} | -\hat{p}(\bar{e}_t) \le C \}$ can be computationally challenging. Conceptually similar to the work in \cite{cleaveland2023conformal}, which we summarized in Sidebar \textbf{Normalization Constants via Mixed Integer Linear Complementarity Programming}, the authors in  \cite{stutz2021learning} integrate the shape of the abstraction directly into the loss function when training a neural network. However, this may result in abstractions with complex shapes that are not amenable for downstream  formal verification and control. Hence, one may  instead be interested in obtaining efficient shapes, e.g., by fitting convex shapes such as ellipsoids or polytopes to the error $\|e_\tau-\hat{e}_{\tau|t}\|$, see \cite{tumu2023multi}. Other  work looked at the problem of multi-variate regression, see \cite{thurin2025optimal}.  More generally, minimum volume abstractions (also referred to as prediction or covering sets) for multimodal distributions were considered in \cite{tumu2023multi,braun2025minimum,gao2025volume}.

\textbf{Other techniques for statistical abstractions in dynamic environments.} Constructing statistical abstractions in dynamic environments has received a lot of attention recently. The authors in \cite{sun2022copula} propose an alternative approach by learning and calibrating copulas, which are cumulative distribution functions of $e_1,\hdots,e_T$ that have a marginal representation of $e_t$ for each $t\in\{1,\hdots,T\}$. The authors in \cite{marques2024quantifying} showed how to construct statistical abstractions of  dynamical systems with aleatoric and epistemic uncertainty. In \cite{zhou2024conformalized}, adaptive conformal prediction techniques (recall Sidebar \textbf{Conformal Prediction under Distribution Shift}) are combined with  nonconformity scores similar to  \eqref{eq:R_max1} and \eqref{eq:R_max2}  to obtain abstractions of the form \eqref{eq:cp_open_} and \eqref{eq:cp_closed_} that adapt to ``difficult to predict'' trajectories. All aforementioned approaches require calibration datasets of multiple trajectories as per Assumption \ref{ass3}. This limitation was addressed in \cite{lee2024single} within a single trajectory approach under the assumption that samples of $e_t$ converge to a stationary distribution and are ``approximately'' independent when sufficiently separated in time. To compute statistical abstractions, our exposition has focused on deterministic trajectory predictors $\mu$. Without much change, we can incorporate stochastic trajectory predictors, e.g., generative and probabilistic predictors such as diffusion models, Kalman filters, or decision transformers. The authors in \cite{zecchin2024forking} and \cite{wang2022probabilistic} present a nonconformity score that takes multiple predictions into account by computing the minimum distance between these predictions and the ground truth value. Such an approach can  capture multi-modal behavior, e.g., robots that may turn left or right.

\subsection{Safe Control in Dynamic Environments} 

We can now design a control sequence $u_0,\hdots,u_{T-1}$ that solves Problem \ref{prob1}. Motivated by the statistical abstractions in equations \eqref{eq:cp_open_} and \eqref{eq:cp_closed_}, we  define the uncertainty sets
\begin{align*}
    \mathcal{B}_{\tau|t}:=\{\bar{e}_\tau\in\mathbb{R}^{n_e}|\|\bar{e}_\tau- \hat{e}_{\tau|t}\|\le C_{\tau|t}\}.
\end{align*}

In the remainder, we follow \cite{lindemann2022safe} to design a receding horizon control strategy where we solve an optimization problem at each time $t$ to obtain the  control input $u_t$.

\textbf{The optimization problem. } The optimization problem will use all  information available at time $t$ and output a sequence of control inputs $u_{t|t},\hdots,u_{T-1|t}$ and  states $x_{t+1|t},\hdots,x_{T|t}$. We then apply the first control input $u_t:=u_{t|t}$ to the system. Specifically, for a prediction horizon of $H>0$, consider the optimization problem
\begin{subequations}\label{eq:open_loop}
\begin{align}
    &\min_{(u_{t|t},\hdots,u_{T-1|t})} J(x,u)\label{eq:cosss} &\\
     \text{s.t.}\;\;& x_{\tau+1|t}=f_x(x_{\tau|t},u_{\tau|t}), &\tau\in\{t,\hdots,T-1\} \label{eq:dyn}\\
     & x_{t|t}=x_t&\\
     & \min_{\bar{e}_\tau \in \mathcal{B}_{\tau|t}} c(x_{\tau|t},\bar{e}_\tau)\ge 0,&\tau\in\{t+1,\hdots,t+H\}\label{eq:constC_2}\\
     & u_{\tau|t} \in \mathcal{U},x_{\tau+1|t} \in \mathcal{X},&\tau\in\{t,\hdots,T-1\} \label{eq:incon}
\end{align}
\end{subequations}
where $J:\mathbb{R}^{n_x(T+1)}\times \mathbb{R}^{n_uT}\to\mathbb{R}$ is a user-defined cost function that is defined over the state and input trajectories $x:=(x_0,\hdots,x_t,x_{t+1|t},\hdots,x_{T|t})$ and $u:=(u_0,\hdots,u_{t-1},u_{t|t},\hdots,u_{T-1|t})$, i.e., over past and future states and control inputs. The sets $\mathcal{U}\subseteq \mathbb{R}^{n_u}$ and $\mathcal{X}\subseteq \mathbb{R}^{n_x}$ are input and state constraints that we can add to our problem. While the cost function as well as dynamics, state, and input constraints in equations \eqref{eq:cosss}, \eqref{eq:dyn}, and \eqref{eq:incon} are standard in finite-horizon optimal control problems, the safety constraint in \eqref{eq:constC_2} is new and arises as the dynamic environment $e$ is uncertain. In fact, it is a robust safety constraint that enforces that the constraint $c$ at time $\tau$ is enforced for all environment states within the abstraction $\mathcal{B}_{\tau|t}$. This, in turn, will guarantee that $\text{Prob}(c(x_{\tau|t},e_\tau)\ge 0)\ge 1-\delta$. However, the robust safety constraint makes the optimization problem \eqref{eq:open_loop} a bilevel optimization problem. If the constraint function $c$ is Lipschitz continuous in its second argument, with Lipschitz constant $L$, we can instead use a computationally more efficient reformulation as 
\begin{align}\label{eq:reform_lip}
    c(x_{\tau|t},\hat{e}_{\tau|t})\ge LC_{\tau|t},&\tau\in\{t+1,\hdots,t+H\}.
\end{align}
It is easy to verify that satisfaction of the constraint in \eqref{eq:reform_lip} is a sufficient condition for satisfaction of the constraint in \eqref{eq:constC_2}, see e.g., \cite[Theorem 2]{lindemann2022safe} for intuition. For convenience, we summarize this optimization problem as
\begin{subequations}\label{eq:open_loop_}
\begin{align}
    &\min_{(u_{t|t},\hdots,u_{T-1|t})} J(x,u)\label{eq:cosss_} &\\
     \text{s.t.}\;\;& x_{\tau+1|t}=f_x(x_{\tau|t},u_{\tau|t}), &\tau\in\{t,\hdots,T-1\} \label{eq:dyn_}\\
     & x_{t|t}=x_t&\\
     & c(x_{\tau|t},\hat{e}_{\tau|t})\ge LC_{\tau|t},&\tau\in\{t+1,\hdots,t+H\}\label{eq:constC_2_}\\
     & u_{\tau|t} \in \mathcal{U},x_{\tau+1|t} \in \mathcal{X},&\tau\in\{t,\hdots,T-1\}. \label{eq:incon_}
 \end{align}
\end{subequations}

\textbf{Open-loop and closed-loop controllers. } Having formulated the optimization problems in equations \eqref{eq:open_loop} and \eqref{eq:open_loop_}, we obtain a simple and easy to understand open-loop controller at time $t:=0$ if we set $H:=T$.  
\begin{theorem}\label{thm:3}
    Given a test trajectory $e^{(0)}\sim\mathcal{D}_e$ and the statistical abstraction  \eqref{eq:cp_open_}, then  a solution $u_\tau:=u_{\tau|0}$ of the optimization problem  \eqref{eq:open_loop} (or \eqref{eq:open_loop_} if $c$ is Lipschitz continuous) at time $t:=0$ with $H:=T$ is such that  
    \begin{align*}
        \text{Prob}(c(x_t,e_t^{(0)})\ge 0,\forall t\in \{1,\hdots,T\})\ge 1-\delta.
    \end{align*}
\end{theorem}
The proof of this result follows immediately and is intuitive by the construction of the statistical abstractions in \eqref{eq:cp_open_}. While this result provides a solution to Problem \ref{prob1}, Theorem \ref{thm:3} assumes that the optimization problem in \eqref{eq:open_loop} (or \eqref{eq:open_loop_}) is feasible at time $t:=0$. This may not always be the case, especially for long task horizons $T$, inaccurate trajectory predictors $\mu$, or distributions $\mathcal{D}_e$ with larger variance which may lead to large values of $C_{\tau|0}$, rendering the constraint \eqref{eq:constC_2} (or \eqref{eq:constC_2_}) more restrictive. Another direct consequence of this is that the open-loop controller may be conservative. As motivated in the beginning, a better approach is a receding horizon control strategy.
\begin{theorem}\label{thm:4}
    Given a test trajectory $e^{(0)}\sim\mathcal{D}_e$ and the statistical abstraction \eqref{eq:cp_closed_}, then the solutions $u_t:=u_{t|t}$ of the optimization problem \eqref{eq:open_loop} (or \eqref{eq:open_loop_} if $c$ is Lipschitz continuous)  at times $t\in\{0,\hdots,T-1\}$ with  $H>0$ are such that  
    \begin{align*}
        \text{Prob}(c(x_t,e_t^{(0)})\ge 0,\forall t\in \{1,\hdots,T\})\ge 1-\delta.
    \end{align*}

    \textbf{Proof:} By assumption, the optimization problem \eqref{eq:open_loop} (or \eqref{eq:open_loop_}) is feasible at each time $t\in \{0,\hdots,T-1\}$ resulting in the control input $u_t:=u_{t|t}$. As the control input is recomputed and applied iteratively, only the constraint \eqref{eq:constC_2} (or \eqref{eq:constC_2}) for $\tau=t+1$, i.e., one-step ahead, becomes relevant. In fact, due to constraint \eqref{eq:constC_2} (or \eqref{eq:constC_2}) and the statistical abstraction \eqref{eq:cp_closed_}, it then follows that $\text{Prob}\big(c(x_{t+1|t},e_{t+1})\ge 0, \forall t\in\{0,\hdots,T-1\}\big)\ge 1-\delta$. This implies that $\text{Prob}\big(c(x_{t+1},e_{t+1})\ge 0, \forall t\in\{0,\hdots,T-1\}\big)\ge 1-\delta$, proving the main result. \qed
\end{theorem}

The result in Theorem \ref{thm:4} is more subtle than Theorem \ref{thm:3} as it only requires  statistical abstractions for one-step ahead predictions, i.e., abstractions of the form \eqref{eq:cp_closed_}. The intuition from the  proof is that the control input $u_t$ is computed and applied to the system iteratively at each time $t\in\{0,\hdots,T-1\}$, enforcing that at the next time $t+1$ the constraint $\text{Prob}(c(x_{t+1},e_{t+1})\ge 0)\ge 1-\delta$ holds. This insight motivates greedy control strategies by selecting small prediction horizons $H$, as long as the optimization problem remains feasible (more on this below). Additionally, this motivates versions of the optimization problem in \eqref{eq:open_loop} (or \eqref{eq:open_loop_}) that incorporate the knowledge that prediction regions $C_{\tau|t}$ for $\tau> t+1$ will be updated  until $\tau=t+1$, which is the time when $C_{\tau|t}$ becomes relevant for obtaining the guarantees in Theorem \ref{thm:4}. In other words, instead of the constraint in \eqref{eq:constC_2}, we can enforce 
\begin{align*}
    \min_{\bar{e}_\tau \in \bar{\mathcal{B}}_{\tau|t}}  &c(x_{\tau|t},\bar{e}_\tau)\ge 0,\tau\in\{t+1,\hdots,t+H\}\\
    \bar{\mathcal{B}}_{\tau|t}&:=\{\bar{e}_\tau\in\mathbb{R}^{n_e}|\|\bar{e}_\tau- \hat{e}_{\tau|t}\|\le C_{\tau|\tau-1}\}
\end{align*}
without losing the guarantees in Theorem \ref{thm:4}. Note that we here simply used the uncertainty set $\bar{\mathcal{B}}_{\tau|t}$ instead of ${\mathcal{B}}_{\tau|t}$ by using $C_{\tau|\tau-1}$ instead of $C_{\tau|t}$. In practice, the set $C_{\tau|\tau-1}$ is expected to be less conservative than $C_{\tau|t}$. Similarly, instead of the constraint in \eqref{eq:constC_2_}, we can enforce
\begin{align*}
     c(x_{\tau|t},\hat{e}_{\tau|t})\ge LC_{\tau|\tau-1},&\tau\in\{t+1,\hdots,t+H\}.
\end{align*}

\textbf{Recursive feasibility. }Theorem \ref{thm:4}  assumes that the optimization problem  \eqref{eq:open_loop} (or \eqref{eq:open_loop_}) is feasible at each time $t\in\{0,\hdots,T-1\}$. Recursive feasibility ensures that a feasible optimization problem at the initial time $t:=0$ implies feasibility of the optimization problem for all future times $t>0$. Standard receding horizon control approaches with time-invariant state constraints enforce recursive feasibility using terminal constraints and costs, see e.g., \cite{chen1998quasi,mayne2000constrained}.  However, enforcing recursive feasibility of the optimization problem  \eqref{eq:open_loop} (or \eqref{eq:open_loop_}) is challenging as the constraint set depends on the a-priori unknown predictions $\hat{e}_{\tau|t}$. While deriving a full solution to this issue is an open problem, recursive feasibility can be enforced in this setting for shrinking horizon control schemes, see \cite{stamouli2024recursively} for  details. In a shrinking horizon control scheme, we  solve the optimization problem \eqref{eq:open_loop} (or \eqref{eq:open_loop_}) at time $t:=0$ with $H:=T$, and then continue solving \eqref{eq:open_loop} (or \eqref{eq:open_loop_}) iteratively at times $t>0$ by shrinking the task and the prediction horizons $T$ and $H$, respectively. In practice, recursive feasibility can also be achieved by adding slack variables to the constraints in \eqref{eq:constC_2} (or \eqref{eq:constC_2_}).

We next illustrate the difference between open-loop and closed-loop controllers with an example taken from \cite{lindemann2022safe}, and we refer to \cite{lindemann2022safe,tonkens2023scalable,yu2023signal} for more case studies.

\begin{example}
    The case study in \cite{lindemann2022safe} considers a mobile robot, modelled using bicycle dynamics, that has to avoid three pedestrians, modelled by a distribution $e\sim \mathcal{D}_e$ that is implicitly defined by the ORCA simulator \cite{van2008reciprocal}. Statistical abstractions of the form \eqref{eq:cp_open_} and \eqref{eq:cp_closed_} are obtained using a long short-term memory network and the ``naive approach'' from Section \textbf{Statistical Abstractions of Dynamic
Environments}. For one experiment, Figure \ref{fig:scen1} shows the results of the open-loop controller (bottom) and the closed-loop  controller (top). The statistical abstractions are shown in green and are safely avoided in both cases.  As the abstraction in the closed-loop case is  updated iteratively, the robot trajectory is less conservative than in  the open-loop case. We refer the reader to \cite{lindemann2022safe} for a detailed empirical analysis of the statistical validity of the approach. 
    \begin{figure*}
\centering
\vspace{-0.15cm}
\includegraphics[scale=0.375]{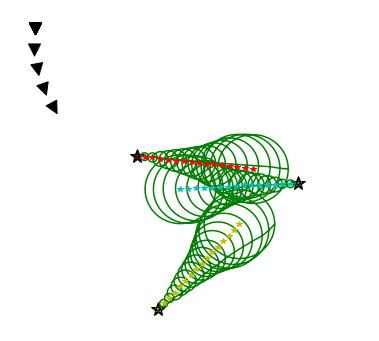}
\includegraphics[scale=0.375]{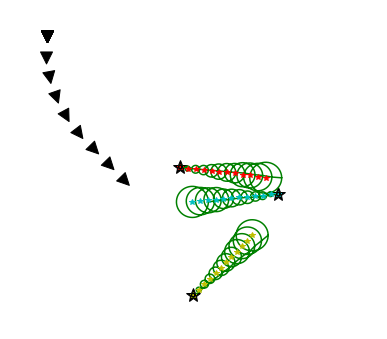}
\includegraphics[scale=0.375]{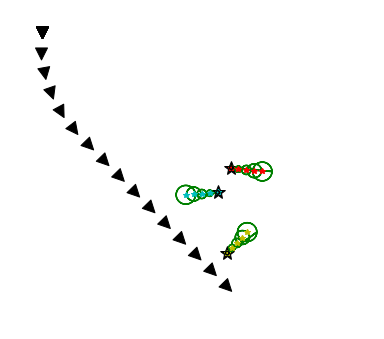}\\\vspace{-0.35cm}
\includegraphics[scale=0.375]{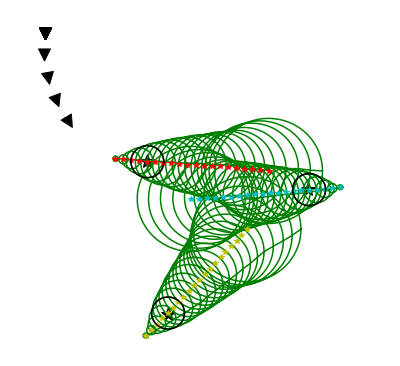}\vspace{-0.5cm}
\includegraphics[scale=0.375]{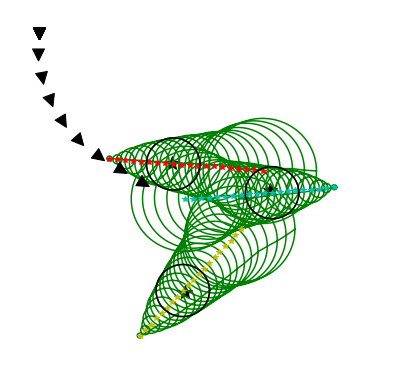}
\includegraphics[scale=0.375]{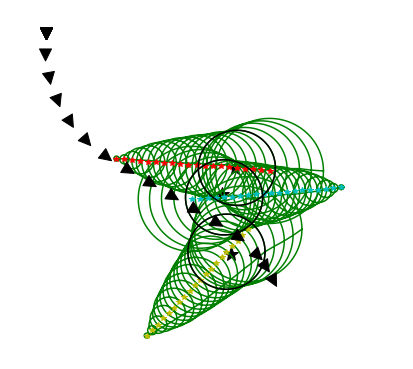}
\caption{The top (bottom) three figures show the closed-loop (open-loop) robot trajectory as black triangles at times $4$, $8$, and $15$. The actual pedestrian trajectories are indicated by black stars. The trajectory predictions are indicated by red, blue, and yellow stars with the corresponding uncertainty sets shown in green. Figure taken from \cite{lindemann2022safe}.}
\label{fig:scen1}
\vspace{-0.5cm}
\end{figure*}

\end{example}

Before moving on to the next sub-section, we discuss how the presented approach can be generalized to environments that consist of multiple agents and to the case where there is interaction between the $x$ and $e$ sub-systems. 

\textbf{Dealing with multiple agents. } The environment state $e_t$ may have structure that we can leverage to obtain more accurate and informative prediction sets and to improve efficiency in solving the optimization problem \eqref{eq:open_loop}. One such case is when the environment consists of $N$ uncontrollable dynamic agents, e.g., remote-controlled robots or humans. In this case, the environment state is described by the vector $e_t:=(e_{t,1},\hdots,e_{t,N})$ where $e_{t,n}$ denotes the state of agent $n$. Correspondingly, let us denote by $\hat{e}_{\tau|t,n}$ the prediction of $e_{\tau,n}$, i.e., the state of agent $n$ at time $\tau$, made at time $t$. With minor modification of  the nonconformity scores \eqref{eq:R_max1} and \eqref{eq:R_max2}, we  define the nonconformity scores
\begin{align}
    R^{(i)}_\text{OL}&:=\max_{(\tau,n)\in \{1,\hdots,T\}\times\{1,\hdots,N\}} \alpha_{\tau|0,n}\|e^{(i)}_{\tau,n}-\hat{e}^{(i)}_{\tau|0,n}\|\label{eq:R_max1_}\\
    R^{(i)}_\text{CL}&:=\max_{(\tau,n)\in\{0,\hdots,T-1\}\times\{1,\hdots,N\}} \alpha_{t+1|t,n}\|e^{(i)}_{t+1,n}-\hat{e}^{(i)}_{t+1|t,n}\|\label{eq:R_max2_}
\end{align}
for the open-loop and the closed-loop controllers. Compared to \eqref{eq:R_max1} and \eqref{eq:R_max2}, we additionally take the maximum over all individual agent prediction errors. The normalization constants $\alpha_{\tau|0,n}>0$ and $\alpha_{t+1|t,n}>0$ can be found following the same idea as discussed before. Similarly to Theorem \ref{thm:2}, we then obtain the statistical abstractions
\begin{align*}
         \text{Prob}( &||e_{\tau,n}- \hat{e}_{\tau|0,n}|| \le C_{\tau|0,n}, \\
         &\forall (\tau,n) \in \{1, \dots, T\}\times\{1,\hdots,N\}) \ge 1-\delta\\
         \text{Prob}(&||e_{t+1,n} - \hat{e}_{t+1|t,n}|| \le C_{t+1|t,n}, \\
         &\forall (t,n) \in \{0, \dots, T-1\}\times\{1,\hdots,N\} )\ge 1-\delta  
\end{align*}
where $C_{\tau|0,n}:={C}_\text{OL}/\alpha_{\tau|0,n}$ and $C_{t+1|t,n}:={C}_\text{CL}/\alpha_{t+1|t,n}$ with $C_\text{OL}:=\text{Quantile}_{1-{\delta}}( R^{(1)}_\text{OL}, \hdots, R^{(K)}_\text{OL}, \infty )$ and $C_\text{CL}:=\text{Quantile}_{1-{\delta}}( R^{(1)}_\text{CL}, \hdots, R^{(K)}_\text{CL}, \infty )$ using the nonconformity scores  \eqref{eq:R_max1_} and \eqref{eq:R_max2_}, respectively. Following this construction, we then use the structured uncertainty sets
\begin{align*}
    \mathcal{B}_{\tau|t}:=\{\bar{e}_\tau\in\mathbb{R}^{n_e}|\forall n\in\{1,\hdots,N\},\|\bar{e}_{\tau,n}- \hat{e}_{\tau|t,n}\|\le C_{\tau|t,n}\}
\end{align*}
within the optimization problem \eqref{eq:open_loop}. The structure that is now present in $\mathcal{B}_{\tau|t}$ usually simplifies solving \eqref{eq:open_loop}  as the constraint $c$ is defined in terms of individual agent states in practice which allows rewriting the constraint \eqref{eq:constC_2}. 

\textbf{Dealing with agent interactions. } So far, we considered the system in  \eqref{eq:LEAS_model_} where the behavior of the environment $e$ does not depend on the behavior of the control system $x$, and hence not on the control input $u$. In some applications, however, there may be interaction in the sense that the distribution $\mathcal{D}_e$ of the environment $e$ depends on the trajectory $x$ such that $e\sim \mathcal{D}_e(x)$. For instance, in robot navigation the environment state $e_t$ may follow the dynamics $e_{t+1}=f_e(e_t,v_t)+g_e(e_t,x_t)$ where the function $g_e:\mathbb{R}^{n_e}\times\mathbb{R}^{n_x}\to\mathbb{R}^{n_e}$ models (social) interaction such as repulsive forces for collision avoidance \cite{rimon1990exact,dimarogonas2007rendezvous,csenbacslar2023dream,helbing1995social}.  In such cases, the  interaction term $g_e(e_t,x_t)$ is usually small when the safety constraint $c$ imposes conservative safety margins, e.g., a self-driving car being forced to take conservative control actions that result in socially acceptable trajectories which do not change the behavior of pedestrians. In these situations, we do not expect our control guarantees to degrade dramatically. Indeed, we can measure the effect of, and even compensate for, interaction induced shifts in the distribution of $\mathcal{D}_e$. Consider the scenario where calibration trajectories $e^{(1)},\hdots,e^{(K)}$ are sampled from a nominal distribution $\mathcal{D}_e$ as per Assumption \ref{ass3}, while the test trajectory $e^{(0)}$ is sampled from the distribution $\mathcal{D}_e(x)$ that models interaction with $x$. For instance, the distribution $\mathcal{D}_e$ may describe the dynamics $e_{t+1}=f_e(e_t,v_t)$, while the distribution $\mathcal{D}_e(x)$ may describe the dynamics $e_{t+1}=f_e(e_t,v_t)+g_e(e_t,x_t)$. For this scenario, the authors in \cite{barber2023conformal} provide a lower  bound of $1-\delta-\tilde{\delta}$ for the statistical coverage  where $\tilde{\delta}$ is a function of, and monotonically increases with, the KL-divergence between $\mathcal{D}_e$ and $\mathcal{D}_e(x)$. An alternative approach would be to construct robust statistical abstractions when the f-divergence between $\mathcal{D}_e$ and $\mathcal{D}_e(x)$ is upper bounded (or known) by using robust conformal prediction, see Sidebar \textbf{Conformal Prediction under Distribution Shift}. If the interaction induced distribution shift is difficult to measure, adaptive conformal prediction can be used to design interaction-aware controllers, as e.g., in \cite{dixit2023adaptive,sheng2024safe,yao2024sonic}.

\subsection{Temporal Logic-Constrained Control} 
So far, we considered constraint functions $c$  that were enforced point-wise in time. Such constraints can enforce safety, e.g., avoiding collisions with other agents at all times. We may, however, be interested in enforcing more complex trajectory-wise safety and performance constraints, e.g., repeatedly tracking a specific agent or tracking agents that enter forbidden areas. Such constraints can  be expressed as temporal logic specifications over the outputs $x$ and $e$ of the LEAS. We here use signal temporal logic (STL) which extends predicate  logic, introduced in Sidebar \textbf{LEC Specifications in Predicate Logic}, by additionally reasoning over temporal properties. We refer readers not familiar with STL to Sidebar \textbf{LEAS Specifications in Signal Temporal Logic} for more details and examples. We are now interested in solving the following problem.

\begin{sidebar}{LEAS Specifications in Signal Temporal Logic}
\section[LEAS Specifications in Signal Temporal Logic]{}\phantomsection
   \label{sidebar-proof-CP}
\setcounter{sequation}{0}
\renewcommand{\thesequation}{S\arabic{sequation}}
\setcounter{stable}{0}
\renewcommand{\thestable}{S\arabic{stable}}
\setcounter{sfigure}{0}
\renewcommand{\thesfigure}{S\arabic{sfigure}}
 Temporal logics are used to express complex system specifications, e.g., linear temporal logic \cite{pnueli1977temporal} or signal temporal logic (STL) \cite{maler2004monitoring}. STL is a predicate logic  equipped with additional temporal operators to reason over the temporal behavior of system trajectories. An STL specification $\phi$ is constructed  from atomic constraint functions $h:\mathbb{R}^{n_z}\to \mathbb{R}$ that impose constraints of the form $h(z_t)\ge 0$. We also define the Boolean predicate $\mu:\mathbb{R}^{n_z}\to \{\text{True},\text{False}\}$ as $\mu(z_t)=\text{True}$ if $ h(z_t)\ge 0$ and $\mu(z_t)=\text{False}$ otherwise. We define the syntax of STL over $z$ as 
\begin{align}\label{eq:full_STL}
\phi \; ::= \; \text{True} \; | \; \mu \; | \;  \neg \phi' \; | \; \phi' \wedge \phi'' \; | \; \phi'  U_I \phi''
\end{align}
 where  $U_I$ denotes the until operator with time interval $I\subseteq \mathbb{R}_{\ge 0}$, while the other operators were already discussed in the Sidebar \textbf{LEC Specifications in Predicate Logic}. Intuitively, the until operator $\phi' {U}_I \phi''$ encodes that $\phi'$ has to be true from now on until $\phi''$ becomes true at some future time within the time interval $I$. We can further define eventually and always operators as
 \begin{align*}
F_I\phi'&\coloneqq\top U_I \phi' &\text{ (eventually)},\\
G_I\phi'&\coloneqq\neg F_I\neg \phi' &\text{ (always)}.
\end{align*}
Intuitively, the eventually $F_I \phi'$ (always $G_I \phi'$) operator encodes that $\phi'$ has to be true at some time (for all future times) within the time interval $I$. This intuition carries over into the semantics. Formally, to determine if an STL specification as defined in \eqref{eq:full_STL} is satisfied, we define the Boolean semantics
 	\begin{align*}
	 (z,t) \models \text{True}  &\;\;\;\text{ iff } \;\;\;	\text{holds by definition}, \\
	 ({z},t) \models
	 \mu   &\;\;\;\text{ iff }\;\;\;	h({z}_t)\ge 0\\
	 ({z},t) \models \neg\phi' &\;\;\;\text{ iff }\;\;\; ({z},t) \not\models \phi'\\
	 ({z},t) \models \phi' \wedge \phi'' &\;\;\;\text{ iff }\;\;\; ({z},t) \models \phi' \text{ and } ({z},t) \models \phi''\\
	 ({z},t) \models \phi' U_I \phi'' &\;\;\;\text{ iff }\;\;\; \exists t'' \in (t\oplus I)\cap\mathbb{N} \text{ s.t. }({z},t'')\models \phi'' \text{ and } \\& \qquad \quad \forall t'\in [t,t'']\cap\mathbb{N} \text{, }({z},t') \models \phi'.
	\end{align*}
Note that $(z, t)\models \phi$ indicates that the trajectory $z$ satisfies  $\phi$ at time step $t$. By convention, we use $z\models \phi$ when $t=0$. 

To give a simple example, consider the specification 
\begin{align*}
    \phi:=G_{[0,\infty)} \big(F_{I_1} \|x-e\|\le \epsilon \wedge (\|e-A\|\le \epsilon \implies F_{I_2} \|x-e\|\le \epsilon)\big)
\end{align*} 
that expresses that the system $x$ should repeatedly (with period $I_1$) track agent $e$, while $x$ should immediately (within the time interval $I_2$) track $e$ whenever $e$ enters the forbidden region $A$.

We again define quantitative semantics $\rho^\phi(z,t)$ that reduce checking $(z,t) \models \phi$ to checking positivity of $\rho^\phi$. Let
	\begin{align*}
	\rho^{\top}({z},t)& := \infty,\\
	\rho^{\mu}({z},t)& := h(z_t),\\
	\rho^{\neg\phi'}({z},t) &:= 	-\rho^{\phi'}({z},t),\\
	\rho^{\phi' \wedge \phi''}({z},t) &:= 	\min(\rho^{\phi'}({z},t),\rho^{\phi''}({z},t)),\\
	\rho^{\phi' U_I \phi''}({z},t) &:= \underset{t''\in (t\oplus I)\cap\mathbb{N}}{\text{max}}  \min(\rho^{\phi''}({z},t''),\underset{t'\in [t,t'']\cap\mathbb{N}}{\text{min}}\rho^{\phi'}({z},t') ).
	\end{align*}
As before,  we use $\rho^\phi(z)$ when $t=0$ by convention.

STL specifications are defined over trajectories $z$ with infinite length. In practice, however, one is often interested in bounded STL specifications, which are STL specifications where the time interval $I$ in the STL syntax is  bounded. In this survey, we limit ourselves to bounded STL specifications. To check satisfaction of a bounded STL specification $\phi$, i.e., to check that $(z,t)\models \phi$, only a signal $z$ of finite length is required. The length of this signal, denoted by $T^\phi$, can recursively be computed as
 \begin{align*}
     T^\text{True}&=T^\mu:=0\\
     T^{\neg\phi}&:=T^\phi\\
     T^{\phi'\wedge\phi''}&:=\max(T^{\phi'},T^{\phi''})\\
     T^{\phi' U_I \phi''}&:=\max \{I\cap \mathbb{N}\}+\max(T^{\phi'},T^{\phi''}).
 \end{align*}
 
\end{sidebar}

\begin{probl}\label{prob2}
    Given the system in \eqref{eq:LEAS_model_} and an STL specification $\phi$ over the trajectories of $x$ and $e$, design control inputs $u_0,\hdots,u_{T-1}$ such that $x$ satisfies $\phi$ with a probability no less than $1-\delta$, i.e., such that
    \begin{align}\label{eq:prob_bool}
        \text{Prob}((x,e)\models \phi)\ge 1-\delta.
    \end{align} 
\end{probl}

It was shown in \cite{yu2023signal} that Problem \ref{prob2} can be solved in two different ways by using the Boolean semantics or the quantitative semantics associated with the specification $\phi$, see Sidebar \textbf{LEAS Specifications in Signal Temporal Logic} for details. We summarize the Boolean semantics-based control approach next, hereby closely following \cite{yu2023signal}. 

\textbf{Mixed Integer Linear Encoding.} STL specifications can be encoded as constraints within a mixed integer linear program \cite{raman1,raman2015reactive}. The idea  is to introduce a binary variable $z_0^\phi\in \{0,1\}$ and a set of constraints over $(x,e)$ such that $z_0^\phi = 1$ if and only if $(x,e)\models \phi$. Since the environment  $e$ is stochastic, this encoding cannot be used here. Instead, we introduce a binary variable $\bar{z}_0^\phi\in \{0,1\}$ and a set of constraints over $x$ and the statistical abstractions such that $\bar{z}_0^\phi = 1$ implies that $\text{Prob}((x,e)\models \phi)\ge 1-\delta$. This encoding provides sufficient conditions, and we remark that necessary conditions cannot be derived here due to the non-uniqueness of statistical abstractions. 

Recall that an STL specification $\phi$ is recursively built from  predicates $\mu$ using Boolean and temporal operators. We thus derive  constraints encoding $\phi$ recursively.

\textbf{Predicates. }For each predicate $\mu$ in $\phi$ and for each time $\tau\ge 0$, we introduce a binary variable $\bar{z}_{\tau|t}^{\mu} \in \{0,1\}$. If $\tau \leq t$, then we have observed the value of $e_\tau$ already, and we set $\bar{z}_{\tau|t}^{\mu} = 1$ if $\mu(x_\tau, e_\tau) \geq 0$. If $\tau > t$, then we would like to constrain $\bar{z}_{\tau|t}^{\mu} $ such that  $\bar{z}_{\tau|t}^{\mu} = 1$ implies $\text{Prob}(\mu(x_{\tau|t}, e_\tau) \geq 0) \geq 1 - \delta$, while $\bar{z}_{\tau|t}^{\mu} = 0$ implies $\text{Prob}(\mu(x_{\tau|t}, e_\tau) < 0) \geq 1 - \delta$. We achieve this objective by using the Big-M method to define the constraints
\begin{subequations}\label{eq:encoding_1}
    \begin{align}
    - \min_{\bar{e} \in \mathcal{B}_{\tau|t}} \mu(x_{\tau|t}, \bar{e}) & \leq M (1-\bar{z}_{\tau|t}^{\mu}),\\
    \max_{\bar{e} \in \mathcal{B}_{\tau|t}} \mu(x_{\tau|t}, \bar{e}) & \leq M \bar{z}_{\tau|t}^{\mu} -\epsilon, 
\end{align}
\end{subequations}
where $M$ and $\epsilon$ are sufficiently large and small positive constants, respectively, see \cite{bemporad1999control} for details on how to select $M$ and $\epsilon$. The minimization of $\mu(x_{\tau|t}, \bar{e})$ over  $\bar{e}\in \mathcal{B}_{\tau|t}$ accounts for all states within the statistical abstraction $\mathcal{B}_{\tau|t}$. Note that the integration of the constraints in \eqref{eq:encoding_1} into an optimization problem that aims to compute $u_{0|t},\hdots,u_{T-1|t}$ results in a bilevel optimization problem. The inner optimization problems $\min_{\bar{e} \in \mathcal{B}_{\tau|t}} \mu(x_{\tau|t}, \bar{e})$ and $\max_{\bar{e} \in \mathcal{B}_{\tau|t}} \mu(x_{\tau|t}, \bar{e})$ can be reformulated using its KKT conditions when $\mu$ is a linear function, see \cite{yu2023signal} for details.

\textbf{Boolean and temporal operators. } Boolean and temporal operators can recursively be encoded using the constraint encoding from \cite{raman1,raman2015reactive}, as we illustrate next. 

For the negation $\phi:=\neg \phi'$,  we introduce a binary variable $\bar{z}_{\tau|t}^{\phi} \in \{0,1\}$  such that $\bar{z}_{\tau|t}^{\phi} =1$ if and only if  $\bar{z}_{\tau|t}^{\phi'}=0$. For this, we simply enforce the constraint 
\begin{align}\label{eq:encoding_2}
\bar{z}_{\tau|t}^{\phi}=1-\bar{z}_{\tau|t}^{\phi'}.
\end{align}

For the conjunction $\phi:=\phi'\wedge \phi''$, we introduce a binary variable $\bar{z}_{\tau|t}^{\phi} \!\in\! \{0,1\}$  such that $\bar{z}_{\tau|t}^{\phi} =1$ if and only if  $\bar{z}_{\tau|t}^{\phi'}=\bar{z}_{\tau|t}^{\phi''} =1$. We achieve this by enforcing the constraints
\begin{subequations}\label{eq:encoding_3}
        \begin{align}
        & \bar{z}_{\tau|t}^{\phi} \leq \bar{z}_{\tau|t}^{\phi'},  \\
        & \bar{z}_{\tau|t}^{\phi} \leq \bar{z}_{\tau|t}^{\phi''}, \\
        & \bar{z}_{\tau|t}^{\phi} \geq -1+\bar{z}_{\tau|t}^{\phi'}+\bar{z}_{\tau|t}^{\phi''}. 
    \end{align}
\end{subequations}

For the temporal until operator, we follow a similar procedure. Specifically, note that we can rewrite the until operator $\phi=\phi'\mathbf{U}_{[a,b]}\phi''$ as 
\begin{align}\label{eq:encoding_4}
    \bar{z}_{\tau|t}^{\phi} = \bigvee_{\tau'=\tau+a}^{\tau+b} (\bar{z}_{\tau'|t}^{\phi''} \wedge \bigwedge_{\tau''=\tau}^{\tau'} \bar{z}_{\tau''|t}^{\phi'}).
\end{align}
The conjunction and disjunction operators in equation \eqref{eq:encoding_4} can then  be encoded via the constraints \eqref{eq:encoding_2} and \eqref{eq:encoding_3} (for disjunctions, note that $\phi'\vee \phi''=\neg (\neg \phi' \wedge \neg \phi'')$). 

For an STL specification $\phi$, this encoding now ensures that $\text{Prob}((x,e)\models \phi)\ge 1-\delta$ if $\bar{z}_{0|t}^{\phi} =1$. We summarize this result next and refer the reader to \cite{yu2023signal} for the proof.

\begin{theorem}\label{thm:encoding}
Given an STL specification $\phi$ over trajectories $x$ and $e$ and a recursive encoding $\bar{z}_{0|t}^{\phi}$ via the constraints in equations \eqref{eq:encoding_1}-\eqref{eq:encoding_4}, then $\bar{z}_{0|t}^{\phi} =1$ implies that $(x,e)\models \phi$ holds with a probability no less than $1-\delta$, i.e.,  it holds that
    \begin{align}\label{eq:prob_bool}
        \text{Prob}((x,e)\models \phi)\ge 1-\delta.
    \end{align} 
\end{theorem}

\textbf{Control  synthesis. } We can now solve Problem \ref{prob2}  by computing a sequence of control inputs $u_{t|t},\hdots,u_{T-1|t}$ as the solution of the optimization problem
\begin{subequations}\label{eq:STLopen_loop}
\begin{align}
    &\min_{(u_{t|t},\hdots,u_{T-1|t})} J(x,u)\label{eq:STLcosss} &\\
     \text{s.t.}\;\;& x_{\tau+1|t}=f_x(x_{\tau|t},u_{\tau|t}), &\tau\in\{t,\hdots,T-1\} \label{eq:STLdyn}\\
     & x_{t|t}=x_t&\\
     & \bar{z}_{0|t}^{\phi}=1\label{eq:STLconstC_2}\\
     & \text{Encodings }\eqref{eq:encoding_1}-\eqref{eq:encoding_4} \\
     & u_{\tau|t} \in \mathcal{U},x_{\tau+1|t} \in \mathcal{X},&\tau\in\{t,\hdots,T-1\}. \label{eq:STLincon}
\end{align}
\end{subequations}

Similar to Theorems \ref{thm:3} and \ref{thm:4}, we can now derive open-loop and closed-loop controllers. We summarize these results, which follow from Theorem \ref{thm:encoding}, next.

\begin{theorem}\label{thm:3_}
    Given a test trajectory $e^{(0)}\sim\mathcal{D}_e$ and the statistical abstraction  \eqref{eq:cp_open_}, then  a solution $u_\tau:=u_{\tau|0}$ of the optimization problem  \eqref{eq:STLopen_loop} at time $t:=0$ is such that  
    \begin{align*}
         \text{Prob}((x,e)\models \phi)\ge 1-\delta.
    \end{align*}
\end{theorem}

\begin{theorem}\label{thm:4_}
    Given a test trajectory $e^{(0)}\sim\mathcal{D}_e$ and the statistical abstraction \eqref{eq:cp_closed_}, then the solutions $u_t:=u_{t|t}$ of the optimization problem \eqref{eq:STLopen_loop}  at times $t\in\{0,\hdots,T-1\}$ are such that  
    \begin{align*}
         \text{Prob}((x,e)\models \phi)\ge 1-\delta.
    \end{align*}
\end{theorem}

We remark that recursive feasibility for the closed-loop controller in Theorem \ref{thm:4_} is a more delicate yet interesting problem compared to recursive feasibility in Theorem \ref{thm:4}. A detailed discussion is beyond the scope of this survey paper, and we hence refer the interested reader to \cite{yu2023signal}. 

\textbf{Control synthesis using quantitative semantics. } In our previous exposition, we solved Problem \ref{prob2} using the Boolean semantics associated with $\phi$. However,  Problem \ref{prob2} can also be solved using the quantitative semantics $\rho^\phi:\mathbb{R}^{Tn_x}\times\mathbb{R}^{Tn_e}\to\mathbb{R}$. Indeed, \cite{yu2023signal} shows that one can design controllers such that $\text{Prob}(\rho^\phi(x,e)>0)\ge 1-\delta$ by which it follows that Problem \ref{prob2} is solved. Importantly, one can now choose to maximize $\rho^\phi(x,e)$. While controllers obtained this way are more robust, they are at the same time also computationally more demanding.

Finally, we invite the interested reader to find illustrative simulation examples in \cite{yu2023signal}.

\subsection{State Estimation and Perception}
Up until now, we assumed to have knowledge of the state $z_t:=(x_t,e_t)$  via perfect sensor measurements $y_t=z_t$. Autonomous systems, however, are complex systems that operate in uncertain environments where the states $x_t$ and $e_t$  cannot be observed directly. These states are usually estimated from noisy sensor measurements $y_t=p(z_t,w_t)$ where $p:\mathbb{R}^{n_z}\times \mathbb{R}^{n_w}\to \mathbb{R}^{n_y}$ is the sensor model and $w_t\sim\mathcal{D}_w$ is sensor noise. A common way of obtaining state estimates is via state estimation algorithms \cite{simon2006optimal}. However, as discussed in Sidebar \textbf{Conformalizing the Kalman Filter}, state estimation algorithms are either computationally prohibitive (e.g., the particle filter) or only valid under restrictive assumptions on the system (e.g., the Kalman filter). This motivated us to conformalize the Kalman filter when applied to systems where no guarantees can be  obtained directly. Many of these classic state estimation algorithms require knowledge of the system dynamics $f$, i.e., the $f_x$ and $f_e$ dynamics of the $x$ and $e$ sub-systems, and the sensor model $p$. However, such knowledge is often difficult to obtain in practice, e.g., mathematical models for the motion of pedestrians or for camera sensors are not available. This limitation has further motivated recent work on learning-enabled perception and state estimation algorithms when using complex camera, LiDAR, and radar sensors \cite{zou2023object,redmon2016you}. Using similar ideas as presented in the aforementioned sidebar, we can conformalize even such complex perception and state estimation algorithms.  

\textbf{Statistical perceptual abstractions.} The complexity of learning-enabled perception and state estimation algorithms has lead to the introduction of more general statistical abstractions, here broadly referred to as perceptual abstractions. However, there is another reason why one should be interested in more general abstractions than the ones obtained in Sidebar \textbf{Conformalizing the Kalman Filter}. To see why, note that the observations $y_t=p(x_t,e_t)$ depend on $x_t$ and hence also on the control inputs $u_0,\hdots,u_{t-1}$.\footnote{Sidebar \textbf{Conformalizing the Kalman Filter} assumed access to a trajectory calibration dataset, i.e., trajectories of the autonomous system driven by the control inputs $u_0,\hdots,u_T$.} However, if the control inputs  $u_0,\hdots,u_{t-1}$ depend on the statistical abstraction, which is likely to be the case and the reason why we design an abstraction in the first place, then the nonconformity scores evaluated over the calibration dataset are not independent as before. As a consequence, the statistical abstraction is not valid anymore. Another way of thinking about this is that changing the sequence of control inputs leads to a distribution shift in the distribution of the nonconformity score. To address this challenge, various authors introduced different notions of perceptual abstractions, see e.g.,   \cite{hsieh2022verifying,astorga2023perception,sun2023learning,dean2021guaranteeing,dean2020robust}. 

In what follows, let the state $z\in\mathbb{R}^{n_z}$ be unknown. For a perception and state estimation algorithm that  computes an estimate $\hat{z}(y)$ of $z$ from a measurement $y=p(z,w)$ where $w\sim\mathcal{D}_w$,  a perceptual abstraction is a set-valued function $\mathcal{P}:\mathbb{R}^{n_y}\times\mathbb{R}^{n_z}\to 2^{\mathbb{R}^{n_z}}$ such that  $z\in \mathcal{P}(y,\hat{z}(y))$.\footnote{The definition of a perceptual abstraction can  be extended to include sequences of measurements instead of one measurement.} As can easily be seen, constructing perceptual abstractions is challenging due to the complexity of the system, the sensors, and  learning-enabled perception and state estimation algorithms. This challenge motivates the definition of statistical perceptual abstractions for which it  holds that
\begin{align*}
	\text{Prob}\big(z\in \mathcal{P}(y,\hat{z}(y))\big)\ge 1-\delta.
\end{align*}

Such guarantees can be obtained with conformal prediction  for compact domains $\mathcal{Z}\subseteq\mathbb{R}^{n_z}$ in the state space when we are able to generously sample multiple calibration datasets from $\mathcal{Z}$. The idea is to construct a grid of $\mathcal{Z}$ and to perform a conformal prediction step for each grid point. Specifically, we construct an $\epsilon$-net $\bar{\mathcal{Z}}$ of $\mathcal{Z}$ where  $\epsilon>0$ is a gridding parameter, i.e., we construct a finite set $\bar{\mathcal{Z}}$  so that for each $z\in\mathcal{Z}$ there exists $z_j\in \bar{\mathcal{Z}}$ such that $\|z-z_j\|\le \epsilon$. For this purpose,  gridding  or randomized algorithms can be used that sample from $\mathcal{Z}$ \cite{vershynin2018high}. We can now apply a conformal prediction step for each grid point $z_j\in\bar{\mathcal{Z}}$, i.e., we compute the nonconformity score
\begin{align}
  R_j^{(i)}:=\|\hat{z}(y_j^{(i)})-z_j\|
\end{align}
over a calibration dataset $y_j^{(i)}:=p(z_j,w^{(i)})$ for $i\in\{1,\hdots,K\}$ where $w^{(i)}$ is independently sampled from $\mathcal{D}_w$, i.e., $w^{(i)}\sim\mathcal{D}_w$. For a test datapoint $y_j^{(0)}:=p(z_j,w^{(0)})$ with $w^{(0)}\sim\mathcal{D}$, we then immediately know that 
\begin{align}\label{eq:nonconf_e}
    \text{Prob}\big(\|\hat{z}(y_j^{(0)})-z_j\|\le C_j\big)\ge 1-\delta
\end{align}
by the choice of $C_j:= \text{Quantile}_{1-\delta}( R^{(1)}_j, \hdots, R^{(K)}_j, \infty )$. We note that equation \eqref{eq:nonconf_e} quantifies uncertainty of the uncertain estimate $\hat{z}(y_j^{(0)})$ as we fixed the state $z_j$  a-priori. 

To revert this argument and obtain a prediction region for an unknown state $z\in\mathcal{Z}$ from an estimate $\hat{z}(y)$ where $y=p(z,w)$ is a noisy measurement with $w\sim\mathcal{D}_w$, we have to ensure that equation \eqref{eq:nonconf_e} holds for all states $z\in\mathcal{Z}$ (instead of only $z_j\in\bar{\mathcal{Z}}$). To do so, the authors in \cite{yang2023safe} use the properties of  the $\epsilon$-net $\bar{\mathcal{Z}}$ along with assumptions on the continuity of the sensor map $p$ and the estimate $\hat{z}$. Specifically, if  $p$ and  $\hat{z}$ have Lipschitz constants $\mathcal{L}_p$ and $\mathcal{L}_{\hat{z}}$, respectively, then for any state $z\in\mathcal{Z}$ it holds that
\begin{align*}
    \text{Prob}\Big(\|\hat{z}(y)-z\|\le\sup_j C_j+(\mathcal{L}_{p}\mathcal{L}_{\hat{z}}+1)\epsilon\Big)\ge 1-\delta.
\end{align*}

Variations of statistical perceptual abstractions were obtained in \cite{hsieh2022verifying,astorga2023perception,sun2023learning,dean2021guaranteeing,dean2020robust} using similar reasoning via combinations of $\epsilon$-nets and statistical techniques such as Hoeffding’s inequality. Statistical perceptual abstractions enable the design of robust control techniques that take perceptual uncertainty into account. While initial progress was made, as e.g., in  \cite{dean2021guaranteeing,dean2020robust,yang2023safe,li2024safe,waite2025state}, the design of controllers with end-to-end safety guarantees is a challenging and open problem. Lastly, we note that  existing work assumes that ground truth information of $z_j$ is available in the form of calibration data. In practice, such an assumption is limiting unless precise state estimates can be obtained or when realistic data with ground truth information can be collected, e.g., from photorealistic simulators. In these cases, the sim2real gap can be addressed using robust uncertainty quantification techniques such as robust conformal prediction (see Sidebar \textbf{Conformal Prediction under Distribution Shift}).

\textbf{Beyond state estimation. } So far, we focused on the design of statistical perceptual abstractions for state estimation and downstream control. However, perception uncertainty can be quantified and incorporated at the task planning level. One line of work considers large language models and conformal prediction for planning in uncertain environments.  The authors in \cite{ren2023robots} use large language models to provide preferences for a set of  high-level actions, e.g., ``go to region A and then B'', ``go to region B and then A'', $\hdots$. Conformal prediction is here  used to constructs sets that contain all actions that conform with the natural language specification of the user with a given confidence. In \cite{ren2024explore}, the same authors explore a setting in which an agent actively explores an uncertain environment to answer a question about the state of the environment. Large vision-language models produce actions for exploration, and conformal prediction is  used until an answer to the question about the environment is found with a desired confidence. Similarly to Theorem \ref{thm:2}, the authors in  \cite{ren2023robots,ren2024explore} construct multi-step prediction sets. Importantly, these works assume that the behavior of the environment is independent from the behavior of the control system, similarly to the system considered in \eqref{eq:LEAS_model_}. This way, the focus is not directly on computing control inputs, but  on high-level task planning. In contrast, the authors in \cite{dixit2024perceive} consider a control system that depends on an environment that is drawn from a distribution of potential environments, and a method is proposed to modify an existing perception system using conformal prediction so that a given controller satisfies a desired task. 

\subsection{Related Work: Safe Control Techniques}

There is a broad range of safe control techniques, commonly known under the umbrella of constrained control. Popular techniques include model predictive control \cite{garcia1989model,chen1998quasi,morari1999model,mesbah2016stochastic}, control barrier functions \cite{ames2016control,ames2019control,xu2015robustness,clark2021control,yaghoubi2020risk}, and funnel control \cite{bechlioulis2008robust,ilchmann2002tracking,mehdifar2023control,sui2020novel}, with respective extensions to temporal logic specifications in \cite{lindemann2018control,lindemann2020barrier},  \cite{lindemann2021funnel,lindemann2019feedback}, and \cite{raman1,raman2015reactive,farahani2018shrinking}. These control techniques are model-based and typically only provide safety guarantees when the system dynamics are simple and known. Model-based controllers are thus hard to design for complex, possibly learning-enabled systems. This has motivated the use of data-driven and machine learning techniques, e.g., reinforcement learning \cite{levine2016end,ibarz2021train}, imitation learning \cite{argall2009survey,torabi2018behavioral}, or learned safe controllers \cite{srinivasan2020synthesis,taylor2020learning,jin2020neural,dean2021guaranteeing}, even with access to expert demonstrations \cite{robey2020learning,lindemann2021learning,lindemann2024learning}. While these techniques  demonstrate oftentimes good performance in practice, they typically lack  verifiable safety guarantees. 

Toward designing verifiably safe controllers for such complex, often uncertain systems, recent efforts have focused on combining statistical techniques with data-driven control design. While data-driven control techniques are not new within the control field, consider e.g., system identification  \cite{ljung1998system} and adaptive control \cite{ioannou2012robust,sastry2011adaptive}, the integration with statistical techniques enables finite sample guarantees and sample complexity analysis, see e.g.,  \cite{ziemann2023tutorial,tsiamis2023statistical,matni2019self} for recent tutorials. As there are many recent developments in this direction, we proceed by first reviewing conformal prediction-based control techniques followed by surveying other statistical control techniques.

There has been a lot of recent work and progress in solving  control problems in dynamic environments with conformal prediction. Along the same lines of what we presented in Sections \textbf{Statistical Abstractions of Dynamic
Environments} and \textbf{Safe Control in Dynamic Environments}, the authors in \cite{tonkens2023scalable} design controllers using statistical abstractions via conformalized copulas, as originally presented in \cite{sun2022copula}. Conceptually similar is also the work in \cite{chen2021reactive} where an abstraction is built without the use of a trajectory predictor $\mu$. In a different direction, the authors in \cite{yeh2024end,kiyani2025decision} consider end-to-end optimization approaches that find statistical abstractions that are optimal for decision making. In a different setting, conformal prediction has been used to design predictive safety filters for reinforcement learning controllers in dynamic environments \cite{strawn2023conformal}. The underlying assumption in these works is the existence of an independent and identically distributed dataset as per Assumption \ref{ass3}. In \cite{dixit2023adaptive,sheng2024safe,yao2024sonic,driggsinteraction}, the authors relax this assumption and instead use adaptive conformal prediction (see Sidebar \textbf{Conformal Prediction under Distribution Shift}) to build statistical abstractions. These abstractions do not come with guarantees of the form \eqref{eq:cp_open_} (or \eqref{eq:cp_closed_}), but enjoy asymptotic probabilistic guarantees and again enable the design of open-loop and closed-loop controllers. In a similar direction, out-of-distribution detection for dynamic environments and control adaptation is presented in~\cite{contreras2024out}. A version of adaptive conformal prediction, called  intermittent quantile tracking, has also been used in the setting of interactive imitation learning \cite{zhao2024conformalized_}.

While everything that we presented so far was modular by the use of statistical abstractions, the authors in \cite{lekeufack2023conformal} present conformal decision theory and calibrate decisions directly without constructing abstractions. The main idea is that the controller is parameterized by a risk parameter that is updated via an adaptive conformal prediction update rule. Importantly, this update  is applied to the risk parameter and not to the quantile as in adaptive conformal prediction. This method can guarantee that an accumulated loss is asymptotically bounded with probability one.  In yet another direction, the works in \cite{foffano2023conformal,taufiq2022conformal,kuipers2024conformal} propose  conformal off-policy evaluation to compute prediction intervals  for the value function of a reinforcement learning controller using calibration data gathered under a different controller. This requires dealing with the distribution shift between these two controllers which is handled via weighted conformal prediction (see Sidebar \textbf{Conformal Prediction under Distribution Shift}). More recently, conformal prediction was used for control design of (partially) known or uncertain systems, e.g., for nonlinear systems with Lyapunov and barrier functions  \cite{zhang2025conformal,hsu2025statistical,tayal2025cp,wei2025conformal,ms2025cped} or for linear systems \cite{vlahakis2024conformal,vlahakis2025conformal}. We also point the reader to the work in \cite{chee2024uncertainty} where robust controllers are designed based on conformal uncertainty estimates of learned system dynamics models. \textcolor{black}{Multi-agent control problems with conformal prediction were recently considered in \cite{gupta2023cammarl,kuipers2024conformal,wang2024safe,huriot2024safe}.} Lastly, we mention the works in \cite{wang2023conformal,wang2024safe} and \cite{sun2024conformal} where conformal prediction is used to guarantee safety of language-instructed and diffusion model-based planners, respectively. 

A special instantiation of a dynamic environment is when humans are present. In \cite{lidard2024risk}, human intentions are predicted via latent intent prediction algorithms which are subsequently conformalized for interactive planning. The key idea here is to ask the human for help if the uncertainty, computed with conformal prediction, is too large. In a different setting in \cite{zhao2024conformalized}, adaptive conformal prediction is used to obtain confidence estimates when mapping human inputs to high-dimensional robot actions, e.g., in the case of assistive teleoperation of robotic manipulators.

Before moving to the next section, we briefly summarize other statistical techniques for safe control design. Prominent techniques include the use of Gaussian process-based system modeling for control design, see e.g., \cite{beckers2019stable,deisenroth2013gaussian,jain2018learning,liu2018gaussian}, as well as scenario optimization for solving robust or chance constrained control problems, see e.g., \cite{calafiore2006scenario,campi2009scenario,calafiore2012robust,schildbach2014scenario,do2024probabilistically}. We also mention that PAC-Bayes theory was used in \cite{majumdar2021pac,majumdar2018pac} for designing controllers that  provably generalize to new environments. Lastly, we point the reader to control design under risk constraints, see e.g., \cite{akella2024risk,wang2022risk,safaoui2021risk,dixit2021risk,ahmadi2021risk}.

\section{Offline Verification of LEAS with Conformal Prediction}

Model-based control design techniques usually provide correctness and safety guarantees under specific assumptions on the system. In Section \textbf{Control Synthesis for LEAS with Conformal Prediction}, we surveyed a hybrid of model-based control techniques with statistical uncertainty quantification techniques. Along the way, we  encountered various assumptions that may not, or only approximately, hold in practice, e.g., having exact knowledge of the state or encountering dynamic environments whose behavior is independent of the control system. Violation of these assumptions necessitates offline and possibly online verification, see Sidebar \textbf{Spectrum of Formal Verification and Control Techniques} for a distinction between offline and online verification. More generally, learning-enabled controllers may not come with any formal safety guarantees, e.g., most reinforcement or imitation learning techniques. We thus present offline and online verification techniques with conformal prediction in the remainder of this survey.

Throughout the following sections, we consider LEASs as in equation \eqref{eq:LEAS_model} under a controller  of the form \eqref{eq:LEAS_control}. For convenience, we recall the system as
\begin{subequations}\label{eq:LEAS_model___}
\begin{align}
        z_{t+1}&=f(z_t,u_t,v_t),\\
    y_t&=p(z_t,w_t),\\
    u_t&=\pi(\{y_s\}_{s=0}^t),
\end{align}
\end{subequations}
where we adapt the same notation  as in Section \textbf{Control Synthesis for LEAS with Conformal Prediction}. However, as opposed to Section \textbf{Control Synthesis for LEAS with Conformal Prediction}, we now assume that the controller $\pi$ is fixed and given to us, i.e., an arbitrary learning-enabled controller which may not achieve its objective and has to be verified. Importantly, we do not need to limit ourselves to systems of the form \eqref{eq:LEAS_model___} as long as we can collect a calibration dataset that consists of system trajectories (formally stated in Assumption \ref{ass2} below). Such trajectories can typically  be collected from a simulator or  the real system without having explicit knowledge of \eqref{eq:LEAS_model___}.

\subsection{Challenges in LEAS Verification}

The analysis of the system in \eqref{eq:LEAS_model___} is challenging for two main reasons. First, the sensor map $p$ is hard to model, e.g., in the case of cameras, and a description of $p$ is usually not available. In some cases, e.g., for legged locomotion or in social navigation tasks, we may even have  issues identifying a dynamics map $f$. More crucially though, even if good models of $f$ and $p$ are available, the complexity of the LEAS makes the verification of the LEAS in \eqref{eq:LEAS_model___} difficult and computationally challenging.

In Section \textbf{Challenges in LEC Verification}, we provided a detailed discussion on challenges (specifically, scalability and conservatism) that model-based verification techniques face for LECs. These  challenges further amplify when attempting to verify LEASs of the form \eqref{eq:LEAS_model___} which iteratively use complex LECs (often even using LECs more complex than simple feedforward neural networks) within their perception-action loops due to compounding errors and increased computational complexity. Indeed, existing model-based verification techniques for LEASs extend complete and non-complete LEC verification techniques and perform reachability analysis of the LEAS via semidefinite programming \cite{hu2020reach,fazlyab2020safety,chen2023one}, Lyapunov/barrier functions \cite{chen2020learning,mazouz2022safety}, and abstractions and hybrid system models \cite{huang2019reachnn,tran2020nnv,dutta2019reachability,ivanov2019verisig,ivanov2020verifying}, among others. While these techniques have been successful at verifying smaller problem instances,  e.g., autonomous race cars \cite{ivanov2019verisig,ivanov2020case}, cruise controller and emergency braking systems \cite{tran2020nnv,tran2019safety}, autonomous robots \cite{sun2019formal}, or aircraft collision avoidance systems \cite{bak2021second,bak2022neural}, their applicability to real-world LEASs is still limited. Compositional techniques can alleviate the complexity challenges in larger verification problems, see e.g., \cite{dreossi2019compositional}. We emphasize, however, that model-based verification techniques are challenged on resource-constrained platforms and in online verification and control where efficient algorithms are needed, suggesting the use of efficient statistical techniques for probabilistically sound LEAS verification. Such an approach does not even require knowledge of the functions $f$ and $p$.

\subsection{Probabilistically Sound LEAS Verification}
In the remainder, we verify input-output properties of LEASs using ideas that are conceptually similar to the ideas that we presented for verifying input-output properties of LECs in Section \textbf{LEC Verification with Conformal Prediction}. Indeed, a LEAS can be thought of as having inputs $(z_0,\{v_t\}_{t=0}^{T-1},\{w_t\}_{t=0}^{T-1})$ and outputs $z:=(z_0,\hdots,z_T)$ with task horizon $T>0$. In contrast to Section \textbf{LEC Verification with Conformal Prediction}, we now want to reason over input and output trajectories $u$ and $z$. By modifying equation \eqref{eq:LECS_verifiy_}, which we stated for LECs before, we can instead formulate our goal for LEASs verification as
\begin{align}\label{eq:LEASS_verifiy_}
    (z_0,v,w)\sim\mathcal{D}   \implies \text{Prob}(z\models \phi_\text{out})\ge 1-\delta
\end{align} 
where $\phi_\text{out}$ prescribes an output property for the trajectory $z$ of \eqref{eq:LEAS_model___}. We assume that the output property $\phi_\text{out}$ is defined over the first $T+1$ values of $z$, i.e., over $z_0,\hdots,z_T$. In equation \eqref{eq:LEASS_verifiy_}, the input distribution is $\mathcal{D}:=\mathcal{D}_{z}\times \mathcal{D}_{v}^T\times \mathcal{D}_{w}^T $ where $\mathcal{D}_{v}^T$ and $\mathcal{D}_{w}^T$ denote the $T$-fold probability measure of $\mathcal{D}_{v}$ and $\mathcal{D}_{w}$, respectively. Similarly to before, we assume that $\mathcal{D}$ has support over the input set   $\mathcal{C}_\text{in}:=\{(z_0,v,w)\in\mathbb{R}^{n_z+T(n_v+n_w)}|(z_0,v,w)\models\phi_\text{in}\}$ where $\phi_\text{in}$ prescribes an input property for  $(z_0,v,w)$. Throughout this section, we make a similar assumption as Assumption \ref{ass1} in Section \textbf{LEC Verification with Conformal Prediction}, but now assuming that we have a set of independent trajectories from the system in \eqref{eq:LEAS_model___}.

\begin{assumption}\label{ass2}
    We have independently sampled  a calibration dataset of $K$ inputs from the distribution $\mathcal{D}$, i.e., we have access to samples $(z^{(i)}_0,\{v^{(i)}_t\}_{t=0}^{T-1},\{w^{(i)}_t\}_{t=0}^{T-1})\sim \mathcal{D}$ for $i\in\{1,\hdots,K\}$. We use these to obtain a  calibration dataset of $K$ trajectories $z^{(i)}:=(z^{(i)}_0,z^{(i)}_1,\hdots,z^{(i)}_T)$ via the dynamics in \eqref{eq:LEAS_model___}.
\end{assumption}


\textbf{Reachability of LEASs. } Assume that we are given an input set $\mathcal{C}_\text{in}\subseteq\mathbb{R}^{n_z+T(n_v+n_w)}$ and an output set $\mathcal{C}_\text{out}\subseteq\mathbb{R}^{n_z(T+1)}$. By slight abuse of notation, we let $\mathcal{C}_\text{out}[t]\subseteq\mathbb{R}^{n_z}$ denote the projection of $\mathcal{C}_\text{out}$ onto the reachable set at time $t$.  From here, we can essentially proceed similarly as in Section \textbf{Probabilistically Sound LEC Verification} (as summarized in Corollary \ref{cor1}) to verify reachability of $\mathcal{C}_\text{out}$, i.e., to verify $\text{Prob}(z^{(0)}\in\mathcal{C}_\text{out})\ge 1-\delta$, by defining the nonconformity score
\begin{align}\label{R_reach_LEAS}
R^{(i)}:=\min_{t\in\{0,\hdots,T\}}\text{dist}\big(z^{(i)}_t,\mathcal{C}_\text{out}[t]\big)
\end{align}
where $\text{dist}(\cdot)$ is a distance function defined in the same way as in Section \textbf{Probabilistically Sound LEC Verification} to measure the distance between the output $z_t$ and the output set $\mathcal{C}_\text{out}[t]$ at time $t$. By direct application of Lemma \ref{lem:1}, we obtain the following result.
\begin{corollary}\label{cor1_____}
    Given an output set  $\mathcal{C}_\text{out}\subseteq\mathbb{R}^{n_z(T+1)}$, the distribution $\mathcal{D}$ with support over an input set $\mathcal{C}_\text{in}\subseteq\mathbb{R}^{n_z+T(n_v+n_w)}$, and a test trajectory $z^{(0)}:=(z^{(0)}_0,z^{(0)}_1,\hdots,z^{(0)}_T)$ obtained via the dynamics in \eqref{eq:LEAS_model___} from input samples $(z^{(0)}_0,\{v^{(0)}_t\}_{t=0}^{T-1},\{w^{(0)}_t\}_{t=0}^{T-1})\sim \mathcal{D}$, we have that
\begin{align*}
    \text{Prob}(\min_{t\in\{0,\hdots,T\}}\text{dist}\big(z^{(0)}_t,\mathcal{C}_\text{out}[t]\big)\le C)\ge 1-\delta,
\end{align*}
where  $C:=\text{Quantile}_{1-\delta}( R^{(1)}, \hdots, R^{(K)}, \infty )$ with $R^{(i)}$ being defined in equation \eqref{R_reach_LEAS}. 
\end{corollary}

By the definition of the signed distance function and taking the minimum distance to $\mathcal{C}_\text{out}$ over all times $t\in\{0,\hdots,T\}$, we obtain a positive verification answer if $C< 0$. In this case, the absolute value of $C$ again indicates how robustly the reachability specification is satisfied.

We also remark that the reachability analysis performed here was of qualitative nature by checking if the trajectory $z^{(0)}$ remains within the set $\mathcal{C}_\text{out}$ with high confidence. Previously, in Section \textbf{Control Synthesis for LEAS with Conformal Prediction}, we showed how to perform a quantitative analysis by computing a minimal set $\mathcal{C}_\text{out}$ that satisfies $\text{Prob}(z^{(0)}\in\mathcal{C}_\text{out})\ge 1-\delta$.

\textbf{General Verification of LEASs. } We would  like to verify more general input-output properties of LEASs than  reachability, where we are conveniently handed the input and output sets $\mathcal{C}_\text{in}$ and $\mathcal{C}_\text{out}$. Such properties can  again be expressed as STL specifications, and we remind the reader of Sidebar \textbf{LEAS Specifications in Signal Temporal Logic} for an introduction. Similar to predicate logic specifications in Section \textbf{Probabilistically Sound LEC Verification}, the sidebar instructs how we can construct a real-valued performance function\footnote{The length of the task horizon $T$ can easily be computed from the specification $\phi_\text{out}$ as instructed in Sidebar \textbf{LEAS Specifications in Signal Temporal Logic}.} $\rho^{\phi_\text{out}}:\mathbb{R}^{Tn_z}\to\mathbb{R}$ that satisfies the following soundness property
\begin{align*}
    \rho^{\phi_\text{out}}(z)>0 \; \implies \; z\models\phi_\text{out}.
\end{align*}
As before, larger values of $\rho^{\phi_\text{out}}(z)$ are beneficial as they indicate robustness against perturbation in $z$, see \cite{donze2010robust,fainekos2009robustness}. 

From here, we can again proceed in the same way as in Section \textbf{Probabilistically Sound LEC Verification} (as summarized in Corollary \ref{cor2}) to verify an STL specification $\phi_\text{out}$, i.e., to verify $\text{Prob}(-C\le \rho^{\phi_\text{out}}(z^{(0)}))\ge 1-\delta$, which we briefly summarize next.
\begin{corollary}\label{cor2_}
    Given an STL  specification $\phi_\text{out}$ with sound performance function $\rho^{\phi_\text{out}}:\mathbb{R}^{n_y}\to\mathbb{R}$, the distribution $\mathcal{D}$, and a test trajectory $z^{(0)}:=(z^{(0)}_0,z^{(0)}_1,\hdots,z^{(0)}_T)$ obtained via the dynamics in \eqref{eq:LEAS_model___} from input samples $(z^{(0)}_0,\{v^{(0)}_t\}_{t=0}^{T-1},\{w^{(0)}_t\}_{t=0}^{T-1})\sim \mathcal{D}$, we have
\begin{align*}
    \text{Prob}(-C\le \rho^{\phi_\text{out}}(z^{(0)}))\ge 1-\delta.
\end{align*}
where  $C:=\text{Quantile}_{1-\delta}( R^{(1)}, \hdots, R^{(K)}, \infty )$ with $R^{(i)}:=- \rho^{\phi_\text{out}}(z^{(i)})$. 
\end{corollary}
Finally, to illustrate what we learned so far, we present a case study on the verification of a cart-pole system with a learning-enabled feedback controller in Sidebar \textbf{Verifying a Learning-Enabled Cart-Pole System}.

\begin{figure*}
    \centering
    \begin{subfigure}[t]{0.31\textwidth}
        \includegraphics[width=\textwidth]{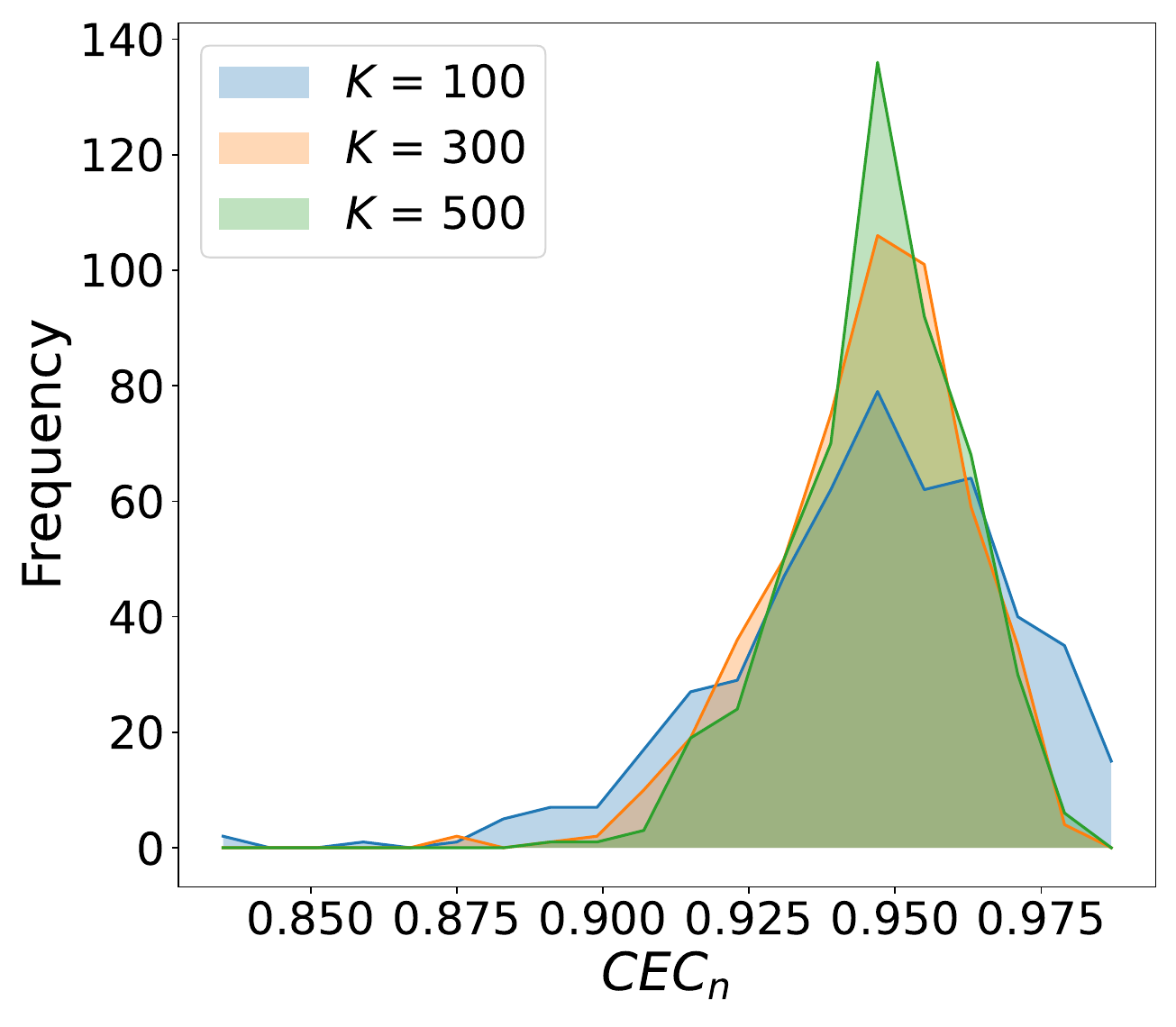}
        \caption{Histogram of the conditional empirical coverage $CEC_n$ over all $N$ experiments.}
        \label{fig:histogram_cecs_example_2}
    \end{subfigure}
    \hspace{2mm}
    \begin{subfigure}[t]{0.31\textwidth}
        \includegraphics[width=\textwidth]{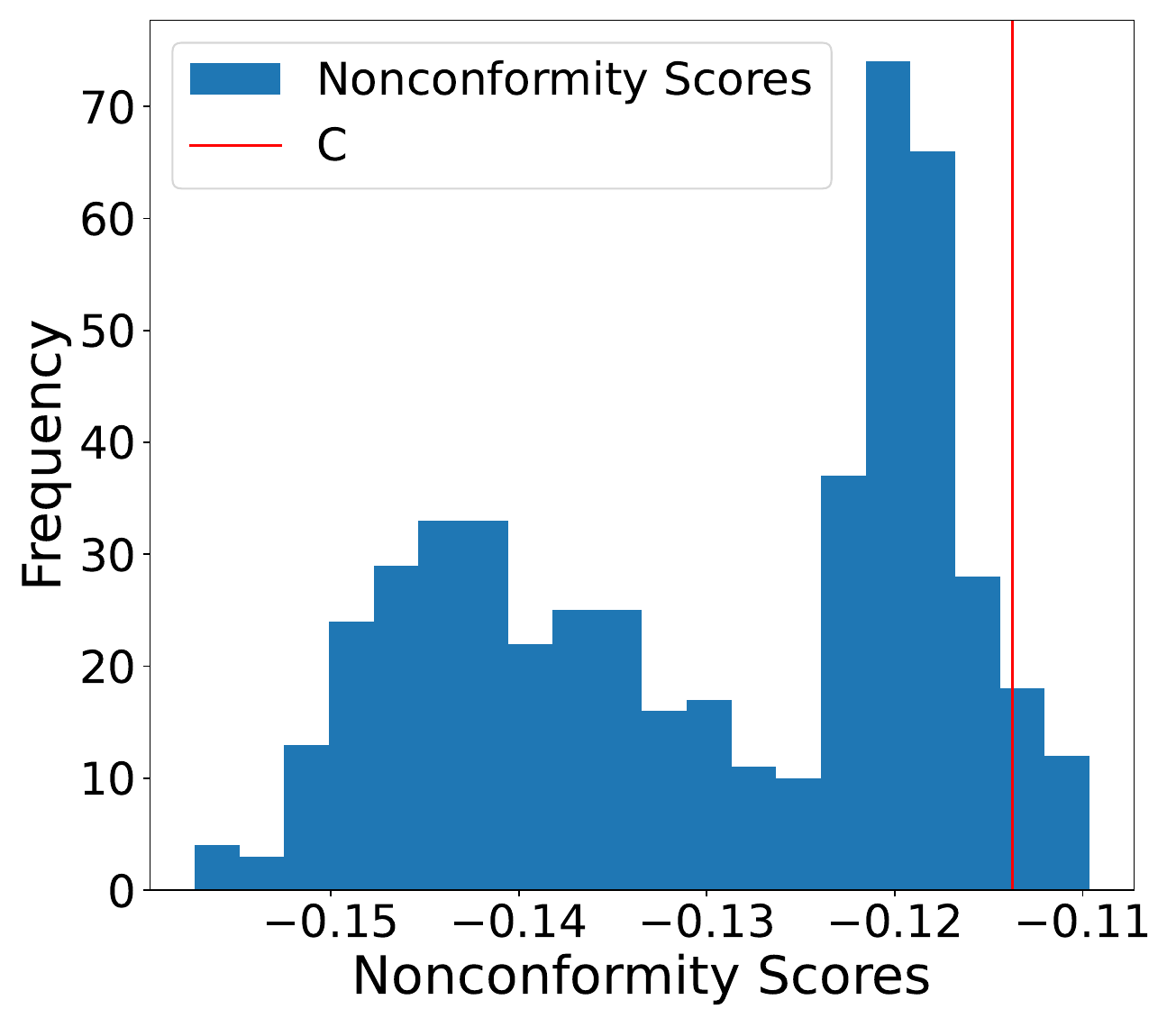}
        \caption{Histogram  of the nonconformity score $R^{(i)}$ over $K:=500$ calibration datapoints (for one of the $N$ experiments).}
        \label{fig:histogram_nonconformities_example_2}
    \end{subfigure}
    \hspace{2mm}
    \begin{subfigure}[t]{0.31\textwidth} 
        \includegraphics[width=\textwidth]{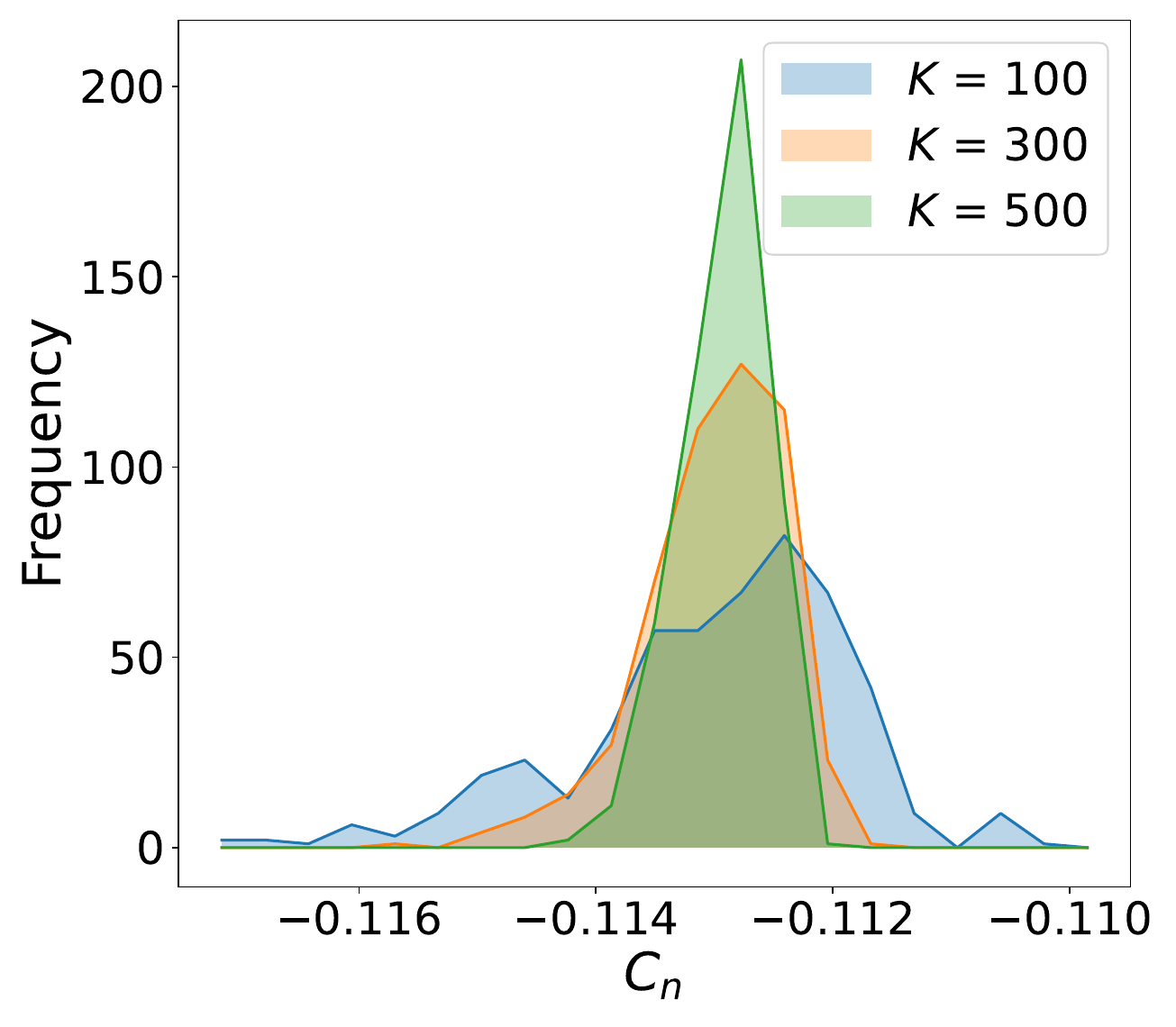}
        \caption{Histogram of the bound $C_n$ over all $N$ experiments.}
        \label{fig:histogram_cs_example_2}
    \end{subfigure}
    \caption{Experimental Results for Sidebar \textbf{Verifying a Learning-Enabled Cart-Pole System}.}
\end{figure*}

\begin{sidebar}{Verifying a Learning-Enabled Cart-Pole System}
\section[Verifying a Learning-Enabled Cart-Pole System]{}\phantomsection
   \label{sidebar-proof-CP}
\setcounter{sequation}{0}
\renewcommand{\thesequation}{S\arabic{sequation}}
\setcounter{stable}{0}
\renewcommand{\thestable}{S\arabic{stable}}
\setcounter{sfigure}{0}
\renewcommand{\thesfigure}{S\arabic{sfigure}}
   We want to verify the correctness of a learning-enabled state-feedback controller for a cart-pole system \cite{barto1983neuronlike,florian2007correct}. The goal of the controller is to stabilize a pole placed on top of a frictionless cart. The state of the system is $z_t := (z^p_t, z^v_t, z^\theta_t, z^\omega_t) \in \mathbb{R}^4$ where $z^p_t$, $z^v_t$, $z^\theta_t$, and $z^\omega_t$ indicate  cart position, cart velocity, angle of the pole (w.r.t. the upright position), and angular velocity of the pole, respectively. We assume that $z_0$ is uniformly sampled with $z_0\sim\mathcal{U}((-0.05, 0.05)^4)$. We train a neural network controller \cite{mnih2013playing} that outputs the force applied to the cart, here with a discrete action space pushing the cart to the left or to the right. This controller can thus be viewed as a bang-bang controller. We simulate the cart-pole system in the OpenAI Gym \cite{brockman2016openai}.

    We want to verify that, at all times $t \le T:=228$, the angle of the pole  $z^\theta_t$ deviates minimally from the upright position and the cart stays close to the origin. For this specification, we can construct the following sound performance function\footnote{The corresponding STL specification is written as $\phi_\text{out} \models G_{[0, T]}(\|z^\theta\| \le \delta_\theta \wedge \|z^p\| \le \delta_p)$.}
     \begin{align*}
        \rho^{\phi_\text{out}}(z) \coloneqq \min_{t \in \{0, \hdots, T\}}(\min(\delta_\theta - \|z^\theta_t\|, \delta_p - \|z^p_t\|))
    \end{align*}
    where $\delta_\theta \coloneqq 0.2$ and $\delta_p \coloneqq 4.5$ denote  permissible thresholds for angle and position displacements, respectively. 

     We set the failure probability to $\delta \coloneqq 0.05$ and use the nonconformity score $R^{(i)}:=-\rho^{\phi_\text{out}}(z^{(i)})$ as discussed before. By this choice, we know that $\text{Prob}(-C\le \rho^{\phi_\text{out}}(z^{(0)}))\ge 1-\delta$. We conduct $N \coloneqq 500$ experiments for calibration set sizes of $K \in \{100, 300, 500\}$. Again, we first  verify   statistical validity empirically  and compute the empirical coverage $EC$ according to equation \eqref{eq:emp_coverage} as $0.95$, $0.956$, and $0.952$, respectively.  We also plot the conditional empirical coverage $CEC_n$ according to equation \eqref{eq:cond_emp_coverage} with $J := 400$ in Figure \ref{fig:histogram_cecs_example_2}. We plot the histogram of the nonconformity scores of the calibration data from one experiment in Figure \ref{fig:histogram_nonconformities_example_2}.  To illustrate the verification results, we compute $\sum_{n=1}^NC_n/N$ which is the average of the bound $C_n$ over all $N$ experiments, where we recall from equations \eqref{eq:emp_coverage} and \eqref{eq:cond_emp_coverage} that $C_n:=C(R^{(1)}_{n},\hdots,R^{(K)}_{n})$. For $K \in \{100, 300, 500\}$, we obtain the values $-0.11283$, $-0.11283$, and $-0.11277$, respectively. We can thus conclude that the system satisfies the specification with a confidence no less than $1-\delta$. In Figure \ref{fig:histogram_cs_example_2}, we also plot the histogram of $C_n$ over all $N$ experiments. 
\end{sidebar}

\subsection{Related Work: Statistical Verification Techniques}

In Section \textbf{Challenges in LEAS Verification}, we described computational challenges that arise in the verification of LEASs when using model-based approaches. Probabilistic and parametric model checking techniques, see \cite{bianco1995model,hansson1994logic,katoen2005markov,kwiatkowska2007stochastic, kwiatkowska2011prism} and \cite{daws2005symbolic, hahn2010param, hahn2011probabilistic, hahn2011synthesis}, respectively, enable the verification of stochastic LEASs, but are subject to the same computational challenges. Statistical verification techniques, on the other hand, offer computationally tractable alternatives, e.g., via  conformal prediction as demonstrated in the previous sections.  ´

\textbf{Reachability. } Conformal prediction was first used for reachability analysis in \cite{bortolussi2019neural}, with extensions to partially observable systems \cite{cairoli2021neural} and distributed systems \cite{cairoli2023conformal}. The idea here is to conformalize a learned predictor that predicts whether or not the system trajectory, starting from a given initial condition, is safe with respect to an unsafe set. While these techniques provide probabilistic safety guarantees, they do not provide reachable sets that can be used for downstream tasks, such as control synthesis in Section \textbf{Control Synthesis for LEAS with Conformal Prediction}. The authors in \cite{hashemi2023data,hashemi2024statistical,kwon2024conformalized} obtain probabilistic reachable sets by using conformal prediction and reachability analysis techniques for neural network surrogate models of the underlying unknown system. In \cite{tebjou2023data}, reachable sets are obtained via the use of Christoffel functions \cite{devonport2021data} that are subsequently conformalized. Computing probabilistic reachable sets can also be cast as a chance constrained optimization problem that can be solved via scenario optimization \cite{devonport2020estimating, dietrich2024nonconvex}. Here, reachable sets are parameterized and the optimal parameters are found by solving a chance constrained optimization problem. Conformal prediction and scenario optimization were used in \cite{lin2024verification} to quantify correctness of learned solutions to Hamilton-Jacobi reachability problems. Lastly, we note that techniques for computing probabilistic reachable sets share similarities with techniques for computing statistical abstractions, as we defined in Section \textbf{Statistical Abstractions of Dynamic Environments}. The difference mainly lies in computational complexity as statistical abstractions for control should be computed efficiently.


\textbf{General verification. } Conformal prediction was used to verify more general system properties, such as temporal logic specification. The authors in \cite{qin2022statistical,qin2024statistical}  learn a surrogate model for the performance function $\rho^{\phi_\text{out}}$ of a temporal logic specification $\phi_\text{out}$, which they subsequently conformalize to obtain probabilistic satisfaction guarantees. The works in \cite{lindemann2023conformal,zhao2024robust,cairoli2023conformal} propose online verification algorithms which we will discuss in more detail in Section \textbf{Online Verification of LEAS with Conformal Prediction}.

\begin{sidebar}{Conformal Prediction and Qualitative SMC Techniques}
\section[Conformal Prediction and Qualitative SMC Techniques]{}\phantomsection
   \label{sidebar-proof-CP}
\setcounter{sequation}{0}
\renewcommand{\thesequation}{S\arabic{sequation}}
\setcounter{stable}{0}
\renewcommand{\thestable}{S\arabic{stable}}
\setcounter{sfigure}{0}
\renewcommand{\thesfigure}{S\arabic{sfigure}}
\textbf{Conformal prediction} can provide a qualitative SMC technique with the nonconformity score $R^{(i)}:=-\mathbb{1}(z^{(i)} \models \phi_{out})$ where $z^{(i)} \sim \mathcal{D}$ are again $K$ independent and identically sampled calibration trajectories.  By solving the expression $\text{Quantile}_{1-\delta}( R^{(1)}, \hdots, R^{(K)}, \infty )=-1$ for some $\delta\in (0,1)$, we know that $\text{Prob}( \mathbb{1}(z^{(0)} \models \phi_{out})\ge1)\ge 1-\delta$ holds, which is equivalent to  $\text{Prob}(z^{(0)} \models \phi_{out})\ge 1-\delta$. If we additionally find the smallest $\delta\in(0,1)$ that satisfies $\text{Quantile}_{1-\delta}( R^{(1)}, \hdots, R^{(K)}, \infty )=-1$, e.g., via line search, we obtain a  quantitative SMC technique that provides an estimate of $\text{Prob}(z^{(0)} \models \phi_{out})$. In comparison to existing SMC techniques in the literature, which we describe below, it is important to keep in mind that conformal prediction can  provide both marginal as well as calibration conditional guarantees.

Let now $p := \text{Prob}(z^{(0)} \models \phi_{out})$ denote the unknown task satisfaction probability.  In \textbf{hypothesis testing}, we  construct two hypotheses $H_0: p \ge p_0$ and $H_1: p \le p_1$ where $(p_1, p_0) \subseteq (0, 1)$ is a user-defined ``indifference'' region, i.e., a  region that is small enough for us to ignore.  We consider the experiment where we accept $H_1$ if $\sum_{i = 1}^K\mathbb{1}(z^{(i)} \models \phi_{out}) \le \xi$, while we accept $H_0$ if $\sum_{i = 1}^K\mathbb{1}(z^{(i)} \models \phi_{out}) > \xi$ where $\xi$ is a decision threshold. Our task is  to find a sampling plan consisting of  $(K, \xi)$. An optimal sampling plan $(K,\xi)$ minimizes the number of calibration trajectories $K$ while maintaining a low probability of type-I error (accepting $H_1$ when $H_0$ holds), denoted as $\alpha$, and a low probability of type-II error (accepting $H_0$ when $H_1$ holds), denoted as $\beta$. For fixed values of $p_1,p_0,\alpha,\beta$, this problem can be formulated as an optimization problem and solved via binary search. We refer the interested reader to \cite{grubbs1949designing, younes2004verification} for details and discussions on subleties of the methods.   It is intuitively clear that $K$ has to become greater as $\alpha$ and $\beta$ become smaller and as the indifference region $(p_1,p_0)$ shrinks. An optimal plan obtained this way (referred to as single sampling plan) uses a fixed $K$, while one can obtain verification results faster with early stopping via sequential probability ratio testing~\cite{wald1992sequential}.


A qualitative SMC technique can also be obtained via the \textbf{Clopper-Pearson bound} \cite{wang2019statistical,zarei2020statistical}. Consider a probability threshold $p_0$ and the estimate $\hat{p} \coloneqq \sum_{i = 1}^K\mathbb{1}(z^{(i)} \models \phi_{out})/K$ of $p:= \text{Prob}(z^{(0)} \models \phi_{out})$. If $\hat{p} \ge p_0$, it holds that the probability that $p \ge p_0$ is lower bounded by $1 - \alpha_{clopper}$\footnote{Note that these are again calibration conditional guarantees where the outer probability is over all calibration trajectories.} where $\alpha_{clopper}$ is a function of $K$ and $p_0$ as discussed in detail in \cite{zarei2020statistical}.  Concretely, $\alpha_{clopper}$ can be found via computing the cumulative distribution function of a beta distribution defined over the parameters $K$, $p_0$, and $\sum_{i = 1}^K\mathbb{1}(z^{(i)} \models \phi_{out})$. Clopper-Pearson bounds are tighter than  bounds obtained via Hoeffding's inequality. We note that guarantees for the opposite direction can be derived, i.e., if $\hat{p} < p_0$, the probability that $p < p_0$ is lower bounded by $1 - \alpha_{clopper}$. 
\end{sidebar}

Statistical model checking (SMC) has been used as  an umbrella term for a set of statistical verification techniques, see e.g., \cite{larsen2016statistical, agha2018survey, legay2019statistical, zarei2020statistical} for an overview. However, SMC techniques in the literature are different from the techniques that we presented here (and also different from techniques in \cite{qin2022statistical,qin2024statistical,lindemann2023conformal,zhao2024robust,cairoli2023conformal}) due to (1) the  goal being to verify task satisfaction, i.e., verifying that  $z\models \phi_\text{out}$, as opposed to checking quantitative performance properties, i.e., finding bounds $\rho^*$ such that  $\rho^{\phi_\text{out}}(z)\ge \rho^*$, and (2) the use of statistical techniques that are different from conformal prediction (details are provided below). 

SMC techniques can broadly be divided into two categories: qualitative and quantitative. For a given failure probability $\delta\in(0,1)$, qualitative techniques aim to check whether or not $\text{Prob}(z\models \phi_\text{out})\ge 1-\delta$ holds. Quantitative techniques, on the other hand, aim to estimate $\text{Prob}(z\models \phi_\text{out})$ directly. Existing techniques for qualitative SMC use hypothesis testing \cite{ sen2004statistical, legay2019statistical,younes2004verification}, e.g., via single sampling plans \cite{grubbs1949designing,younes2004verification} or sequential probability ratio tests \cite{wald1992sequential}, and more recently Clopper-Pearson bounds \cite{wang2019statistical, zarei2020statistical}. In Sidebar \textbf{Conformal Prediction and Qualitative SMC Techniques}, we provide a short summary of existing qualitative SMC techniques and show, in comparison, how a qualitative (and even quantitative) technique based on conformal prediction can be designed. Quantitative SMC techniques, on the other hand, rely on random sampling of trajectories to directly estimate $\text{Prob}(z \models \phi_{out})$. Alongside, quantitative SMC techniques use  concentration inequalities, such as the Chernoff-Hoeffding inequality, to obtain calibration conditional coverage guarantees for $\text{Prob}(z \models \phi_{out})$, see e.g., \cite{legay2019statistical}. There are various extensions of these techniques that we want to briefly mention for the interested reader, e.g., the use of importance sampling \cite{kahn1953methods, zhao2016accelerated, kim2016improving, uesato2018rigorous, o2018scalable} and importance splitting \cite{kahn1951estimation, cerou2007adaptive, jegourel2013importance, jegourel2014effective, norden2019efficient} in the case of rare events when $\text{Prob}(z \models \phi_{out})$ is small.

Lastly, we mention statistical techniques that solve other verification problems. We briefly touched upon risk metrics before as an alternative to probabilistic coverage guarantees provided by conformal prediction. The statistical verification of LEASs in terms of their risk, as quantified using risk metrics, was presented in \cite{chapman2021risk, lindemann2022temporal, akella2022scenario, akella2022sample, lindemann2023risk}. Statistical verification techniques for system conformance that check the similarity of two stochastic systems were proposed using the Kolmogorov-Smirnov test and conformal prediction, see \cite{wang2021probabilistic} and \cite{qin2023conformance}, respectively.


\section{Online Verification of LEAS with Conformal Prediction}

Model-based verification techniques, e.g., model checking, exhaustively check all possible system behaviors for correctness. This exhaustive search is a computational bottleneck in practice, usually limiting the applicability of these techniques to low fidelity models that may not capture the system dynamics sufficiently well. On the other hand, the search space in online verification is much smaller allowing the use of high fidelity models. In Section \textbf{Offline Verification of LEAS with Conformal Prediction}, we saw that statistical verification techniques, specifically those using conformal prediction, can alleviate some of these computational challenges present in model-based verification. However, these techniques (similar to the control techniques presented in Section \textbf{Control Synthesis for LEAS with Conformal Prediction}) provide probabilistic guarantees meaning that there is still a failure probability (of at most $\delta$). More specifically, a statistical verification technique may certify a system to be safe a-priori, e.g., with a probability of $1-\delta:=0.999$, but we may encounter one of the  $\delta=0.001$ fraction of  realizations that can result in unsafe system behavior when executing the system. This again motivates  online verification as a complementary safety mechanism.

\textbf{Predictive online verification.} Online verification techniques check if all extensions of a  partial trajectory, e.g., a trajectory observed online at time $t$, satisfy a specification. As discussed in Sidebar \textbf{Spectrum of Formal Verification and Control Techniques}, online verification can be conservative or inconclusive since no predictions about the system behavior are used. Having efficient and possibly learning-enabled prediction algorithms in hand, and knowing how to efficiently quantify uncertainty, we are  here instead interested in predictive online verification of LEASs. Consider again a LEAS as in equation \eqref{eq:LEAS_model___}, which we recall for convenience as
\begin{subequations}\label{eq:LEAS_model__}
\begin{align}
        z_{t+1}&=f(z_t,u_t,v_t),\\
    y_t&=p(z_t,w_t),\\
    u_t&=\pi(\{y_s\}_{s=0}^t),
\end{align}
\end{subequations}
where we adapt the same notation as before. Furthermore, we assume throughout this section that we have a  calibration set  of trajectories from \eqref{eq:LEAS_model__} available, similarly to  Assumption \ref{ass2} in Section \textbf{Offline Verification of LEAS with Conformal Prediction}.
\begin{assumption}\label{ass4}
    We have independently sampled  a calibration dataset of $K$ inputs from the distribution $\mathcal{D}$, i.e., we have access to samples $(z^{(i)}_0,\{v^{(i)}_t\}_{t=0}^{T-1},\{w^{(i)}_t\}_{t=0}^{T-1})\sim \mathcal{D}$ for $i\in\{1,\hdots,K\}$. We use these to obtain a  calibration dataset of $K$ trajectories $z^{(i)}:=(z^{(i)}_0,z^{(i)}_1,\hdots,z^{(i)}_T)$ via the dynamics in \eqref{eq:LEAS_model__}.
\end{assumption}

Additionally, we are given a sound performance function $\rho^{\phi}:\mathbb{R}^{Tn_z}\to \mathbb{R}$ that describes the performance of the LEAS with respect to a specification $\phi$ that we evaluate for the trajectory $z$ of the LEAS. Such specifications can again conveniently be expressed in a temporal logic formalism such as signal temporal logic, recall Sidebar \textbf{LEAS Specifications in Signal Temporal Logic}. 

Informally, given an LEAS and a specification $\phi$, we  want to calculate the probability that the system trajectory $z:=(z_0,\hdots,z_T)$  satisfies $\phi$ based on the partial trajectory $z_{\text{obs}}:=(z_0,\hdots,z_t)$ observed at time $t$. Note  that the trajectory $z$ can be written as $z=(z_{\text{obs}},z_{\text{un}})$ where knowledge of $z_{\text{un}}:=(z_{t+1},\hdots,z_{T})$ is not available at time $t$. Formally, we aim to solve the following problem.
\begin{probl}\label{prob3}
    Given the system in \eqref{eq:LEAS_model__}, a specification $\phi$ over the trajectory $z$ with sound performance function $\rho^{\phi}:\mathbb{R}^{Tn_z}\to \mathbb{R}$, and the trajectory  $z_{\text{obs}}:=(z_0,\hdots,z_t)$ observed at time $t$, compute a lower performance bound $\rho^*$ such that
    \begin{align*}
        \text{Prob}(\rho^{\phi}(z)\ge \rho^*)\ge 1-\delta.
    \end{align*}
\end{probl}

\begin{figure*}
	\centering
	\includegraphics[scale=0.23]{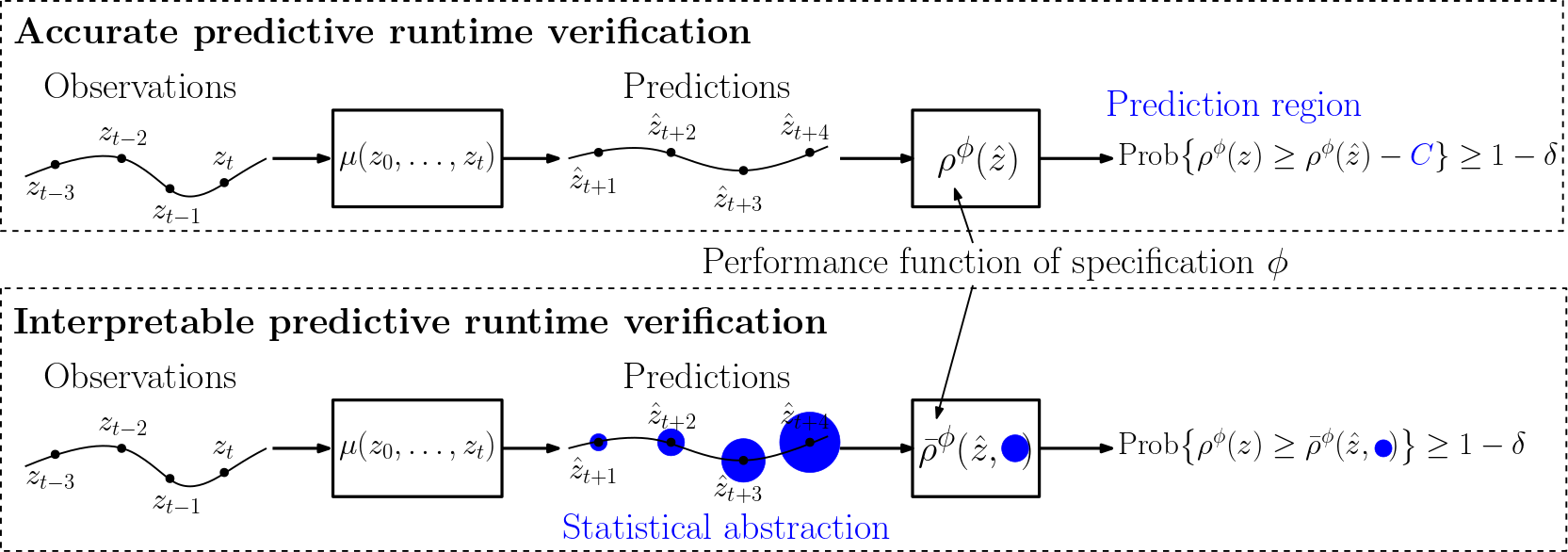}
	\caption{The top (bottom) figure shows the accurate (interpretable) predictive online verification algorithm. Both algorithms use past observations $(z_0,\hdots,z_t)$ to predict future states $(\hat{z}_{t+1},\hat{z}_{t+2},\hdots)$. The accurate algorithm applies conformal prediction  to the nonconformity score $\rho^{\phi}(\hat{z})-\rho^{\phi}(z)$ to directly obtain a probabilistic lower bound for $\rho^\phi(z)$. The interpretable algorithm constructs a statistical abstraction, as introduced in Section \textbf{Statistical Abstractions of Dynamic Environments}, and uses these to compute the worst case performance function $\bar{\rho}^\phi(\hat{z})$ which is a probabilistic lower bound for $\rho^\phi(z)$. Figure  taken from~\cite{lindemann2023conformal}. }
	\label{fig:prv1}
\end{figure*}

The intuition of the performance bound  $\rho^*$ is similar to the bounds that we derived for verification algorithms in Sections \textbf{LEC Verification with Conformal Prediction} and \textbf{Offline Verification of LEAS with Conformal Prediction}, i.e., $\rho^*>0$ is a sufficient condition to guarantee that $\text{Prob}(z\models \phi)\ge 1-\delta$ holds. 

\textbf{Accurate and interpretable predictive online verification. } We present two predictive online verification algorithms, closely following \cite{lindemann2023conformal}. The idea, similar to Section \textbf{Statistical Abstractions of Dynamic Environments}, is to use trajectory predictors $\mu:\mathbb{R}^{(t+1)n_z}\to\mathbb{R}^{(T-t)n_z}$ to predict future states $(z_{t+1},\hdots,z_T)$ from past and present states $(z_0,\hdots,z_t)$. However, different from Section \textbf{Statistical Abstractions of Dynamic Environments}, we now predict the state of the LEAS and not only the state of the environment. We denote predictions of $z_\tau$ for future times $\tau>t$ by $\hat{z}_{\tau|t}$, i.e., we have
\begin{align*}
	(\hat{z}_{t+1|t},\hdots,\hat{z}_{T|t})=\mu(z_0,\hdots,z_t).
\end{align*} 

We would now  like to use these predictions to compute $\rho^{\phi}(z)$. However, as future states $z_{\text{un}}$ are uncertain and as the predictions may not be accurate, we need to construct prediction regions for $\rho^{\phi}(z)$ by quantifying uncertainty  using conformal prediction. We first present an accurate predictive online verification algorithm that  constructs such prediction regions directly by  relating the nonconformity score to the performance function $\rho^{\phi}(z)$. As a consequence, this algorithm provides accurate prediction regions, i.e., tight lower bounds $\rho^*$. However, this algorithm is lacking interpretability in the sense that no information is provided as to why a specification is satisfied or violated. For the case that the specification $\phi$ expresses a signal temporal logic specification, we thus present an interpretable predictive online verification algorithm. This algorithm constructs statistical abstractions for future states $z_{\text{un}}$ first, following the same idea as in Theorems \ref{thm:1} and \ref{thm:2} in Section \textbf{Statistical Abstractions of Dynamic Environments}, and then uses these  to obtain a prediction region for the performance function $\rho^{\phi}(z)$. We refer the reader to Figure \ref{fig:prv1} for an illustration that is taken from \cite{lindemann2023conformal}.

\subsection{Accurate Predictive Online Verification}
As in Section \textbf{Statistical Abstractions of Dynamic Environments}, we compute the predictions 
\begin{align*}
(\hat{z}_{t+1|t}^{(i)},\hdots,\hat{z}_{T|t}^{(i)}):=\mu(z_0^{(i)},\hdots,z_t^{(i)}) 
\end{align*} 
for all available  calibration data $i\in\{1,\hdots,K\}$. Using these predictions, we define the predicted trajectory
  \begin{align*}
      \hat{z}^{(i)}:=(z_0^{(i)},\hdots,z_t^{(i)},\hat{z}_{t+1|t}^{(i)},\hdots,\hat{z}_{T|t}^{(i)})
  \end{align*}
  which is the concatenation of the partial trajectory $z_0^{(i)},\hdots,z_t^{(i)}$ and the predictions of future states $\hat{z}_{t+1|t}^{(i)},\hdots,\hat{z}_{T|t}^{(i)}$. We can now use the predicted trajectory to compute the predicted performance $\rho^{\phi}(\hat{z}^{(i)})$. Let us then compute the nonconformity score 
\begin{align}\label{eq:prv_direct}
    R^{(i)}:=\rho^{\phi}(\hat{z}^{(i)})-\rho^{\phi}(z^{(i)})
\end{align}
as the difference between the predicted and the true performance. We then immediately get the following result. 
\begin{theorem}\label{thm:5}
    Given an output property $\phi$ with sound performance function $\rho^{\phi}:\mathbb{R}^{Tn_z}\to \mathbb{R}$ and a test trajectory $z^{(0)}$  obtained via the dynamics in \eqref{eq:LEAS_model__} from samples $(z^{(0)}_0,\{v^{(0)}_t\}_{t=0}^{T-1},\{w^{(0)}_t\}_{t=0}^{T-1})\sim \mathcal{D}$, then we have that  
    \begin{align*}
        \text{Prob}(\rho^{\phi}(z^{(0)})\ge \rho^{\phi}(\hat{z}^{(0)})-C)\ge 1-\delta
    \end{align*}
    where $C:=\text{Quantile}_{1-\delta}( R^{(1)}, \hdots, R^{(K)}, \infty )$ with $R^{(i)}$ being defined in \eqref{eq:prv_direct}.
\end{theorem}

The proof follows immediately by the definition of the nonconformity score in equation \eqref{eq:prv_direct}. The interpretation of this result is that we obtain a probabilistic lower bound for the performance $\rho^{\phi}(z^{(0)})$ of the test trajectory in terms of the predicted performance $\rho^{\phi}(\hat{z}^{(0)})$ which is calibrated by $C$. In this way, the accurate predictive online verification algorithm solves  Problem \ref{prob3} with $\rho^*:=\rho^{\phi}(\hat{z}^{(0)})-C$. We provide an illustrative simulation example, and a comparison with the interpretable predictive online verification algorithm (presented next), at the end of the next sub-section.

\subsection{Interpretable Predictive Online Verification}
Similar to the statistical abstractions in equations \eqref{eq:cp_open_} and \eqref{eq:cp_closed_} that we used for constructing open-loop and closed-loop controllers, we construct statistical abstractions here. However, we are interested in abstractions of the form
\begin{align}\label{eq:cp_monitor_}
         \text{Prob}(||z_{\tau}^{(0)} &- \hat{z}_{\tau|t}^{(0)}|| \le C_{\tau|t}, \forall \tau \in \{t+1, \dots, T\}) \ge 1-\delta, 
\end{align}
i.e., in predictions regions for $\tau$-step ahead predictions made at time $t$ for the test trajectory $z^{(0)}$. The challenge lies again in computing non-conservative bounds $C_{\tau|t}$ from the calibration data $z^{(1)},\hdots, z^{(K)}$. We discussed several techniques for this purpose in Section \textbf{Statistical Abstractions of Dynamic Environments}, specifically in Theorems \ref{thm:1} and \ref{thm:2}, that can be easily applied here.

\textbf{Worst case performance function. }Having obtained  statistical abstractions of the form \eqref{eq:cp_monitor_} via a valid choice of $C_{\tau|t}$, we now aim to calculate the worst case performance of the function $\rho^{\phi}$ over the uncertainty sets
\begin{align*}
    \mathcal{B}_{\tau|t}:=\{\bar{z}_\tau\in\mathbb{R}^{n_e}|\|\bar{z}_\tau- \hat{z}_{\tau|t}^{(0)}\|\le C_{\tau|t}\}.
\end{align*}
 In an ideal world where we can solve nonconvex optimization problems, this worst case performance is computed as
 \begin{align*}
     \min_{\bar{z}_{t+1}\in \mathcal{B}_{\tau|t},\hdots,\bar{z}_{T}\in \mathcal{B}_{T|t}}\rho^{\phi}(\bar{z})
 \end{align*}
 where $\bar{z}:=(z_0,\hdots,z_t,\bar{z}_{t+1},\hdots,\bar{z}_{T})$ is the concatenation of the partial trajectory $z_0,\hdots,z_t$ and the worst case values $\bar{z}_{t+1},\hdots,\bar{z}_{T}$ from the uncertainty sets $\mathcal{B}_{\tau|t},\hdots, \mathcal{B}_{T|t}$. Unfortunately, this worst case performance is hard to compute, motivating us to construct a lower bound  $\bar{\rho}^{\phi}$ (referred to as the worst case performance function) that can be computed recursively over the structure of the STL specification $\phi$, i.e., for predicates, Boolean, and temporal operators.

\textbf{Predicates. } We compute $\bar{\rho}^{\mu}(\hat{z}^{(0)},\tau)$ for predicates $\mu$ as
 \begin{align*}
     \bar{\rho}^{\mu}(\hat{z}^{(0)},\tau)& := 
     \begin{cases}
     h(z_\tau^{(0)}) &\text{ if } \tau\le t\\
     \min_{\bar{z}_\tau\in \mathcal{B}_{\tau|t}} h(\bar{z}_\tau) &\text{ otherwise. }
     \end{cases}
 \end{align*}
  The intuition here is that we know the value of $z_\tau^{(0)}$, and hence also the value of the performance function ${\rho}^{\mu}(z^{(0)},\tau)= h(z_\tau^{(0)})$, at time $\tau\le t$. For times $\tau>t$, we know  that $\text{Prob}(z_\tau^{(0)} \in \mathcal{B}_{\tau|t})\ge 1-\delta$ since the statistical abstraction in \eqref{eq:cp_monitor_} is valid which motivates us to compute $\bar{\rho}^{\mu}(\hat{z}^{(0)},\tau):=\min_{\bar{z}_\tau\in \mathcal{B}_{\tau|t}} h(\bar{z}_\tau)$. As a consequence, it is easy to see that $\bar{\rho}^{\mu}(\hat{z}^{(0)},\tau)$ is a lower bound of ${\rho}^{\mu}(z^{(0)},\tau)$ with probability no less than $1-\delta$, i.e., that 
  \begin{align*}
      \text{Prob}(\rho^\mu(z^{(0)},\tau)\ge \bar{\rho}^{\mu}(\hat{z}^{(0)},\tau))\ge 1-\delta. 
  \end{align*}

  Solving $\min_{\bar{z}_\tau\in \mathcal{B}_{\tau|t}} h(\bar{z}_\tau)$ can generally be a non-convex optimization problem. For convex predicate functions $h$, however, it is easy to see that  $\min_{\bar{z}_\tau\in \mathcal{B}_{\tau|t}} h(\bar{z}_\tau)$ is a convex optimization problem that can efficiently  be solved. For Lipschitz continuous functions $h$ with Lipschitz constant $L_h$, we obtain an efficient lower bound as
  \begin{align*}
      \min_{\bar{z}_\tau\in \mathcal{B}_{\tau|t}} h(\bar{z}_\tau)\ge h(\hat{z}_{\tau|t})-L_hC_{\tau|t}.
  \end{align*}

  \textbf{Boolean and temporal operators. } We use the probabilistic lower bound $\bar{\rho}^{\mu}(\hat{z}^{(0)},\tau)$ to recursively derive probabilistic lower bounds for Boolean and temporal operators. This requires  that the specification $\phi$ contains no negations as deriving a lower bound for $\rho^{\neg\mu}(z^{(0)},\tau)$ would require knowledge of an upper bound for $\rho^{\mu}(z^{(0)},\tau)$, see Sidebar \textbf{LEAS Specifications in Signal Temporal Logic}. This is without loss of generality as every STL specification $\phi$ can be re-written into an equivalent STL specification without negations by instead using disjunction ($\vee$), eventually ($F_I$), and always operators ($G_I$), see \cite{sadraddini2015robust} for details. 

\begin{figure*}
\centering
\begin{subfigure}[t]{0.62\textwidth}
    \includegraphics[width=\textwidth]{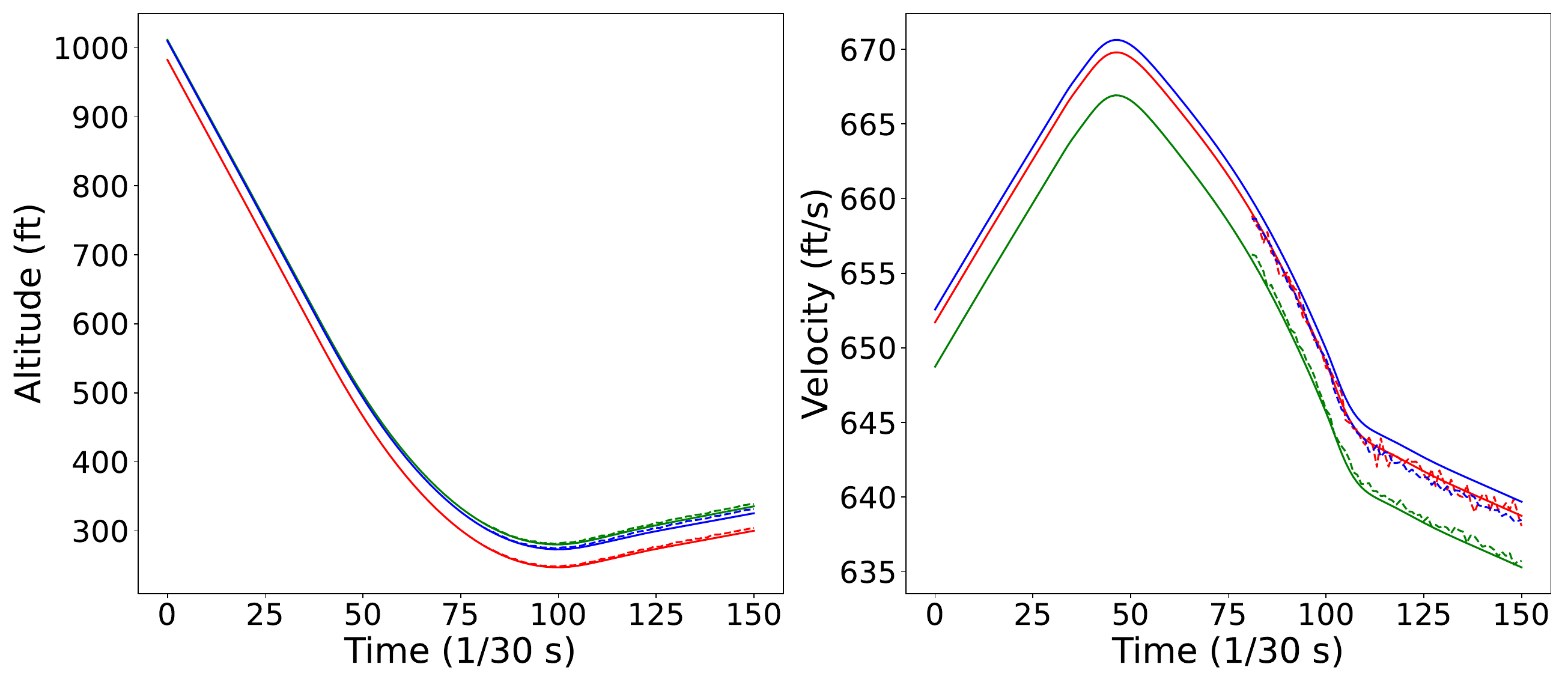}
    \caption{Example Trajectories from $\mathcal{D}$.}
    \label{fig:trajectories_example_5}
\end{subfigure}
\hspace{8.4mm}
\begin{subfigure}[t]{0.30\textwidth} 
    \includegraphics[width=\textwidth]{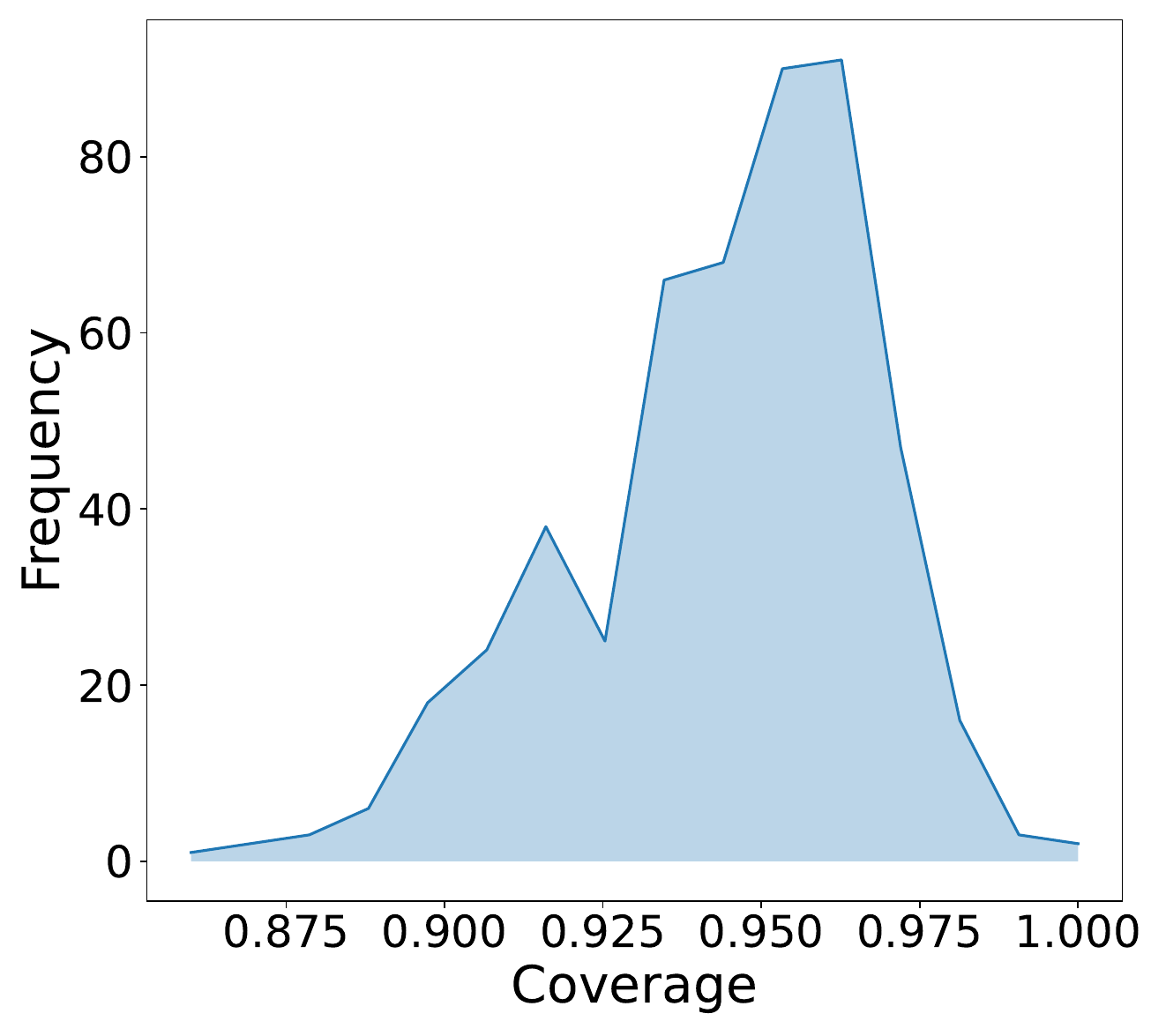}
    \caption{Histogram of $\rho^\phi(z^{(0)}) \ge \rho^*$ over all $N$ experiments for the accurate method.}
    \label{fig:direct_coverage_example_5}
\end{subfigure}
\hspace{2mm}
\begin{subfigure}[t]{0.31\textwidth}
    \includegraphics[width=\textwidth]{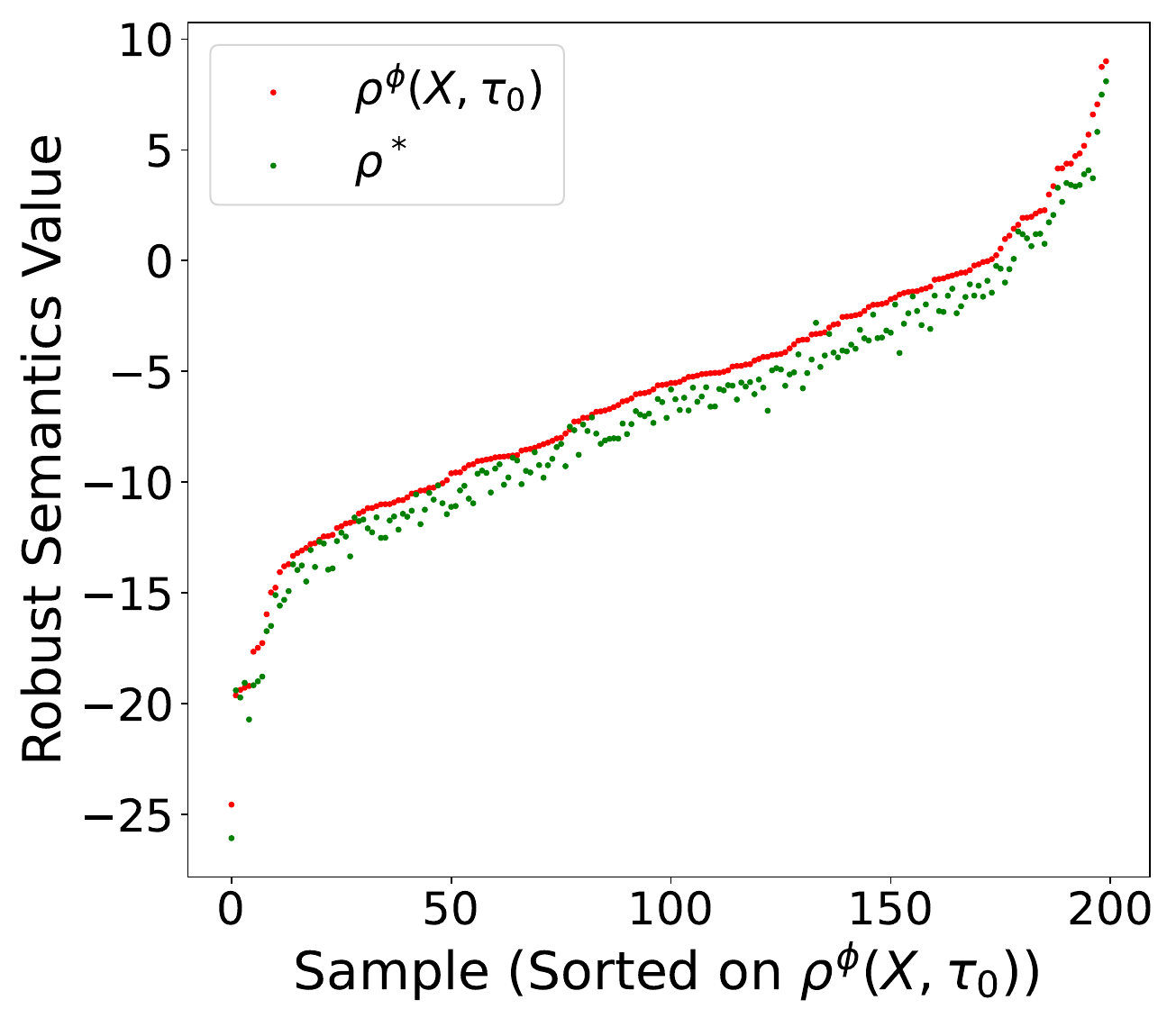}
    \caption{Ground truth performance $\rho^\phi(z)$ and lower performance bound $\rho^*$ from the accurate method.}
    \label{fig:direct_robustness_example_5}
\end{subfigure}
\hspace{2mm}
\begin{subfigure}[t]{0.31\textwidth}
    \includegraphics[width=\textwidth]{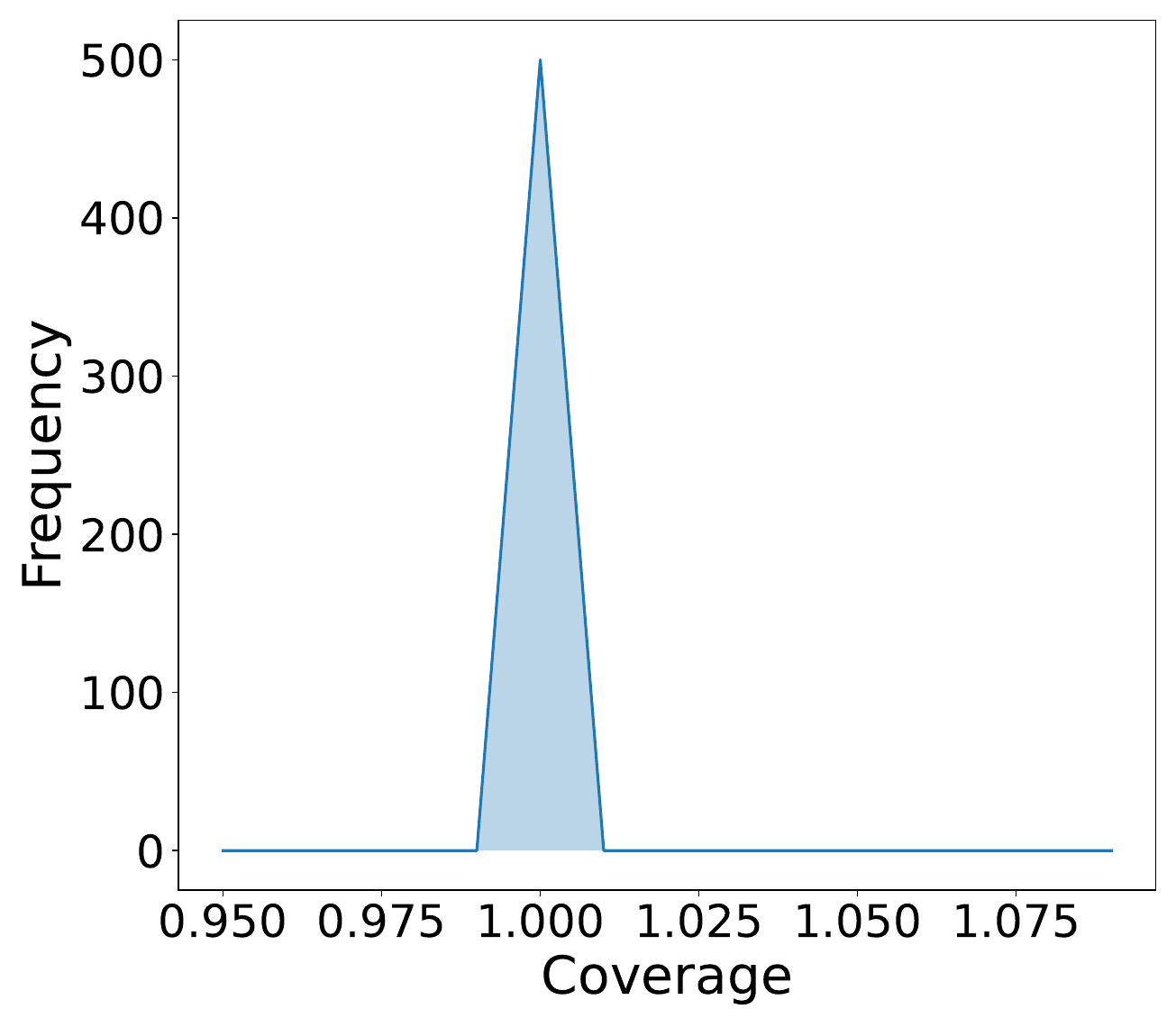}
    \caption{Histogram of $\rho^\phi(z^{(0)}) \ge \rho^*$ over all $N$ experiments for the interpretable method.}
    \label{fig:indirect_coverage_example_5}
\end{subfigure}
\hspace{2mm}
\begin{subfigure}[t]{0.31\textwidth}
    \includegraphics[width=\textwidth]{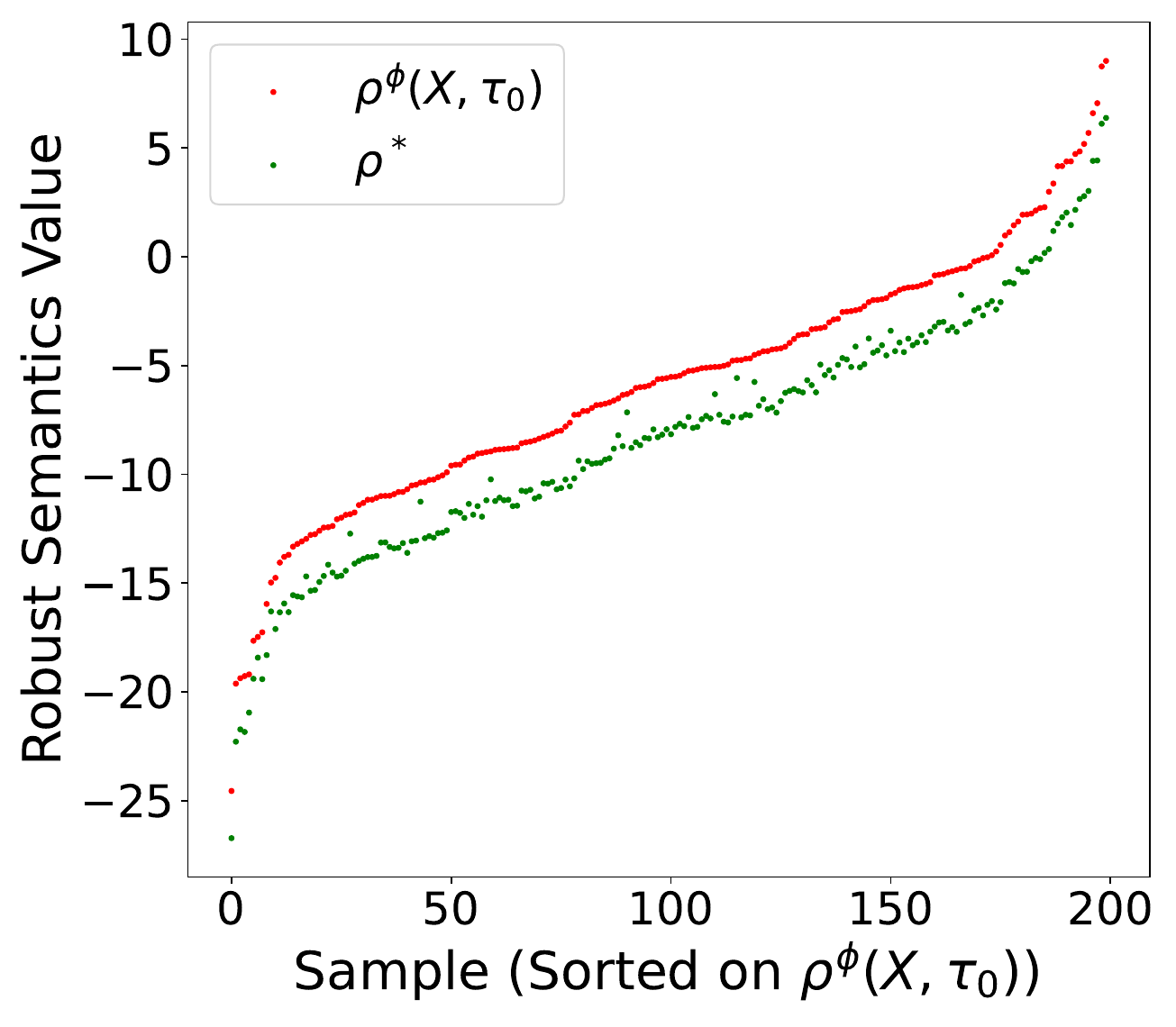}
    \caption{Ground truth performance $\rho^\phi(z)$ and lower performance bound $\rho^*$ from the interpretable method.}
    \label{fig:indirect_robustness_example_5}
\end{subfigure}
\caption{Experimental Results for Example \ref{ex:prv}}
\end{figure*}
  
  The worst case performance function for temporal and Boolean operators is defined in the same way the performance function is defined. Specifically, following Sidebar \textbf{LEAS Specifications in Signal Temporal Logic}, we have
  \begin{align*}
	\bar{\rho}^{\top}(\hat{z},t)& := \infty,\\
	\bar{\rho}^{\phi' \wedge \phi''}(\hat{z},t) &:= 	\min(\bar{\rho}^{\phi'}(\hat{z},t),\bar{\rho}^{\phi''}(\hat{z},t)),\\
     \bar{\rho}^{\phi' \vee \phi''}(\hat{z},t) &:= 	\max(\bar{\rho}^{\phi'}(\hat{z},t),\bar{\rho}^{\phi''}(\hat{z},t)),\\
     \bar{\rho}^{F_I \phi}(\hat{z},t) &:= \underset{t'\in (t\oplus I)\cap\mathbb{N}}{\text{max}}  \bar{\rho}^{\phi}(\hat{z},t')\\
     \bar{\rho}^{G_I \phi}(\hat{z},t) &:= \underset{t'\in (t\oplus I)\cap\mathbb{N}}{\text{min}}  \bar{\rho}^{\phi}(\hat{z},t')\\
	\bar{\rho}^{\phi' U_I \phi''}(\hat{z},t) &:= \underset{t''\in (t\oplus I)\cap\mathbb{N}}{\text{max}}  \min(\bar{\rho}^{\phi''}(\hat{z},t''),\underset{t'\in [t,t'']\cap\mathbb{N}}{\text{min}}\bar{\rho}^{\phi'}(\hat{z},t') ).
	\end{align*}

We remark that the computation of the worst case performance function for Boolean and temporal operators is computationally extremely efficient, e.g., it does not involve solving an optimization problem as in the case of predicates. By construction, the value of $\bar{\rho}^{\phi}(\hat{z}):=\bar{\rho}^{\phi}(\hat{z},0)$ gives us a probabilistic lower bound of ${\rho}^{\phi}({z})$.
 \begin{theorem}\label{thm:6}
 Given an STL specification $\phi$ with sound performance function $\rho^{\phi}:\mathbb{R}^{Tn_z}\to \mathbb{R}$, a valid statistical abstraction of the form \eqref{eq:cp_monitor_}, and a test trajectory $z^{(0)}$  obtained via the dynamics in \eqref{eq:LEAS_model__} from samples $(z^{(0)}_0,\{v^{(0)}_t\}_{t=0}^{T-1},\{w^{(0)}_t\}_{t=0}^{T-1})\sim \mathcal{D}$, then we have that  
    \begin{align*}
        \text{Prob}({\rho}^{\phi}({z}^{(0)})\ge \bar{\rho}^{\phi}(\hat{z}^{(0)}))\ge 1-\delta.
    \end{align*}
\end{theorem}

The interpretable predictive online verification algorithm hence solves Problem \ref{prob3} with $\rho^*:= \bar{\rho}^{\phi}(\hat{z})$.

\textbf{Comparing the presented predictive online verification algorithms. } In comparison with the accurate predictive online verification algorithm, the interpretable predictive online verification algorithm can be more conservative when the statistical abstraction in \eqref{eq:cp_monitor_} is conservative. Another reason why the interpretable version may be more conservative is due to the computation of $\bar{\rho}^{\mu}(\hat{z}^{(0)},\tau)$ which involves a worst case argument over the statistical abstraction for all predicates $\mu$ within $\phi$ and for all times $\tau>t$. Conservatism may also arise when a nonconvex optimization problem needs to be solved for computing $\bar{\rho}^{\mu}(\hat{z}^{(0)},\tau)$. On the other hand, the values of $\bar{\rho}^{\mu}(\hat{z}^{(0)},\tau)$ provide interpretability. We illustrate both predictive online verification algorithms in an example next.

\begin{example}\label{ex:prv}
    Consider the aircraft simulator from \cite{heidlauf2018verification} in which a highly nonlinear F-16 aircraft  with its controller is modeled by a $16$-dimensional state $z_t$. For verification, we focus on the altitude $z^h_t$ (in ft) and the  speed $z^s_t$ (in ft/s). The aircraft trajectories are generated from the simulator with initial conditions drawn from a normal distribution, i.e.,  $(z^h_0,z^s_0) \sim \mathcal{N}(1000, 10^2)\times \mathcal{N}(650, 5^2)=:\mathcal{D}$. We  consider the specification
    \begin{align*}
        \phi \coloneqq G_{[0, T]}(z^h \ge \zeta_1 \wedge(z^h < \zeta_2 \implies z^s \le \zeta_3))
    \end{align*}
    where $T \coloneqq 150$, $\zeta_1 \coloneqq 100$, $\zeta_2 \coloneqq 300$, and $\zeta_3 \coloneqq 650$. In words, this specification requires that the aircraft (1)  always has a safe altitude ($z^h \ge \zeta_1$), and (2) maintains a sufficiently low speed ($z^s \le \zeta_3$) whenever it is close to the ground ($z^h < \zeta_2$). We next consider online verification at time $t \coloneqq 80$. 
    
    We start by training a long short-term memory network. For illustration, we show $3$ sampled trajectories (in solid lines) and their predictions (in dashed lines) in Figure \ref{fig:trajectories_example_5}.  We then set the failure probability to $\delta \coloneqq 0.05$. To verify  statistical validity empirically, we conduct $N \coloneqq 500$ experiments with $K:=700$ calibration trajectories and $J:=200$ test trajectories each. 

    \textbf{Accurate method. } We use the nonconformity score $R^{(i)}$ as per equation \eqref{eq:prv_direct}. By this choice, we know that $\text{Prob}(\rho^\phi(z^{(0)}) \ge \rho^\phi(\hat{z}^{(0)}) - C)\ge 1-\delta$ from Theorem \ref{thm:5}.  We then compute the empirical coverage $EC$ according to equation \eqref{eq:emp_coverage} as $0.934$. We also plot the histogram of the conditional empirical coverage $CEC_n$ according to equation \eqref{eq:cond_emp_coverage} over all $N$ experiments in Figure \ref{fig:direct_coverage_example_5}. Note that $CEC_n$ is equivalent to the conditional empirical coverage of $\rho^\phi(z^{(0)}) \ge \rho^\phi(\hat{z}^{(0)}) - C$.\footnote{Specifically, note that $CEC_n$ in equation \eqref{eq:cond_emp_coverage} is equivalent to  $\sum_{j=1}^{J} \mathbb{1}\big(\rho^\phi(z^{(0)}_{nj} \ge \rho^\phi(\hat{z}_{nj}^{(0)})- C_n\big)/{J}$
    where ${z}_{nj}^{(0)}$ is the $j$th test trajectory of the $n$th experiment, while $C_n$ is as in \eqref{eq:cond_emp_coverage}.}  As expected, the histogram centers around $1 - \delta$.  For one experiment, we also show in Figure \ref{fig:direct_robustness_example_5} the ground truth performance $\rho^\phi(z^{(0)})$ with the associated lower performance bound $\rho^* := \rho^\phi(\hat{z}^{(0)}) - C$ for all test trajectories $z^{(0)}$.  It can be seen that the lower bound is tight, specifically in comparison with the interpretable method that we present next.
    
    \textbf{Interpretable method. } We use a statistical abstraction of the form \eqref{eq:cp_monitor_} and obtain the bounds $C_{\tau|t}$ by following the ``single nonconformity score approach'' discussed in Section \textbf{Statistical Abstractions of Dynamic Environments}. We also define the worst case performance function $\bar{\rho}^\phi(\bar{z}^{(0)})$ as instructed before Theorem \ref{thm:6}. By this choice, we know that $\text{Prob}(\rho^\phi(z^{(0)}) \ge \bar{\rho}^\phi(\hat{z}^{(0)}))\ge 1-\delta$ from Theorem \ref{thm:6}.  To empirically validate this statement, we compute the conditional empirical coverage of $\rho^\phi(z^{(0)}) \ge \bar{\rho}^\phi(\hat{z}^{(0)})$, which is now not equivalent to $CEC_n$  in equation \eqref{eq:cond_emp_coverage} (unlike in the case for the accurate method).\footnote{Formally, we compute $\sum_{j=1}^{J} \mathbb{1}\big(\rho^\phi(z_{nj}^{(0)}) \ge \bar{\rho}^\phi(\hat{z}_{nj}^{(0)})\big)/{J}$.} We plot the corresponding histogram over all $N$ experiments in Figure \ref{fig:indirect_coverage_example_5}. For one experiment, we also show in Figure \ref{fig:indirect_robustness_example_5}  the ground truth performance $\rho^\phi(z^{(0)})$ with the associated lower performance bound $\rho^* := \bar{\rho}^\phi(\hat{z}^{(0)})$ for all test trajectories $z^{(0)}$. As mentioned before, the intepretability of the method comes at the cost of conservatism, visible here by coverage much larger than $1-\delta$ and a lower bound that is not as tight as for the accurate method. 
    
    

\end{example}

\begin{figure*}
\centering
\begin{subfigure}[t]{0.31\textwidth}
    \includegraphics[width=\textwidth]{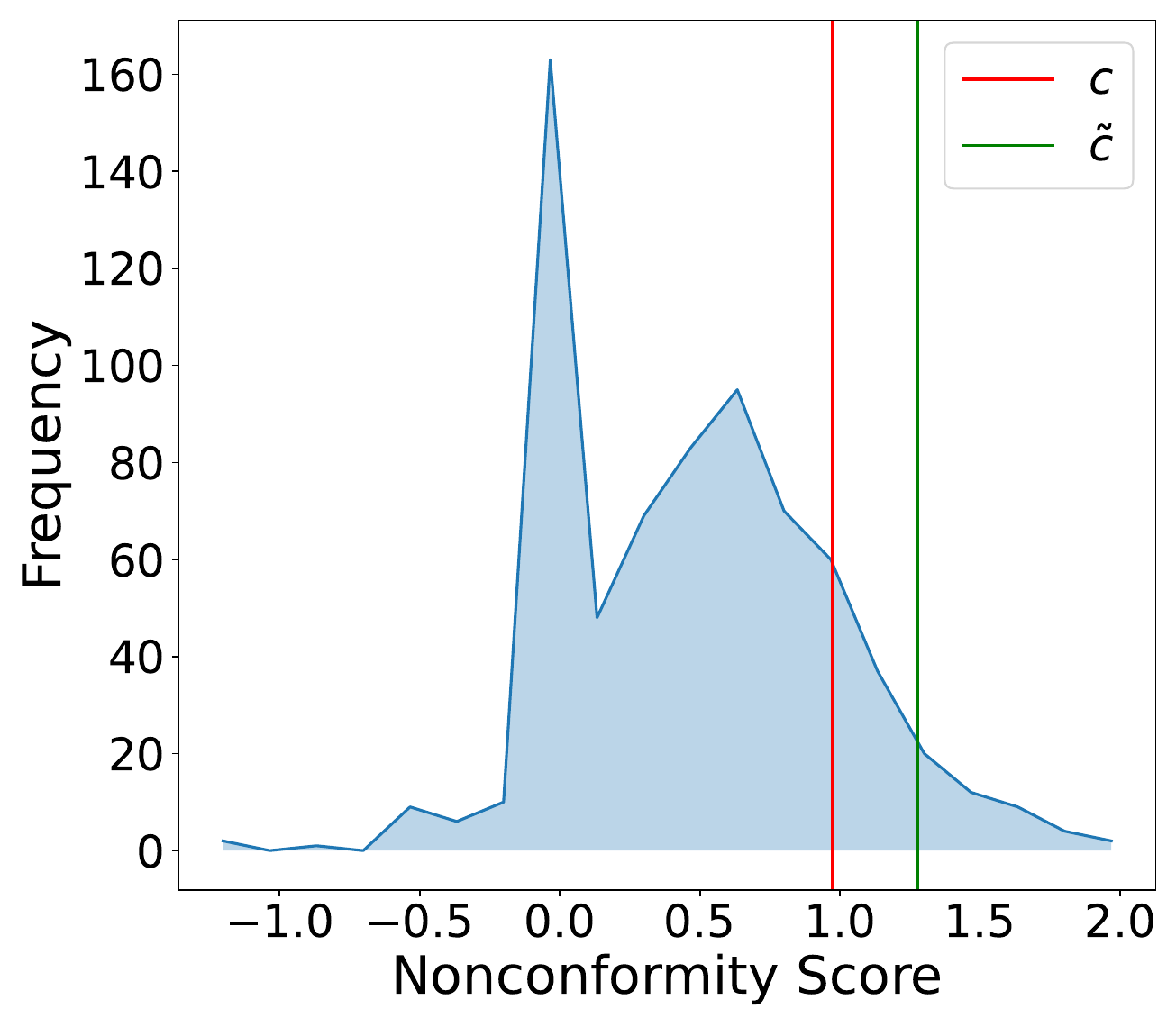}
    \caption{Histogram  of the nonconformity scores $R^{(i)}$ in \eqref{eq:prv_direct}.}
    \label{fig:direct_nonconformities_example_6}
\end{subfigure}
\hspace{2mm}
\begin{subfigure}[t]{0.31\textwidth} 
    \includegraphics[width=\textwidth]{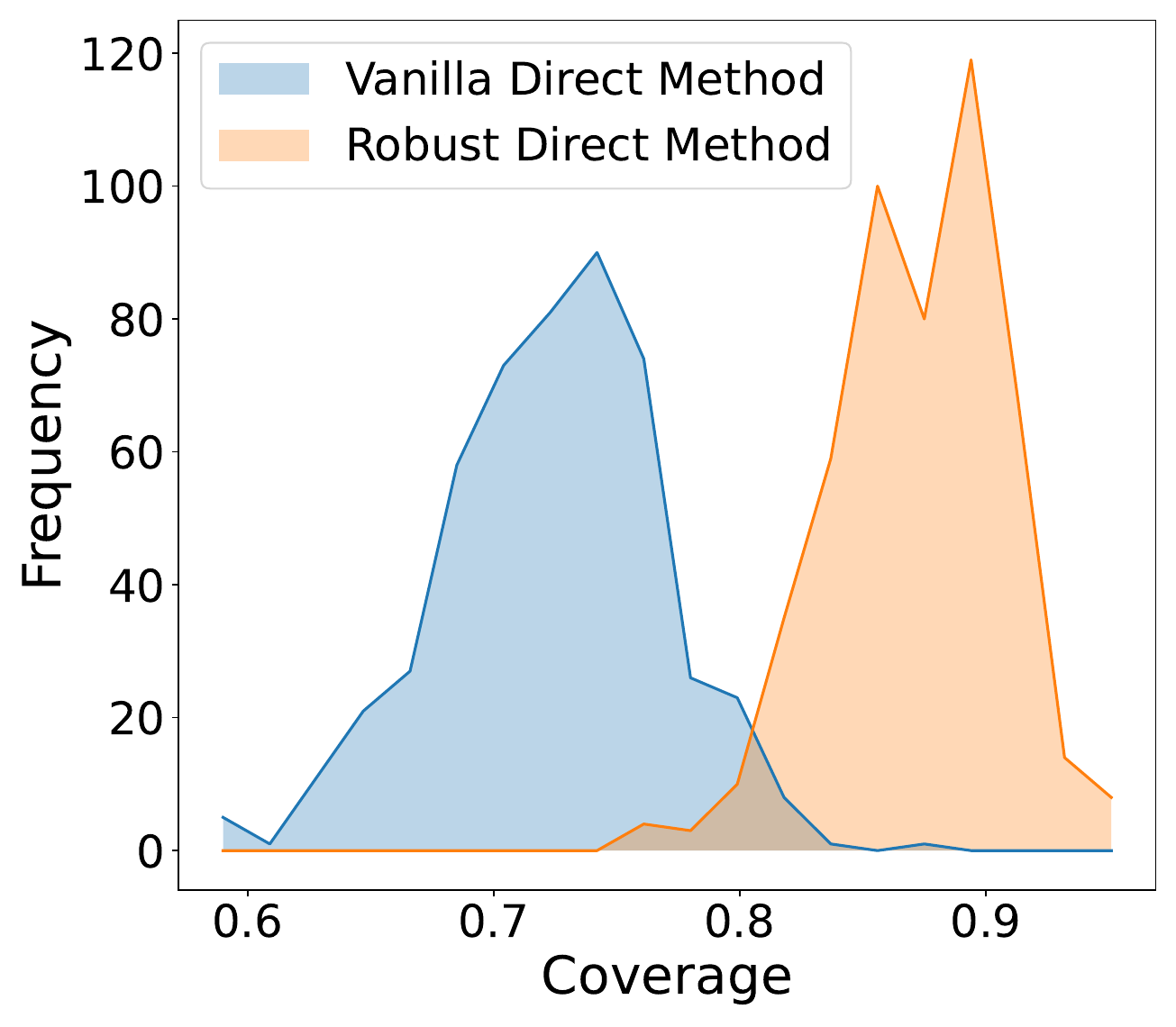}
    \caption{Histogram of $\rho^\phi(z^{(0)}) \ge \rho^*$ over all $N$ experiments for the non-robust and robust accurate methods.}
    \label{fig:direct_coverage_example_6}
\end{subfigure}
\hspace{2mm}
\begin{subfigure}[t]{0.31\textwidth}
    \includegraphics[width=\textwidth]{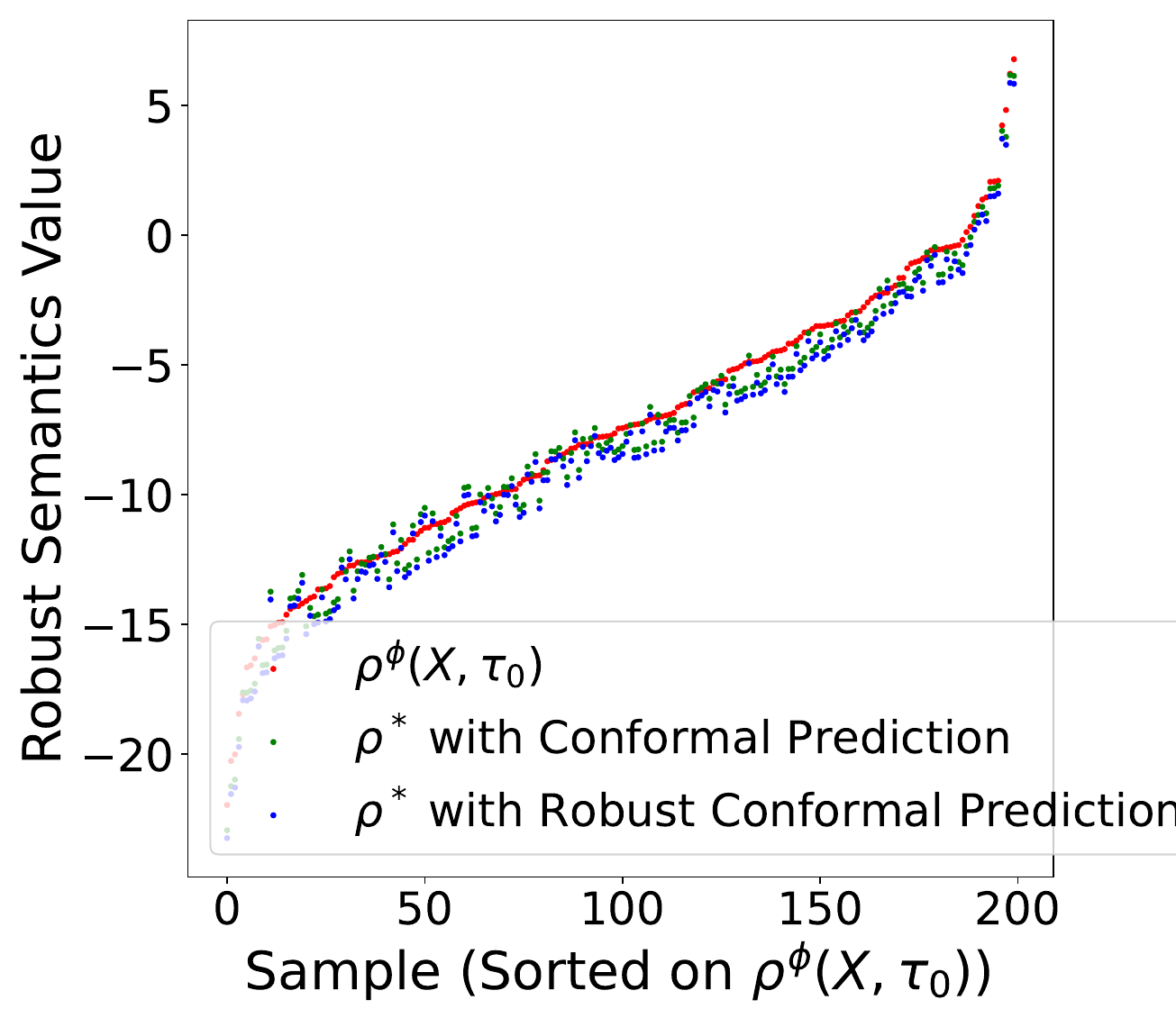}
    \caption{Ground truth performance $\rho^\phi(z)$ and lower performance bound $\rho^*$ from the accurate methods.}
    \label{fig:direct_robustness_example_6}
\end{subfigure}
\hspace{2mm}
\begin{subfigure}[t]{0.31\textwidth}
    \includegraphics[width=\textwidth]{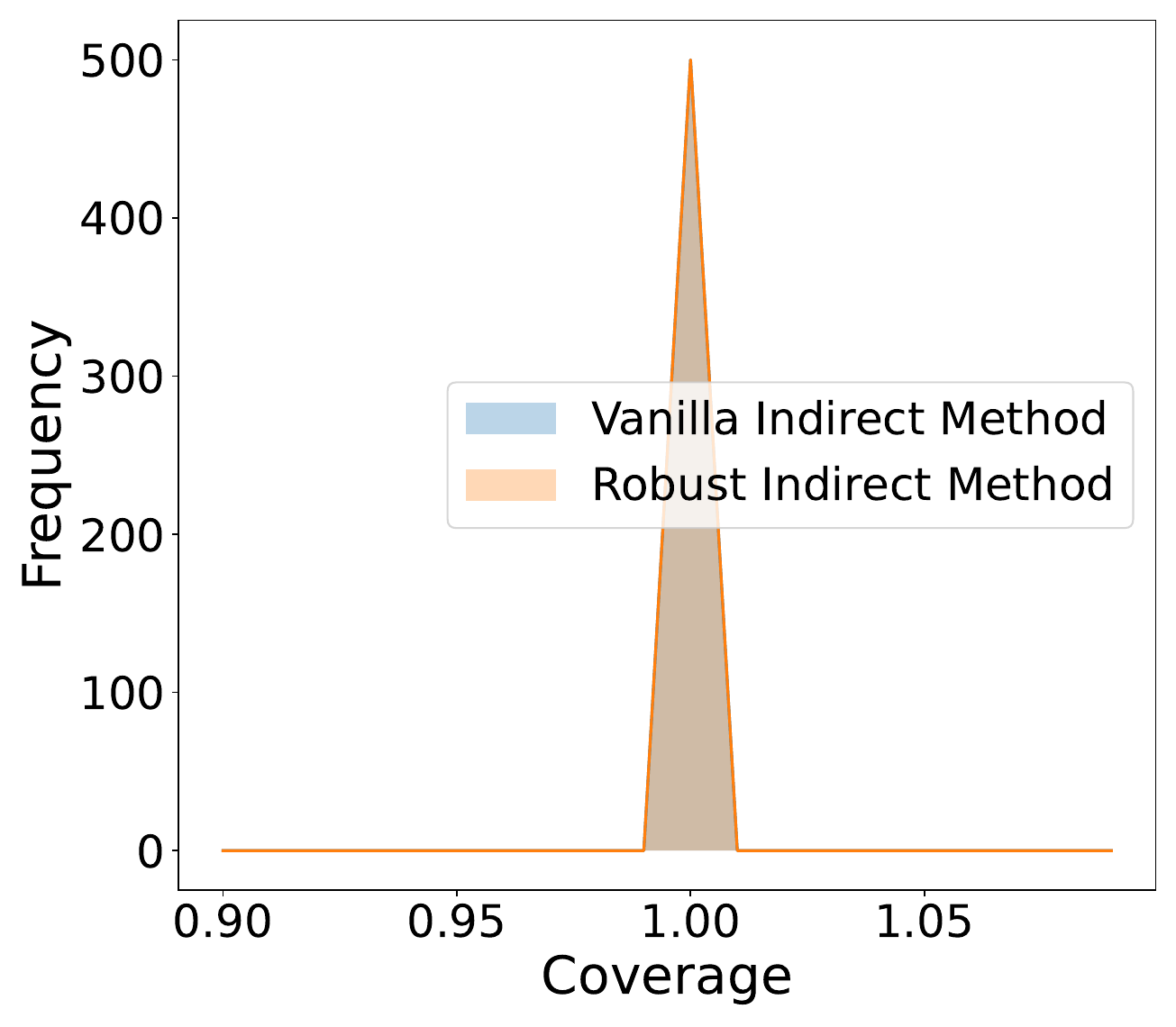}
    \caption{Histogram of $\rho^\phi(z^{(0)}) \ge \rho^*$ over all $N$ experiments for the non-robust and robust interpretable method.}
    \label{fig:indirect_coverage_example_6}
\end{subfigure}
\hspace{2mm}
\begin{subfigure}[t]{0.31\textwidth}
    \includegraphics[width=\textwidth]{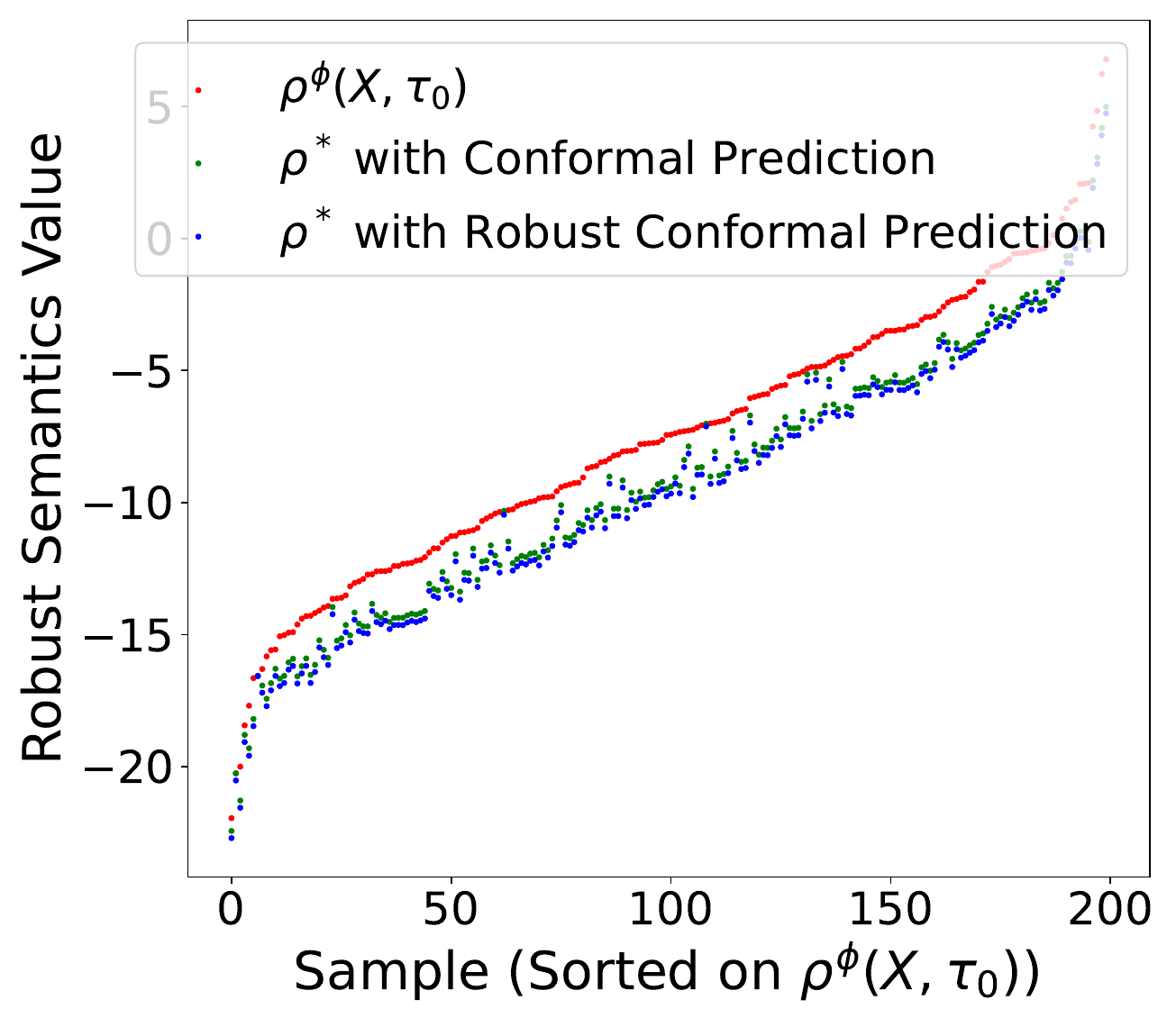}
    \caption{Ground truth performance $\rho^\phi(z)$ and lower performance bound $\rho^*$ from the interpretable methods.}
    \label{fig:indirect_robustness_example_6}
\end{subfigure}
\caption{Experimental Results for Example \ref{ex:robust_prv}}
\end{figure*}

\subsection{Predictive Online Verification under Distribution Shift}

Both predictive online verification algorithms rely on a calibration dataset of trajectories $z^{(1)},\hdots,z^{(K)}$ as per Assumption \ref{ass4}. In practice, these trajectories are typically generated by a simulator or a digital twin model. A simulator implicitly models a distribution over trajectories. However, the system that is deployed in the real world may encounter a different distribution over trajectories due to the sim2real gap, i.e., the actual distribution of  trajectories in a deployment setting may be shifted from the distribution assumed by the simulator. Sometimes, one may be able to collect trajectories from the real system. However, the distribution of these trajectories may still be different from the  distribution encountered during deployment, e.g., when the system encounters new scenarios. To account for such distribution shifts, we follow \cite{zhao2024robust} and construct predictive online verification algorithms  by using robust conformal prediction, recall Sidebar \textbf{Conformal Prediction under Distribution Shift}. 

\textbf{Quantifying distribution shifts. }Let $z^{(1)},\hdots,z^{(K)}\sim\mathcal{Z}$ be trajectories sampled from a calibration distribution $\mathcal{Z}$, i.e., as in Assumption \ref{ass4} where the calibration trajectory $z^{(i)}$ is generated by the dynamics in \eqref{eq:LEAS_model__} from an input sample $(z^{(i)}_0,\{v^{(i)}_t\}_{t=0}^{T-1},\{w^{(i)}_t\}_{t=0}^{T-1})\sim \mathcal{D}$. In contrast, let $z^{(0)}\sim\mathcal{Z}_0$ now be a test trajectory sampled from a deployment distribution $\mathcal{Z}_0$, i.e., the trajectory $z^{(0)}$ is generated by dynamics different from \eqref{eq:LEAS_model__} and an input sample $(z^{(0)}_0,\{v^{(0)}_t\}_{t=0}^{T-1},\{w^{(0)}_t\}_{t=0}^{T-1})\sim \mathcal{D}_0$ where $\mathcal{D}_0$ may also not be the same as $\mathcal{D}$. The starting point in \cite{zhao2024robust} is an assumption on the difference between trajectories from calibration and deployment distribution.
\begin{assumption}\label{ass5}
 	The deployment and calibration distributions $\mathcal{Z}_0$ and $\mathcal{Z}$ are such that $D_f(\mathcal{Z}_0,\mathcal{Z})\le \epsilon$ where $D_f(\cdot)$ is an $f$-divergence and $\epsilon>0$ measures the distribution shift. 
 \end{assumption}

Assumption \ref{ass5} means that the distribution of the test trajectory $z^{(0)}\sim \mathcal{Z}_0$ is unknown but $\epsilon$-close to the distribution of the calibration trajectories $z^{(1)},\hdots,z^{(K)}\sim\mathcal{Z}$. Closeness here is defined in terms of an $f$-divergence, e.g., the total variation distance or the KL divergence. We remark that similar assumptions on the knowledge of an upper bound of the closeness of systems is common in robust control, see e.g., \cite{freeman2008robust,zhou1996robust}. In practice, the distribution shift $\epsilon$ can either be estimated empirically, see e.g., \cite{cauchois2024robust}, or it can be a tuning parameter that induces robustness margins.

\textbf{Robust predictive online verification. } We are now interested in solving Problem \ref{prob3} under Assumption \ref{ass5}. Let thus $R^{(i)}$ be a nonconformity score, i.e., a general function of $z^{(i)}$. As a consequence, there exist a calibration and a deployment distribution $\mathcal{R}$ and $\mathcal{R}_0$ such that $R^{(i)}\sim\mathcal{R}$ for $i\in\{1,\hdots, K\}$ and $R^{(0)}\sim\mathcal{R}_0$. Importantly, it holds that $D_f(\mathcal{Z}_0,\mathcal{Z})\le \epsilon$ implies $D_f(\mathcal{R}_0,\mathcal{R})\le \epsilon$ by the data processing inequality. For this reason, we can obtain an accurate and an interpretable predictive online verification algorithm in the same way as before, but now using robust conformal prediction. We summarize these results next, starting with the accurate verification algorithm.
\begin{corollary}\label{cor3}
        Given an output property $\phi$ with sound performance function $\rho^{\phi}:\mathbb{R}^{Tn_z}\to \mathbb{R}$ and a test trajectory $z^{(0)}\sim\mathcal{Z}_0$ that satisfies Assumption \ref{ass5}, then we have that  
    \begin{align*}
        \text{Prob}(\rho^{\phi}(z^{(0)})\ge \rho^{\phi}(\hat{z}^{(0)})-\tilde{C})\ge 1-\delta
    \end{align*}
    where $\tilde{C}:=\text{Quantile}_{1-\tilde{\delta}}( R^{(1)}, \hdots, R^{(K)})$ as in equation \eqref{eq:C_tilde} with $R^{(i)}$ being defined in \eqref{eq:prv_direct} and $\tilde{\delta}$ being defined in \eqref{eq:tilde_delta}.
\end{corollary}

For the construction of an interpretable predictive online verification algorithm, we first have to construct a valid robust statistical abstraction of the form 
\begin{align}\label{eq:cp_monitor__}
         \text{Prob}(||z_{\tau}^{(0)} &- \hat{z}_{\tau|t}^{(0)}|| \le \tilde{C}_{\tau|t}, \forall \tau \in \{t+1, \dots, T\}) \ge 1-\delta, 
    \end{align}
i.e., \eqref{eq:cp_monitor__} has to hold for any $z^{(0)}\sim\mathcal{Z}_0$ that satisfies Assumption \ref{ass5}. We can compute $\tilde{C}_{\tau|t}$ in the same way as $C_{\tau|t}$ in \eqref{eq:cp_monitor_}, but by replacing $\delta$ with $\tilde{\delta}$ from \eqref{eq:tilde_delta}.
 \begin{corollary}\label{cor4}
      Given an STL specification $\phi$ with sound performance function $\rho^{\phi}:\mathbb{R}^{Tn_z}\to \mathbb{R}$, a valid robust statistical abstraction of the form \eqref{eq:cp_monitor__}, and a test trajectory $z^{(0)}\sim\mathcal{Z}_0$ that satisfies Assumption \ref{ass5}, then we have that  
    \begin{align*}
        \text{Prob}({\rho}^{\phi}({z})\ge \bar{\rho}^{\phi}(\hat{z}))\ge 1-\delta.
    \end{align*}
 \end{corollary}

\begin{example}\label{ex:robust_prv}
    Recall Example \ref{ex:prv} where we obtained calibration trajectories generated from the F-16 aircraft simulator with initial conditions drawn from $\mathcal{N}(1000, 10^2)\times \mathcal{N}(650, 5^2)=:\mathcal{D}$. In contrast, however, we here consider that test trajectories are generated with initial conditions drawn from $\mathcal{N}(998, 10^2)\times \mathcal{N}(651, 5^2)=:\mathcal{D}_0$.  The remaining setting (specification $\phi$, time $t$, predictor $\mu$) is the same as in Example \ref{ex:prv}. We use the total variation distance to measure the distribution shift, i.e., to measure $D_f(\mathcal{D}_0, \mathcal{D})$. We estimate $D_f(\mathcal{D}_0, \mathcal{D})$ from data to be $\approx0.129$, and correspondingly set $\epsilon := 0.129$.\footnote{We refer the reader to \cite{zhao2024robust} for details on how the total variation $D_f(\mathcal{D}_0, \mathcal{D})$ can be estimated from data samples.}
    

In this example, we set the failure probability to $\delta \coloneqq 0.2$. To verify  statistical validity empirically, we  conduct $N \coloneqq 500$ experiments with $K:=700$ calibration trajectories (sampled from $\mathcal{D}$) and $J:=200$ test trajectories (sampled from $\mathcal{D}_0$). Our goals here are to (1) illustrate the correctness of the robust predictive online algorithms from Corollaries \ref{cor3} and \ref{cor4}, and (2) compare to the vanilla  predictive online algorithms from Theorems \ref{thm:5} and \ref{thm:6} which will not achieve valid coverage.

 
\textbf{Accurate method. } We use the nonconformity score $R^{(i)}$ as per equation \eqref{eq:prv_direct}. By this choice, we know that $\text{Prob}(\rho^\phi(z^{(0)}) \ge \rho^\phi(\hat{z}^{(0)}) - \tilde{C})\ge 1-\delta$ from Corollary \ref{cor3} where $\tilde{C}$ follows from equation \eqref{eq:C_tilde} (as opposed to  $C$ following from equation \eqref{eq:C_computation} in the vanilla case). For one experiment, we plot the histogram of the nonconformity scores and the values of $\tilde{C}$ and $C$ in Figure \ref{fig:direct_nonconformities_example_6} for comparison. As expected, we notice that $\tilde{C} > C$. We next plot the histograms of the conditional empirical coverages of $\rho^\phi(z^{(0)}) \ge \rho^\phi(\hat{z}^{(0)}) - \tilde{C}$ and $\rho^\phi(z^{(0)}) \ge \rho^\phi(\hat{z}^{(0)}) - C$ over all $N$ experiments in Figure \ref{fig:direct_coverage_example_6}. It can be seen that the vanilla method cannot achieve $1-\delta$ coverage due to the distribution shift, while the robust method achieves a coverage greater than $1-\delta$,  accounting for the distribution shift. For one experiment, we also show in Figure \ref{fig:direct_robustness_example_6} the ground truth performance $\rho^\phi(z^{(0)})$ with the associated lower performance bound $\rho^* := \rho^\phi(\hat{z}^{(0)}) - \tilde{C}$ for all test trajectories $z^{(0)}$. The lower bound appears tight despite the conservatism inherent to robust methods.

\textbf{Interpretable method. }We use a robust statistical abstraction of the form \eqref{eq:cp_monitor__} and obtain the bounds $\tilde{C}_{\tau|t}$ by following the ``single nonconformity score approach'' discussed in Section \textbf{Statistical Abstractions of Dynamic Environments}, but now using robust conformal prediction. We again define the worst case performance function $\bar{\rho}^\phi(\bar{z}^{(0)})$ as instructed before Theorem \ref{thm:6}. By this choice, we know that $\text{Prob}(\rho^\phi(z^{(0)}) \ge \bar{\rho}^\phi(\hat{z}^{(0)}))\ge 1-\delta$ from Corollary \ref{cor4}.  To empirically validate this statement, we compute the conditional empirical coverage of $\rho^\phi(z^{(0)}) \ge \bar{\rho}^\phi(\hat{z}^{(0)})$ for the robust and the vanilla method. We plot the corresponding histograms over all $N$ experiments in Figure \ref{fig:indirect_coverage_example_6}. Due to  conservatism of the interpretable method,  valid $1-\delta$ coverage is achieved for both methods. For one experiment, we also show in Figure \ref{fig:indirect_robustness_example_6}  the ground truth performance $\rho^\phi(z^{(0)})$ with the associated lower performance bound $\rho^* := \bar{\rho}^\phi(\hat{z}^{(0)})$ for all test trajectories $z^{(0)}$. The vanilla method provides tighter bounds than the robust method as it does not account for all possible distribution shifts of size $\epsilon$. 
\end{example}

\subsection{Related Work: Online Verification Techniques}

Online verification algorithms check if a trajectory satisfies a temporal logic specification by using nothing more than the part of the trajectory that has already been observed. Standard algorithms, see e.g., \cite{bauer2011runtime,leucker2009brief,cassar2017survey,ho2014online,abate2019monitor}, provide one out of three verification answers: (1) ``satisfied'' if all extensions of the partial trajectory satisfy the specification, (2) ``dissatisfied'' if no extension of the partial trajectory satisfies the specification, and (3) ``inconclusive'' otherwise. Multi-valued temporal logics, such as in \cite{lercher2024using,mascle2020ltl}, extend beyond this setting. For temporal logic specifications that admit a sound performance function $\rho^\phi$, such as signal temporal logic,  there even exist online verification algorithms  that provide a real-valued verification answer, see e.g., \cite{deshmukh2017robust,dokhanchi2014line,bartocci2018specification,jakvsic2018algebraic,visconti2021online,mascle2020ltl}. A real-valued verification answer can be interpreted as the robustness of the partial trajectory. 

Predictive online verification algorithms  use a  model  of the system (which is usually non-deterministic) to obtain less conservative verification answers. Existing algorithms either provide robust satisfaction guarantees for all possible behaviors of the non-deterministic models, see e.g., \cite{yu2024model,ghosh2022offline,abbas2022leveraging}, or probabilistic satisfaction guarantees for stochastic models, see e.g.,  \cite{qin2020clairvoyant,babaee2018predictive,babaee2018mathcal,sistla2011runtime,yoon2019predictive}. The authors in \cite{leucker2012sliding,zhang2012runtime,ma2021predictive} follow  a conceptually similar idea, but integrate model predictions and prediction uncertainties into the semantics of a temporal logic formalism. Predictive online verification algorithms using Bayesian inference approaches were presented in \cite{yoon2021predictive,wilcox2010runtime,jaeger2020statistical}. The aforementioned techniques, however, often only apply to specific system models or do not provide strong probabilistic coverage guarantees as can be obtained with conformal prediction.  

As demonstrated before, conformal prediction can be used to design predictive online verification algorithms that provide probabilistic verification guarantees for LEASs. Conformal prediction was also used  in \cite{cairoli2023conformal} to design predictive online verification algorithms.  The idea in \cite{cairoli2023conformal} is to train a quantile regressor that directly predicts the quantile of the distribution of the performance function $\rho^\phi$. Subsequently, conformalized quantile regression (see Sidebar \textbf{Heteroskedasticity and Conformal Prediction}) is used to capture uncertainty and variability in these predictions. In contrast, in this section we trained trajectory predictors that may generally be harder to obtain and require more engineering effort during training. However, the technique from \cite{cairoli2023conformal} requires to train a new quantile regressor each time the specification $\phi$ changes or a new specification is added. Similarly to the accurate predictive online verification algorithm from Section \textbf{Accurate Predictive Online Verification}, the technique from \cite{cairoli2023conformal} does not provide interpretability in case of failure as opposed to the interpretable predictive online verification algorithms from Section \textbf{Interpretable Predictive Online Verification}. Along the same lines, we briefly point the reader to the work in \cite{zhang2023online} which presents online causation monitoring. Lastly, we  mention the use of conformal prediction for out-of-distribution detection \cite{cai2020real,cai2021inductive,kaur2023codit,bashari2025robust}, anomaly detection \cite{xu2021conformal,laxhammar2014conformal,laxhammar2015inductive}, and failure detection \cite{luo2022sample,sinha2023closing}.

\section{Limitations and Open Problems}
To conclude this  article, we would like to discuss limitations and list open problems that we believe are  important to be addressed in our pursuit of designing safe autonomous systems. This list is by no means exhaustive and  naturally biased by our own views and ideas.

\subsection{Limitations}
\textcolor{black}{\textbf{Scalability. }Conformal prediction involves computing the empirical quantile $\text{Quantile}_{1-\delta}( R^{(1)}, \hdots, R^{(K)}, \infty )$ which, as previously mentioned,  can be done using standard sorting algorithms with  $\mathcal{O}(K \log(K))$ time complexity. While this is   generally a scalable approach, larger datasets (which are needed to achieve a smaller failure probability $\delta$)  lead to larger time complexity which may be limiting in some situations, e.g., when computation is done on embedded hardware. Another bottleneck in practice may be memory and the ability to store large datasets. This can  become particularly challenging when images/videos are used. }

\noindent \textcolor{black}{\textbf{Conservatism. }We recall from Section \textbf{Conformal Prediction in a Nutshell} that conformal prediction provides tight upper and lower bounds for $\text{Prob}(R^{(0)}\le C(R^{(1)},\hdots,R^{(K)}))$ in the sense that 
\begin{align*}
   1-\delta\le  \text{Prob}(R^{(0)}\le C(R^{(1)},\hdots,R^{(K)}))\le 1-\delta+\frac{1}{K+1}
\end{align*}
when a pre-specified nonconformity score $R^{(i)}$ is used. In this survey, we then spend a lot of time on engineering the ``right'' nonconformity score in the hope of obtaining  tight prediction regions for the problem at hand. While the flexibility in designing $R^{(i)}$ is generally an advantage, it is worth recognizing  that we are not guaranteed to obtain the tightest possible prediction region, which motivates further research as discussed in the next section. }

\noindent \textcolor{black}{\textbf{Practical limitations. } There are many real-world challenges that statistical tools in general, and hence also conformal prediction, face. One challenge that we mentioned throughout this survey is related to various forms of distribution shift. While there exist some robust extensions of conformal prediction (see Sidebar \textbf{Conformal Prediction under Distribution Shift}), it is worth noting that all these extensions have their own limitations. Another important challenge in practice is related to proper calibration dataset selection. While companies may have a lot of data, these datasets may not be well curated. Therefore, an important task is in constructing good datasets.}

\subsection{Open Problems}
One major research direction will be the \textbf{tight integration of statistical techniques into the system design} of an autonomous system. Some of the open problems here, as already discussed in this article, are to deal with distribution shifts caused by: (1) sim2real gaps that arise in practice, and (2)  couplings between the controllers and statistical abstractions and hence in dealing with distribution shifts. Another problem of more academic nature is to investigate similarities and differences of the various statistical techniques that we mentioned throughout this survey, and to get a better understanding of when to use which technique. Finally, we want to state our belief  that statistical techniques can be helpful to integrate  foundational and generative models  into the system design.

Another impactful research direction will be to continue \textbf{the design of efficient and accurate statistical and perceptual abstractions}. While we already presented, in great detail,  techniques to construct said abstractions in this article, we believe that there are important improvements  that need to be made. Some of these improvements are the design of statistical abstractions that: (1) integrate contemporary probabilistic predictors (such as diffusion models and decision transformers), (2) accurately capture geometric and temporal pattern of the underlying data-generating distribution, and (3) can  account for multi modality in the data. We also believe that techniques for constructing perceptual abstractions are needed that: (1) are data-efficient, and (2) can provide end-to-end control guarantees. Another important milestone will be the joint design of  statistical and perceptual abstractions.

Another broad research direction is on \textbf{the design of formal verification and control techniques itself}. In practice, autonomous systems are usually organized and designed within layers, e.g., with layers that separately perform behavior planning,  path planning, motion planning, and feedback control. It is unclear in which layers we should be using statistical techniques, and if we should use the same statistical and perceptual abstractions across layers or if individual abstractions should be tailored to specific layers. There are many other open technical questions that are important in practice, but were not addressed in this survey article. How can we guarantee optimality and recursive feasibility of control algorithms? How can we design formal verification and control techniques for long term autonomy, e.g., to accommodate asymptotic properties such as stability and forward invariance? 

The next important research direction that we would like to mention is the \textbf{design of learning-enabled multi-agent systems}. While learning-enabled single-agent systems are fairly well understood by now, it is unclear how to design learning-enabled multi-agent systems due to their size, complex network structure, and data dependencies between agents. New specification languages may be needed to formulate safety requirements, while at the same time the need for scalable and distributed verification and control techniques is apparent. While statistical techniques appear promising towards obtaining scalability, it is unclear how statistical techniques can be distributed.

Lastly, we believe that it will be critical for us as a community to \textbf{showcase practical impact across different application domains  and within other communities}. To achieve this goal, it will be important to provide well-documented and re-usable  toolboxes. We also think that the design of photorealistic and high-fidelity simulators plays a critical role towards accomplishing impressive real-world demonstrations and experiments.

\section{Acknowledgments}
Lars Lindemann, Yiqi Zhao, Xinyi Yu, and Jyotirmoy V. Deshmukh were supported in part by the National Science Foundation through the grants SLES-2417075, SHF-2048094, CNS-1932620, CNS-2039087, FMitF-1837131, CCF-SHF-1932620, Toyota R\&D and Siemens Corporate Research through the USC Center for Autonomy and AI, an Amazon Faculty Research Award, and the Airbus Institute for Engineering Research. George J. Pappas was supported in part by the National Science Foundation through the grant SLES-2331880 and the Air Force Office of Scientific Research through the HYDRA program.

\section{Author Information}
\begin{IEEEbiography}
{\textbf{Lars Lindemann}} is currently an Assistant Professor for Algorithmic Systems Theory in the Automatic Control Laboratory at ETH Zürich. From 2023 to 2025 he was an Assistant Professor in the Thomas Lord Department of Computer Science at the University of Southern California. From 2020 to 2022 he was a Postdoctoral Fellow in the Department of Electrical and Systems Engineering at the University of Pennsylvania. He received his Ph.D. degree in Electrical Engineering from KTH Royal Institute of Technology in 2020.  His research interests include systems and control theory, formal methods, machine learning, and autonomous systems. Professor Lindemann received the Outstanding Student Paper Award at the 58th IEEE Conference on Decision and Control and the Student Best Paper Award (as an advisor) at the 60th IEEE Conference on Decision and Control. He was finalist for the Best Paper Award (as an advisor) at the 2024 International Conference on Cyber-Physical Systems, the Best Paper Award at the 2022 Conference on Hybrid Systems: Computation and Control, and the Best Student Paper Award at the 2018 American Control Conference.
\spacing
\end{IEEEbiography}

\begin{IEEEbiography}
{\textbf{Yiqi Zhao}} received his B.S. degree with Honors in Computer Science from Vanderbilt University, Nashville, TN, USA in 2023. He also completed a secondary major in Mathematics and minors in Electrical Engineering and Data Science at Vanderbilt University. He is currently working towards the Ph.D. degree in Computer Science from the University of Southern California, Los Angeles, CA, USA. His research interests include formal methods, cyber physical systems, systems and control theory, and mathematical optimization. He was finalist for the Best Paper Award at the 2024 International Conference on Cyber-Physical Systems.
\spacing
\end{IEEEbiography}

\begin{IEEEbiography}
{\textbf{Xinyi Yu}} received the M.S. degree from Shanghai Jiao Tong University, Shanghai, China, in 2023, and the B.Eng degree from China University of Petroleum (East China), Qingdao, Shandong, China, in 2020, both in Automation. She is pursuing the Ph.D. degree in Computer Science at the University of Southern California, Los Angeles, CA, USA. Her research interests include formal methods and safety control.
\spacing
\end{IEEEbiography}

\begin{IEEEbiography}
{\textbf{George J. Pappas}} received the Ph.D. degree in electrical engineering and computer sciences from the University of California, Berkeley, Berkeley, CA, USA, in 1998. He is currently the Joseph Moore Professor in and the chair of the Department of Electrical and Systems Engineering, University of Pennsylvania, Philadelphia, PA 19104, USA. He also holds a secondary appointment with the Department of Computer and Information Sciences and the Department of Mechanical Engineering and Applied Mechanics. He is a member of the General Robotics, Automation, Sensing, and Perception Lab and the Penn Research in Embedded Computing and Integrated Systems Engineering Center. He was previously the deputy dean for research with the School of Engineering and Applied Science. His research interests include control theory and, in particular, hybrid systems, embedded systems, cyberphysical systems, and hierarchical and distributed control systems, with applications to unmanned aerial vehicles, distributed robotics, green buildings, and biomolecular networks. He was a recipient of various awards, such as the Antonio Ruberti Young Researcher Prize, the IEEE Control Systems Society George S. Axelby Award, the O. Hugo Schuck Best Paper Award, the George H. Heilmeier Award, the National Science Founda- tion Presidential Early Career Award for Scientists and Engineers, and numerous best student papers awards. He is a Fellow of IEEE. 
\spacing
\end{IEEEbiography}

\begin{IEEEbiography}
{\textbf{Jyotirmoy V. Deshmukh}}  is an Associate Professor of Computer Science in the Thomas Lord Department of Computer Science and the co-director of the USC center for Autonomy and AI. Previously he was a Principal Research Engineer at Toyota motors R \& D. He was a postdoctoral research scholar at the University of Pennsylvania and received his Ph.D. from the University of Texas at Austin. 
\spacing
\end{IEEEbiography}

\bibliography{bib_works.bib}
\bibliographystyle{unsrtnat}

\endarticle
\end{document}